\documentclass{article}

\usepackage{etex}

\usepackage[utf8]{inputenc}		
\usepackage[english]{babel}		
\usepackage[T1]{fontenc}		

\usepackage{geometry}	
\usepackage[fleqn]{empheq}	       
\usepackage{mathptmx}                
\numberwithin{equation}{section}
\usepackage{amssymb}                 
\usepackage[fleqn]{cases}             
\usepackage{mathrsfs}                 
\usepackage{bm}                          
\allowdisplaybreaks

\DeclareSymbolFont{calletters}{OMS}{cmsy}{m}{n}
\DeclareSymbolFontAlphabet{\mathcal}{calletters}

\usepackage[nointegrals]{wasysym}
\usepackage{stmaryrd}
\usepackage{esint}			
\usepackage[misc]{ifsym}
\usepackage{extarrows}

\usepackage[margin=1cm, font={small,it}]{caption}
\usepackage[]{graphicx}		
\usepackage{float}
\usepackage[section]{placeins}
\usepackage{subfig}
\usepackage{wrapfig}		
\usepackage{epstopdf}		
\usepackage{xcolor}

\usepackage{enumitem}

\usepackage{array}			
\usepackage{dcolumn}    
\usepackage{booktabs}
\usepackage{tabu}
\usepackage{tabularx}
\usepackage{multirow}

\usepackage{tikz-cd}
\usetikzlibrary{arrows}
\usepackage{extarrows}

\usepackage{siunitx}

\usepackage[authoryear,round]{natbib}

\usepackage{appendix}	

\usepackage[draft]{pdfcomment}

\usepackage{verbatimbox}
\usepackage{xspace}
\usepackage{widetext}

\usepackage{authblk}

\newcolumntype{.}{D{.}{.}{-1}}
\newcolumntype{R}[1]{>{\raggedright\arraybackslash$\displaystyle}p{#1}<{$}}
\newcolumntype{L}[1]{>{\raggedleft\arraybackslash$\displaystyle}p{#1}<{$}}
\newcolumntype{K}[1]{>{\centering\arraybackslash$\displaystyle}p{#1}<{$}} 
\newcolumntype{C}{>{\centering\arraybackslash$\displaystyle}c<{$}}

\newcolumntype{G}[1]{>{\raggedright\arraybackslash }b{#1}}

\providecommand{\operatorname}[1]{\mathop{\mathrm{#1}}\nolimits}

\newcommand{\intervalle}[4]{\mathopen{#1}#2
  \mathclose{}\mathpunct{};#3
  \mathclose{#4}}
\newcommand{\intervalleff}[2]{\intervalle{[}{#1}{#2}{]}}

\newcommand{\ensemblenombre}[1]{\mathbb{#1}}
\newcommand{\N}{\ensemblenombre{N}}
\newcommand{\Z}{\ensemblenombre{Z}}

\newcommand{\R}{\ensemblenombre{R}}

\DeclareMathOperator{\ci}{\imath}

\newcommand{\diff}{\mathop{}\mathopen{}\mathrm{d}}
\newcommand{\derive}[3][\null]{\dfrac{\diff^{#1}{#2}}{\diff{#3}^{#1}}}              
\newcommand{\dpartial}[3][\null]{\dfrac{\partial^{#1}{#2}}{\partial{#3}^{#1}}}      
\newcommand{\lpartial}[3][\null]{\partial^{#1}{#2}/\partial{#3}^{#1}}               

\newcommand{\vecbf}[1]{\mathbf{#1}}           
\newcommand{\vecgbf}[1]{\boldsymbol{#1}} 

\newcommand{\gordre}[1]{\mathcal{O}\mathopen{}\left(#1\right)}

\newcommand{\Jd}{J_{2}\xspace}

\newcommand{\trunc}[1]{\left[#1\right]}   

\newcommand{\cpoiss}[2]{\left\{ #1 ; #2 \right\}}  

\DeclarePairedDelimiterX{\fracc}[2]{\lbrack}{\rbrack}{\dfrac{#1}{#2}}
\DeclarePairedDelimiterX{\fracp}[2]{\lparen}{\rparen}{\dfrac{#1}{#2}}
\DeclarePairedDelimiterX{\fracabs}[2]{\lvert}{\rvert}{\dfrac{#1}{#2}}

\DeclarePairedDelimiter{\abs}{\lvert}{\rvert}
\DeclarePairedDelimiter{\mean}{\langle}{\rangle}
\DeclarePairedDelimiter{\norm}{\lVert}{\rVert}

\newcommand{\mca}[1]{\mathcal{#1}}
\newcommand{\mcat}[1]{\mca{\widetilde{#1}}}

\renewcommand{\epsilon}{\text{\usefont{OML}{cmr}{m}{n}\symbol{15}}}

\newcommand{\anomelliptic}{\fontfamily{cmr}\selectfont \emph{w}}

\newcommand{\dprime}{\prime\prime}

\newcommand{\dstar}{\star\star}

\newcommand{\ctegravm}{\si{\cubic\metre\per\square\second}}

\newcommand{\radpersec}{\si{\radian\per\second}}

\newcommand{\symbtb}{\prime}
\newcommand{\dutb}[2][\null]{{#2}_{}^{\symbtb\,#1}}
\newcommand{\pucedingk}{}

\makeatletter
\def\my@tag@font{\normalsize}
\def\maketag@@@#1{\hbox{\m@th\normalfont\my@tag@font#1}}
\renewcommand{\eqref}[1]{\textup{(\ref{#1})}}
\makeatother

\def\clap#1{\hbox to 2em{\hss#1\hss}} 
\def\mathclap{\mathpalette\mathclapinternal} 
\def\mathclapinternal#1#2{%
  \clap{$\mathsurround=0pt#1{#2}$}%
}

\newlength{\alignoverwidth}
\newlength{\eqwidth}
\newlength{\eqMpxx}
\newlength{\eqMpxxx}
\newcommand{\alignovereqnum}[1]{%
  \settowidth{\alignoverwidth}{\ensuremath{\overset{{}\eqref{#1}{}}{=}}}%
  \ifthenelse{%
    \lengthtest{\alignoverwidth>\eqMpxx}}{%
    \setlength{\alignoverwidth}{\eqMpxxx}}{%
    \setlength{\alignoverwidth}{\eqMpxx}}%
  \hspace*{0.75\alignoverwidth}\hspace*{-1.5\eqwidth}%
  &%
  \settowidth{\alignoverwidth}{\ensuremath{\overset{{}\eqref{#1}{}}{=}}}%
  \ifthenelse{%
    \lengthtest{\alignoverwidth>\eqMpxx}}{%
    \setlength{\alignoverwidth}{\eqMpxxx}}{%
    \setlength{\alignoverwidth}{\eqMpxx}}%
  \ensuremath{\overset{\mathclap{{}\eqref{#1}{}}}{=}}%
  \hspace*{0.75\alignoverwidth}\hspace*{-1.5\eqwidth}}

\makeatletter%
\newcommand*\DetectFormatOption{%
  \if@twocolumn true\else false\fi}

\makeatother%

\newenvironment{fssizeadapt}{
  \ifthenelse{\boolean{\DetectFormatOption}}{%
    \begin{scriptsize}}{%
      \begin{footnotesize}}}{
      \ifthenelse{\boolean{\DetectFormatOption}}{%
      \end{scriptsize}}{%
    \end{footnotesize}}}

\newcommand{\casefontname}[1]
{
  \ifthenelse{\equal{#1}{1}}{footnotesize}{}
  \ifthenelse{\equal{#1}{3}}{footnotesize}{}
  \ifthenelse{\boolean{\DetectFormatOption}}{scriptsize}{}
}

\begin{document}


\setlength\abovedisplayskip{6pt plus 3pt minus 3pt}		
\setlength\abovedisplayshortskip{0pt plus 2pt}				
\setlength\belowdisplayskip{6pt plus 3pt minus 3pt}		
\setlength\belowdisplayshortskip{4pt plus 2pt minus 2pt}	

\delimitershortfall=-1pt

%
%
%
%
%
%

\title{Analytical theory for highly elliptical orbits including time-dependent perturbations}

\author[1]{G.~Lion\thanks{Guillaume.Lion@obspm.fr}}
\author[2]{G.~Métris\thanks{Gilles.Metris@oca.eu}}
\affil[1]{SYRTE, 
	Observatoire de Paris, 
	PSL Research University, 
	CNRS, 
	Sorbonne Universités,
	UPMC Univ. Paris~06, 
	LNE, 
	61 avenue de l’Observatoire, F-75014 Paris, France}
\affil[2]{Géoazur, 
	Université de Nice Sophia-Antipolis, 
	CNRS (UMR 7329), 
	Observatoire de la Côte d’Azur,
	250 rue Albert Einstein, 
	Sophia Antipolis 06560 Valbonne, 
	France}

\renewcommand\Authands{ and }

\date{}

\maketitle


  \begin{abstract}
  
    Traditional analytical theories of celestial mechanics are not well-adapted when dealing with highly elliptical orbits. On the one hand, analytical solutions are quite generally expanded into power series of the eccentricity and so limited to quasi-circular orbits. On the other hand, the time-dependency due to the motion of the third body (e.g. Moon and Sun) is almost always neglected. We propose several tools to overcome these limitations. Firstly, we have expanded the third-body disturbing function into a finite polynomial using Fourier series in multiple of the satellite's eccentric anomaly (instead of the mean anomaly) and involving Hansen-like coefficients. Next, by combining the classical Brouwer-von Zeipel procedure and the time-dependent Lie-Deprit transforms, we have performed a normalization of the expanded Hamiltonian in order to eliminate all the periodic terms. One of the benefits is that the original Brouwer solution for~$\Jd$ is not modified. The main difficulty lies in the fact that the generating functions of the transformation must be computed by solving a partial differential equation, involving derivatives with respect to the mean anomaly, which appears implicitly in the perturbation. We present a method to solve this equation by means of an iterative process. Finally we have obtained an analytical tool useful for the mission analysis, allowing to propagate the osculating motion of objects on highly elliptical orbits (e > 0.6) over long periods efficiently with very high accuracy, or to determine initial elements or mean elements. Comparisons between the complete solution and the numerical simulations will be presented.

  \end{abstract}



\section{Introduction}
\label{sec:Intro}

{\color{black}
Among the \num{15000} objects listed in the NORAD catalog~\footnote{Available on \url{http://satellitedebris.net}}, about~\num{1400} have highly elliptical orbits (HEO) with an eccentricity greater than $0.5$, mainly in the geostationary transfer orbit (GTO). These are satellites, rocket bodies or any kind of space debris.

For several years, the computation of trajectories is very well controlled numerically.  Numerical methods are  preferred mainly for their convenience and accuracy, especially when making comparisons with respect to the observations or their flexibility whatever the perturbation to be treated. Conversely, analytical theories optimize the speed of calculations, allow to study precisely the dynamics of an object or to study particular classes of useful orbits.

However, the calculation of the HEO can still be greatly improved, especially as regards the analytical theories. Indeed, when we are dealing with this type of orbit, we have to face several difficulties. Due to the fact that they cover a wide range of altitudes, the classification of the perturbations acting on an artificial satellite, space debris, etc. \citep[see][]{Montenbruck_2000aa} changes with the position on the orbit. At low altitude, the quadrupole moment $\Jd$ is the dominant perturbation, while at high-altitude the lunisolar perturbations can reach or exceed the order of the $\Jd$ effect. 
}


One of the issues concerns the expansion of the third-body disturbing function in orbital elements. The importance of the lunisolar perturbations in the determination on the motion of an artificial satellite was raised by~\citet{Kozai_1959aa}. Using a disturbing function truncated to the second degree in the spherical harmonic expansion, he showed that certain long-periodic terms generate large perturbations on the orbital elements, and therefore, the lifetime of a satellite can be greatly affected.
Later, \citet{Musen_1961aa} took into account the third harmonic.
\citet{Kaula_1961aa, Kaula_1966aa} introduced the inclination and eccentricity special functions, fundamental for the analysis of the perturbations of a satellite orbit. This enabled him to give in \citeyear{Kaula_1962aa} the first general expression of the third-body disturbing function using equatorial elements for the satellite and the disturbing body; the function is expanded using Fourier series in terms of the mean anomaly and the so-called Hansen coefficients depending on the eccentricity~$e$ in order to obtain perturbations fully expressed in orbital elements. 
It was noticed by \citet{Kozai_1966aa} that, concerning the Moon, it is more suitable to parametrize its motion in ecliptic elements rather than in equatorial elements. Indeed, in this frame, the inclination of the Moon is roughly constant and the longitude of its right ascending node can be considered as linear with respect to time.
In light of this observation, \citet{Giacaglia_1974aa,Giacaglia_1980aa} established the disturbing function of an Earth's satellite due to the Moon's attraction, using the ecliptic elements for the latter and the equatorial elements for the satellite. Some algebraic errors have been noticed in \citet{Lane_1989aa}, but it is only recently that the expression has been corrected and verified in \citet{Lion_2013ab,Lion_2011ac}.

The main limitation of these papers is that they suppose truncations from a certain order in eccentricity. Generally, the truncation is not explicit because there is no explicit expansion in power of the eccentricity. But in practice, Fourier series of the mean anomaly which converge slowly must be truncated and this relies mainly on the D'Alembert rule \citep[see][]{Brouwer_1961aa} which guarantees an accelerated convergence as long as the eccentricity is small. Because this is indeed the case of numerous natural bodies or artificial satellites, these expansions of the disturbing function are well suited in many situations. 
However, for the orbits of artificial satellites having very high eccentricities, any truncation with respect to the eccentricity is prohibited. \citet{Brumberg_1994aa} investigated this situation. They showed that the series in multiples of the elliptic anomaly $\anomelliptic$, first introduced by \citet{Nacozy_1977aa} and studied later by \citet{Janin_1980aa,Bond_1980aa}, converge faster than the series in multiples of any classical anomaly in many cases. This was confirmed by \citet{Klioner_1997aa}. 
Unfortunately, the introduction of the elliptic anomaly increases seriously the complexity, involving in particular elliptical functions \citep[see e.g.][]{Dixon_1894aa}.  
In the same paper, they provided the expressions of the Fourier coefficients $Y_{s}^{n,m}$ and $Z_{s}^{n,m}$ in terms of hypergeometric functions, coming from the Fourier series expansion of the elliptic motion functions in terms of the true anomaly and of the eccentric anomaly, respectively. 
More discussions and examples can be found in~\cite{Brumberg_1999aa}.

On the other hand, the expansion must be supple enough to define a trade-off between accuracy and complexity for each situation. 
To this end, the use of special functions is well suited to build a closed-form analytical model, like in the theory of \citet[][]{De-Saedeleer_2006aa} for a lunar artificial satellite. Development can be compact, easy to manipulate and the extension of the theory can be chosen for each case by fixing the limits on the summations. The complexity is relegated in the special functions, knowing that efficient algorithms exist to compute them. In short, we shall use the expression of the disturbing function introduced in \citet{Lion_2013ab} and \citet{Lion_2011ac}, mixing mainly the compactness of formulation in exponential form and the convergence of series in eccentric anomaly. 

Besides the question of large eccentricities, the other issue concerns the explicit time-dependency due to the motion of the disturbing body. In the classical analytical theory, this is almost always ignored \citep[see e.g.][]{Roscoe_2013aa} while it should be taken into account  when constructing an analytical solution, in particular by means of canonical transformations. To do this, the key point is to start from a disturbing function using angular variables which are time linear. 
This is precisely the motivation to use ecliptic elements instead of equatorial elements for the Moon perturbation, as explained above. 
In this situation, the PDE (Partial Differential Equation) that we have to solve to construct an analytical theory takes the following form:%
\begin{equation}
\sum_{i \ge 0} \omega_i \dpartial{\mca{V}}{\alpha_i} = {}
\mca{A} \cos \Bigl( \sum_{i \ge 0}   k_i \alpha_i \Bigr)
\quad\Rightarrow\quad
\mca{V} = {} 
\frac{\mca{A}}{\mathchar"1358\limits_{i \ge 0} k_i \omega_i} 
\sin \Bigl( \sum_{i \ge 0}  k_i \alpha_i \Bigr) \;.
\end{equation}
Unfortunately, this mechanics is broken as soon as the fast variable of the satellite motion is no longer the mean anomaly $M$, but the eccentric anomaly $E$. In this case, the equation to solve looks like%
\begin{equation}
\label{eq:EDP_generic}
\omega_{0}\dpartial{\mca{V}}{M} + \sum_{i \ge 1} \omega_i \dpartial{\mca{V}}{\alpha_i} = {}
\mca{A} \cos \Bigl( k_{0} E + \sum_{i \ge 1}   k_{i} \alpha_i \Bigr)
\end{equation}
which admits no exact solution.

In this work, we present a closed-form analytical perturbative theory for highly elliptical orbits overcoming all these limitations. Only the $\Jd$ effect and the third-body perturbations will be considered. 
The paper is organized as follows. In Section \ref{sec:Ham_sys}, we define the hamiltonian system and we focus on the development of the third-body disturbing function. In Section~\ref{sec:Lie_Trans_Approach}, we expose the procedure to normalize the system combining the Brouwer's approach and the Lie-Deprit algorithm including the time dependence. Section~\ref{sec:Lie_Transforms_principle} is devoted to the determination of generating functions to eliminate the short and long periodic terms due to the lunisolar perturbations (Moon and Sun). Especially, we will see how to solve PDE such as~\eqref{eq:EDP_generic} by using an iterative process. In Section~\ref{sec:solution_complete}, we present the complete solution to propagate the orbit at any date: transformations between the mean and osculating elements are given. Finally, numerical tests are carried out in Section~\ref{sec:Num_Tests} to evaluate the performances of our analytical solution.

\section{Hamiltonian formalism}
\label{sec:Ham_sys}

\subsection{Dynamical model}
\label{ssec:Dyn_Model}

In an inertial geocentric reference frame~$(x,y,z)$, we consider the perturbations acting on the Keplerian motion of an
artificial terrestrial satellite (or space debris), induced by the quadrupole moment~$\Jd$ of the Earth and the
point-mass gravitational attraction due to the Moon~($\leftmoon$) and Sun~($\astrosun$).

The motion equations of the satellite derived from the potential~$V$:%
\begin{subequations}
  \begin{align}
    \label{eq:Modele_dynamique}
    \vecgbf{\gamma} 
    & = \vecgbf{\nabla} {V}
    \\
    V 
    & = V_{Kep} + \mca{R}_{\Jd} + \mca{R}_{\leftmoon} + \mca{R}_{\astrosun} \;,
  \end{align}
\end{subequations}
where~$\vecgbf{\gamma}$ is the acceleration vector of the satellite, $\vecgbf{\nabla}$ the gradient operator. 
The first two terms of the potential are related to the Earth's gravity field, with~$V_{Kep}$ the Keplerian term:%
\begin{align}
  \label{eq:Kep_term_rec}
  {V}_{Kep} & = \frac{\mu_{\varoplus}}{r} 
  \;,
\end{align}
and~$\mca{R}_{\Jd}$ the disturbing potential due to the Earth oblatness:%
\begin{equation}
  \label{eq:R_J2_sph}
  \mca{R}_{\Jd} 
  = -\frac{\mu_{\varoplus}}{r} \fracp*{R_{\varoplus}}{r}^{2} \Jd P_{2}(\sin{\phi})
  \;,
\end{equation}
where~$r$ is the satellite's radial distance and~$\phi$ its latitude, $\mu_{\varoplus}$ the geocentric gravitational constant, $R_{\varoplus}$ the mean equatorial radius of the Earth and~$P_{n}(x)$ are the Legendre polynomials of degree~$n$ defined for~$x \in \intervalleff{-1}{1}$.

Designating external bodies (i.e. Moon and Sun) by the prime symbol, the third-body disturbing function~$\dutb{\mca{R}}$ is \citep[][]{Plummer_1960aa, Murray_1999aa}:%
\begin{equation}
  \label{eq:def_pot_3c}
  \dutb{\mca{R}} = \dutb{\mu}
  \left(
    \frac{1}{\norm{\dutb{\vecbf{r}}-\vecbf{r}}}
    -  \frac{\vecbf{r}\cdot\dutb{\vecbf{r}}}{\dutb[3]{r}}
  \right)
  \;,
\end{equation}
with~$\dutb{\mu}$ the third-body gravitational constant, $\vecbf{r}$ and~$\dutb{\vecbf{r}}$ respectively the geocentric
position vector of the artificial satellite and the disturbing body, and~$r$ and~$\dutb{r}$ their associated radial distances.
Since we are interested in the orbits such as~$\dutb{r}/r>1$, $\dutb{\mca{R}}$ can be expressed in power series of~$\dutb{r}/r$ as \citep{Plummer_1960aa, Brouwer_1961aa}:%
\begin{equation}
  \dutb{\mca{R}} \equiv {} 
  \frac{\dutb{\mu}}{\dutb{r}} \sum_{n \geq 2}^{} \left( \frac{r}{\dutb{r}}\right)^n P_n(\cos\Psi) \;.
  \label{eq:R_sph}
\end{equation}
where~$\Psi$ is the elongation of the satellite from the disturbing body.

\subsection{Hamiltonian approach}
\label{ssec:H_formalism}

Introducing the osculating orbital elements: $a$ the semi-major axis, $e$ the eccentricity, $I$ the inclination, $\Omega$ the longitude of the ascending node, $\omega$ the argument of perigee and~$M$ the mean anomaly. 
We define the Delaunay canonical variables~$(\vecbf{y}, \vecbf{Y})$ by%
\begin{subequations}
  \label{eq:eq_Delaunay}
  \begin{align}
    \vecbf{y} & 
    = {}
    \left(
      l = M , g = \omega, H = \Omega 
    \right)^\intercal \;,
    \\
    \vecbf{Y} & = {}
    \left( 
      L = \sqrt{\mu\,a} , G = \eta L, H = G \cos I 
    \right)^\intercal \;,
  \end{align}
\end{subequations}
with~$\eta = \sqrt{1-e^{2}}$.

The orbital dynamics of the satellite motion can be described in the Hamiltonian formalism and treated implicitly as a function of the Delaunay elements:%
\begin{align}
  \label{eq:Ham_sys}
  \mca{H} = {}
  & \mca{H}_{Kep} + \mca{H}_{\Jd} + \mca{H}_{\leftmoon} + \mca{H}_{\astrosun}
  \;,
\end{align}
with
\begin{subequations}
  \begin{align}
    & \mca{H}_{Kep} 
    = {-}\frac{\mu_{\varoplus}}{2\,L^2}
    \\
    & \mca{H}_{\Jd} = {-}\mca{R}_{\Jd} 
    \;,\quad 
    \mca{H}_{\leftmoon} = {-}\mca{R}_{\leftmoon} 
    \;,\quad
    \mca{H}_{\astrosun} = {-}\mca{R}_{\astrosun} 
    \;.
  \end{align}
\end{subequations}

\subsubsection{Oblateness disturbing function}

Using a closed-form representation (see Appendix~\ref{anx:J2_astuces}), the classical perturbation~$\mca{R}_{\Jd}$ can be written%
%
  \begin{subequations}
    \label{eq:RJ2}
    \begin{align}
      \mca{R}_{\Jd} 
      &= \sum_{p=0}^{2} \sum_{q=-3}^{3} 
      \mca{A}_{p,q} 
      \cos \left[ (q+2-2p)\nu + (2-2p)\,\omega \right] 
      \;,
      \\
      \mca{A}_{p,q} 
      & = \Jd R_{\varoplus}^{2}\,\omega_{0}^{2}\,Y_{q}^{-3,0} (e)\,F_{2,0,p}(I) 
    \end{align}
  \end{subequations}
\\
where~$Y_{q}^{n,m}(e)$ are the Fourier coefficients defined in \citet{Brumberg_1995aa, Laskar_2005aa}, and~$F_{n,m,p}(I)$ the inclination functions \citep[see e.g.][]{Izsak_1964aa, Gaposchkin_1973aa, Sneeuw_1992aa, Gooding_2010aa}%
\begin{small}
  \begin{align}
    \label{eq:F_Gaposchkin}
    \begin{split}
      F_{n,m,p} (I) & = {} 
      (-1)^{n-m} \frac{(n+m)!}{2^n (n-p)! p!}
      \\ & \hspace{1em} \times
      \sum_{j = jmin}^{jmax} (-1)^{j} \binom{2p}{j} \binom{2n-2p}{n-m-j}
      \cos^{n+b} \frac{I}{2}  \sin^{n-b} \frac{I}{2} 
    \end{split}
  \end{align}
\end{small}
with~$\, b = m-2p+2j$, $jmax = \min(n-m,2p)$ and~$jmin = \max(0,2p-n-m)$.

\subsubsection{Lunar disturbing function}

In order to be easily handled in our analytical theory, we need a general and compact expression of the third-body disturbing function expressed in terms of the osculating orbital elements or equivalent variables. This could be done by using the equation (5) from \citet[][]{Kaula_1962aa} involving equatorial elements for both the satellite and the disturbing body. But, as noticed by \citet{Kozai_1966aa}, it is more suitable to parametrize the Moon's apparent motion in ecliptic elements. Indeed, the inclination of the Moon is roughly constant in the ecliptic frame and the longitude of the right ascending node can be considered as linear with respect to time. Thus we will assume that the metric elements~$a_{\leftmoon}$, $e_{\leftmoon}$, $I_{\leftmoon}$ (or equivalently $L_{\leftmoon}$, $G_{\leftmoon}$, $H_{\leftmoon}$) are constants and the angular variables~$l_{\leftmoon}$, $g_{\leftmoon}$, $h_{\leftmoon}$ are linear with time,
\begin{align}
  \label{eq:19}
  \vecbf{y}_{\leftmoon} = {} \vecbf{y}_{0,\leftmoon} + \dot{\vecbf{y}}_{\leftmoon} (t-t_{0})   
\end{align}
where $\vecbf{y}_{0,\leftmoon}$ at the epoch $J2000.0$ and the precession rates $\dot{\vecbf{y}}_{\leftmoon}$ are defined in Table~\ref{tab:lunisol_para}.

Such a development can be find in \citep{Giacaglia_1974aa, Giacaglia_1980aa, Lane_1989aa}. However, by comparing their expression with respect to the exact representation of the disturbing function in Cartesian coordinates \eqref{eq:def_pot_3c}, we have noticed that they are incorrect in~ \citet[][]{Lion_2013ab, Lion_2011ac}. In this work, we have demonstrated that the correct solution is%
\begin{align}
  \label{eq:R_moon_HW_exp_Fnmp}
  \begin{split}
    \mca{R}_{\leftmoon} & = {}
    \frac{\mu'}{r'}  \sum_{n \geq 2}^{}
    \sum_{m = -n}^{n} \,
    \sum_{m' = -n}^{n} \,
    \sum_{p = 0}^{n} \,
    \sum_{p' = 0}^{n} 
    \fracp*{r}{r'}^{n} 
    (-1)^{m-m'} \frac{(n-m')!}{(n+m)!} 
    \\
    {} & \hspace{2em} \times
    F_{n,m,p}(I) F_{n,m',p'}(I') U_{n,m,m'}(\epsilon)
    \exp\ci \Theta_{n,m,m',p,p'}^{-}
  \end{split}
\end{align}
or in the trigonometric formulation
\begin{footnotesize}
\begin{align}
\label{eq:R_moon_HW_trig_Fnmp}
\begin{split}
\mca{R}_{\leftmoon} 
& = {} \frac{\mu'}{r'}  
\sum_{n \geq 2}^{} \,
\sum_{m = 0}^{n} \,
\sum_{m' = 0}^{n} \,
\sum_{p = 0}^{n} \,
\sum_{p' = 0}^{n}
\Delta_{0}^{m,m'} (-1)^{m-m'}
\frac{(n-m')!}{(n+m)!} 
\fracp*{r}{r'}^{n} 
F_{n,m,p}(I) F_{n,m',p'}(I') 
\\ & {} \hspace{2em} \times
\left[ U_{n,m,m'}(\epsilon) \cos \Theta_{n,m,m',p,p'}^{-}
+ (-1)^{n-m'} U_{n,m,-m'}(\epsilon) \cos \Theta_{n,m,m',p,p'}^{+} \right] \;,
\end{split}
\end{align}
\end{footnotesize}
\\
with
\begin{subequations}
  \label{eq:R_moon_angles_dev_HW}
  \begin{alignat}{3}
    & \Theta_{n,m,m',p,p'}^{\pm} = {} \Psi_{n,m,p} \pm \Psi'_{n,m',p'} 
    \\
    \label{eq:R_moon_Psi_sat_dev_HW}
    & \Psi_{n,m,p} = (n-2p) (\nu + \omega) + m \Omega \;,
    \\
    \label{eq:R_moon_Psi_3c_dev_HW}
    & \Psi_{n,m',p'}^{\prime} = (n-2p') (\nu' + \omega') + m' \Omega'
  \end{alignat}
\end{subequations}
and
\begin{equation}
\label{eq:Delta}
\Delta_{0}^{m,m'} = {} \frac{(2 - \delta_{0}^{m}) \, (2 - \delta_{0}^{m'})}{2} \;,
\end{equation}
in which~$\delta_{j}^{k}$ is the Kronecker symbol.

The angle~$\epsilon$ is the obliquity of the ecliptic and the~$U_{n,m,m'}(\epsilon)$ are the rotation coefficients \citep[see e.g.][]{Jeffreys_1965aa, Giacaglia_1974aa, Lane_1989aa}%
\begin{small}
  \begin{align}
    \label{eq:Wigner_U}
    U_{n,m,k} (\epsilon) & = 
    (-1)^{n-k} \sum_{r}^{}  (-1)^{\sigma} \binom{n-m}{r} \binom{n+m}{m+k+r} 
    \cos^{a} \fracp*{\epsilon}{2} \, \sin^{2n-a} \fracp*{\epsilon}{2} 
    \;,
  \end{align}
\end{small}
\\
where~$a = 2r+m+k$ and~$r$ is running from~${ \max(0,-k-m)}$ to~$ \min(n-k, n-m)$. Note that these elements are related to the spherical harmonic rotation coefficients, also called the elements of Wigner’s~$d$-matrix \citep[e.g.][]{Wigner_1959aa, Sneeuw_1992aa}:%
\begin{align}
  \label{eq:Wigner_U2d}
  d_{n,m,k} (\epsilon) 
  & =  (-1)^{k-m} \frac{(n-k)!}{(n-m)!} U_{n,m,k} (\epsilon) \;,
\end{align}

Introducing now the elliptic motion functions
\begin{equation}
  \label{eq:_Phi_nk}
  \Phi_{n,k} = {} 
  \fracp*{r}{a}^{n} \exp\ci k\nu \;.
\end{equation}

The disturbing function \eqref{eq:R_moon_HW_exp_Fnmp} still depends on~$r$, $r'$, $\nu$ and $\nu'$ (through~$\theta$ and~$\theta'$). 
To obtain a perturbation fully expressed in orbital elements, the classical way is to introduce expansions in Fourier series of the mean anomaly%
\begin{equation}
  \label{eq:dev_fourier_M}
  \Phi_{n,k} 
  = \sum_{q=-\infty}^{+\infty} X_{q}^{n,k}(e) \exp\ci qM
\end{equation}
where~$X_{q}^{n,k}(e)$ are the well known Hansen coefficients~\citep{Hansen_1853aa, Tisserand_1889aa,Brouwer_1961aa}. 
In the general case, the series~\eqref{eq:dev_fourier_M} always converge as Fourier series, but can converge rather slowly \cite[see e.g.][]{Klioner_1997aa, Brumberg_1999aa}. 
Only in the particular case where~$e$ is small,  the convergence is fast thanks to the d'Alembert property which ensures that~$e^{\abs{k-q}}$ can be factorized in~$X_{q}^{n,k}(e)$. 
That is why the method is reasonably efficient for most of the natural bodies (in particular the Sun and the Moon) but  fails for satellites moving on orbits with high eccentricities. 
In this case, Fourier series of the eccentric anomaly~$E$ \cite[see][]{Brumberg_1994aa} are much more efficient:%
\begin{equation}
  \Phi_{n,k} = \sum_{q=-\infty}^{+\infty} Z_{q}^{n,k}(e) \exp\ci q E \;,
  \label{eq:dev_fourier_E_1}
\end{equation}
In cases where~$0\leq |k| \leq n$, the coefficients~$Z_{q}^{n,k}$ can be expressed in closed-form and the sum over~$q$ is bounded by~$\pm{}n$. Indeed, these are null for~\linebreak[2]{$\abs{q}>n$},%
\begin{equation}
  \label{eq:Z_1_a}
  Z_{q}^{n,k} = {}
  (-1)^{K_{+}} \beta^{K_{+}} (1+\beta^2)^{-n}
  \sum_{s=0}^{s_{max}} 
  \binom{n-k}{s} \binom{n+k}{s+K_{+}} \beta^{2s} \;,
\end{equation}	
with~$\beta = e/(1 + \eta)$, $K_{+} = k-q \ge 0$ and~$s_{max} = \min(n-k, n+k-K_{+})$.

For~\linebreak[4]{$k-q<0$}, we can use the symmetry~$Z_{q}^{n,k} = Z_{-q}^{n,-k}$. Other general expressions and numerical methods to compute these elements can be found in \cite{Klioner_1997aa,Laskar_2005aa,Lion_2013aa}.

Even if this kind of development does not allow to express the disturbing function strictly in orbital elements, the key point is that the required operations (derivation and integration with respect to the mean anomaly) can be easily performed thanks to the relation
\begin{equation}
  \label{eq:20}
  \mathrm{d}l =\frac{r}{a} \mathrm{d}E \;.
\end{equation}

Rewriting the ratio of the radial distances as%
\begin{equation}
  \label{eq:ratio a_ap}
  \frac{1}{r'} \fracp*{r}{r'}^{n} = {} 
  \frac{1}{a'} \fracp*{a}{a'}^n 
  \fracp*{a}{r}  \fracp*{r}{a}^{n+1} \fracp*{a'}{r'}^{n+1}
\end{equation}
in which we have kept a factor~$a/r$ in order to anticipate future calculating steps. Replacing  in~\eqref{eq:R_moon_HW_exp_Fnmp} respectively the elliptic motion functions related to the satellite and the Moon by their representation in Fourier series of~$E$ and~$l_{\astrosun}$, we find the real-valued function can be written in complex form~\cite[see][]{Lion_2013ab,Lion_2011ac}%
\begin{subequations}
  \allowdisplaybreaks[4] 
  \label{eq:R_moon_EO_exp_Fnmp}
  \begin{align}
    \label{eq:R_moon_EO_exp_Fnmp_R}
    & \mca{R}_{\leftmoon} 
    = {} \sum_{n \geq 2}^{} \,
    \sum_{m = -n}^{n} \,
    \sum_{m' = -n}^{n} \, 
    \sum_{p = 0}^{n} \,
    \sum_{p' = 0}^{n} \,
    \sum_{q = -n-1}^{n+1} \, 
    \sum_{q' = -\infty}^{+\infty} \, 
    \mca{R}_{n,m,m',p,p',q,q'} 
    \;,
    \\
    \label{eq:R_moon_EO_exp_Fnmp_Rxxx}
    & \mca{R}_{n,m,m',p,p',q,q'} 
    = {} \frac{a}{r} \mcat{A}_{n,m,m',p,p',q,q'} \exp\ci \, \Theta_{n,m,m',p,p',q,q'}^{-} 
    \;,
    \\
    \label{eq:R_moon_EO_exp_Fnmp_Atxxx}
    & \mcat{A}_{n,m,m',p,p',q,q'} 
    = {} \mca{A}_{n,m,m',p,p',q,q'} U_{n,m,m'}(\epsilon)
    \;,
    \\
    \label{eq:R_moon_EO_exp_Fnmp_Axxx}
    \begin{split}
      & \mca{A}_{n,m,m',p,p',q,q'} 
      = {} \frac{\mu'}{a'}
      \fracp*{a}{a'}^{n}
      (-1)^{m-m'} \frac{(n-m')!}{(n+m)!} 
      F_{n,m,p}  \left( I \right) F_{n,m',p'} (I') 
      \\
      & \hspace{7em} \times Z_{q}^{n+1,n-2p}  \left( e \right) X_{q'}^{-(n+1),n-2p'}  (e')
      \;,
    \end{split}
  \end{align}
\end{subequations}
or into trigonometric form%
\begin{subequations}
  \label{eq:R_moon_EO_trig_Fnmp}
  \begin{align}
    & \mca{R}_{\leftmoon} = {} \sum_{n \geq 2}^{} \,
    \sum_{m = 0}^{n} \,
    \sum_{m' = 0}^{n} \,
    \sum_{p = 0}^{n} \, 
    \sum_{p' = 0}^{n} \,
    \sum_{q = -n-1}^{n+1} \, 
    \sum_{q' = -\infty}^{+\infty} \, 
    \mca{R}_{n,m,m',p,p',q,q'} \;,
    \\
    \begin{split}
      & \mca{R}_{n,m,m',p,p',q,q'} =  {}
      \Delta_{0}^{m,m'} 
      \frac{a}{r} \mca{A}_{n,m,m',p,p',q,q'}
      \left[ U_{n,m,m'} (\epsilon) \cos \Theta_{n,m,m',p,p',q,q'}^{-}
        \vphantom{\Big(} \right. \\ & \hspace{70pt} \left. 
        + \, (-1)^{n-m'} U_{n,m,-m'} (\epsilon)  
        \cos\Theta_{n,m,m',p,p',q,q'}^{+} \right] 
      \;.
    \end{split}
  \end{align}
\end{subequations}
with:%
\begin{subequations}
  \label{eq:R_moon_angles_dev}
  \begin{alignat}{2}
    \label{eq:R_moon_Theta_pm_dev}
    & \Theta_{n,m,m',p,p',q,q'}^{\pm} = {} \Psi_{n,m,p,q} \pm \Psi'_{n,m',p',q'} \;,
    \\
    \label{eq:R_moon_Psi_sat_dev}
    & \Psi_{n,m,p,q} = {} qE + (n-2p) \omega + m \Omega \;,
    \\
    \label{eq:R_moon_Psi_3c_dev}
    & \Psi'_{n,m',p',q'} = {} q' M' + (n-2p') \omega' + m' \Omega' \;.
  \end{alignat}
\end{subequations}

\subsubsection{Solar disturbing function}

Expressed in Hill-Whittaker elements, the more general development for the Sun's disturbing function has been given by \citet[][Eq. 5]{Kaula_1962aa} as:%
\begin{small}
  \begin{align}
  \label{eq:R_sun_HW_exp_Fnmp}
  \mca{R}_{\astrosun} = {}
  \frac{\mu_{\astrosun}}{r_{\astrosun}}  \sum_{n \geq 2}^{}
  \sum_{m = -n}^{n}
  \sum_{p = 0}^{n} \,
  \sum_{p' = 0}^{n} 
  \frac{(n-m)!}{(n+m)!}
  \fracp*{r}{r_{\astrosun}}^{n} 
  F_{n,m,p}(I) F_{n,m,p'}(\epsilon) 
  \exp\ci \Theta_{n,m,p,p'}
  \end{align}
\end{small}
\\
with~$\Theta_{n,m,p,p'} = (n-2p) (\nu + g) + m h - (n-2p') (\nu_{\astrosun} + g_{\astrosun} )$. We assume in our work that the Sun's apparent orbit about the Earth is precessing over the ecliptic plane with linear variations of the angular variables~$g_{\astrosun}$ and~$l_{\astrosun}$, and constant metric elements~$a_{\astrosun}, e_{\astrosun}, I_{\astrosun}$ (or equivalently $L_{\astrosun}, G_{\astrosun}, H_{\astrosun}$):%
\begin{align}
\label{eq:19b}
\vecbf{y}_{\astrosun} = {} \vecbf{y}_{0,\astrosun} + \dot{\vecbf{y}}_{\astrosun} (t-t_{0})   
\end{align}
where $\vecbf{y}_{0,\astrosun}$ at the epoch $J2000.0$ and the precession rates $\dot{\vecbf{y}}_{\astrosun}$ are defined in Table~\ref{tab:lunisol_para}. Because~$I_{\astrosun} = \epsilon$, the ascending node~$\Omega_{\astrosun}$ is not defined.

As done in the previous section, we keep a factor~$a/r$ to anticipate future calculations. Replacing respectively the elliptic motion functions related to the satellite and the Sun by their representation in Fourier series of~$E$ and~$l_{\astrosun}$ gives%
\begin{subequations}
  \allowdisplaybreaks[1]
  \label{eq:R_sun_EO_exp_Fnmp}
  \begin{align}
    & \mca{R}_{\astrosun} = {}
    \sum_{n>=2}^{}
    \sum_{m=-n}^{n}
    \sum_{p=0}^{n}
    \sum_{p'=0}^{n}
    \sum_{q=-(n+1)}^{n+1}
    \sum_{q'=-\infty}^{\infty}  \, 
    \mca{R}_{n,m,p,p',q,q'} \;,
    \\
    & \mca{R}_{n,m,p,p',q,q'} 
    = {} \frac{a}{r} \mca{A}_{n,m,p,p',q,q'}
    \exp i \, \Theta_{n,m,p,p',q,q'} \;,
    \\
    \begin{split}
      & \mca{A}_{n,m,p,p',q,q'} 
      = {} \frac{\mu'}{a'}
      \left( \frac{a}{a'} \right)^{n}
      \frac{(n-m)!}{(n+m)!} 
      F_{n,m,p}  \left( I \right) F_{n,m,p'} \left( \epsilon \right) 
      \\ & \hspace{7em} \times
      Z_{q}^{n+1,n-2p}  (e) X_{q'}^{-(n+1),n-2p'}  (e')
      \;,
    \end{split}
  \end{align}
\end{subequations}
or equivalently in the trigonometric form
\begin{subequations}
  \label{eq:R_sun_EO_cos_Fnmp}
  \begin{align}
    & \mca{R}_{\astrosun} 
    = {} \sum_{n \geq 2}^{} \,
    \sum_{m = 0}^{n} \,
    \sum_{p = 0}^{n} \, 
    \sum_{p' = 0}^{n} \,
    \sum_{q = -n-1}^{n+1} \, 
    \sum_{q' = -\infty}^{+\infty} \, 
    \mca{R}_{n,m,p,p',q,q'} \;,
    \\
    & \mca{R}_{n,m,p,p',q,q'}  = {}
    (2 - \delta_{0}^{m}) \frac{a}{r} 
    \mca{A}_{n,m,p,p',q,q'}
    \cos \Theta_{n,m,p,p',q,q'} \;.
  \end{align}
\end{subequations}
with
\begin{subequations}
  \allowdisplaybreaks[1]
  \label{eq:R_sun_EO_exp_Fnmp_angles}
  \begin{alignat}{2}
    & \Theta_{n,m,p,p',q,q'} 
    = {} \Psi_{n,m,p,q} - \Psi'_{n,p',q'}  
    \;,
    \\
    & \Psi_{n,m,p,q}  
    = {} q E + (n-2p) g + m h  
    \;,
    \\
    & \Psi'_{n,p',q'} 
    = {} q' l_{\astrosun} + (n-2p') g_{\astrosun}
    \;.
  \end{alignat}
\end{subequations}
%


\section{The Lie transforms approach: principle} 
\label{sec:Lie_Trans_Approach}

Consider the Hamiltonian%
\begin{equation}
  \label{eq:10}
  \mca{H}(\vecbf{y},\vecbf{y}',\vecbf{Y},\vecbf{Y}') 
  = \mca{H}_{Kep}(L) + \mca{H}_{\Jd}(l,g,\vecbf{Y}) + \mca{H}_{3b}(\vecbf{y},\vecbf{y}^{\prime},\vecbf{Y},\vecbf{Y}^{\prime})
\end{equation}
with~$\mca{H}_{Kep}$ modeling the keplerian part, $\mca{H}_{\Jd}$ the~$\Jd$ effect and~$\mca{H}_{3b}$ the third-body attraction.

The Delaunay equations are given by%
\begin{align}
  \label{eq:eq_delaunay_ini}
  \derive{\vecbf{y}}{t} = {}\dpartial{\mca{H}}{\vecbf{Y}}
  \;, \quad
  \derive{\vecbf{Y}}{t} = {-}\dpartial{\mca{H}}{\vecbf{y}}
\end{align}
In this section, we present our approach to solve \eqref{eq:eq_delaunay_ini} by means of canonical perturbative methods. This combines (i) the Lie transforms \citep{Deprit_1969aa}, including the time dependence because of the third-body motion and (ii) the Brouwer-von Zeipel method \citep{Brouwer_1959aa}, involving two successive transformations. Firstly, we show how to build the canonical transformation eliminating the short-period mean anomaly~$l$. Then, we normalize the resulting dynamical system with a second transformation  eliminating all the long-period angular variables~$(g,h,l',g',h')$.

Consider a function~$f=f(\theta_{1,2,\ldots,K,\ldots,N})$ depending on~$N$ angular variables~$\theta_{j}$.
Thereafter, we define the averaging value of~$f$ over~$K$ angular variables by:%
\begin{equation}
  \label{eq:def_val_mean_f}
  \mean*{f(\theta_{j})}_{\theta_{1}, \theta_{2}, \ldots, \theta_{K}}
  = {} \frac{1}{(2\pi)^{K}} \underbrace{\int_{0}^{2\pi} \int_{0}^{2\pi} \cdots \int_{0}^{2\pi}}_{K} 
  f(\theta_{j}) \prod_{k = 1}^{K} \diff{\theta_{k}} 
  \;.
\end{equation}

\subsection{Isolating the secular and the periodic perturbations}

To facilitate the determination of the generating functions modeling the short and long period of the system, we proceed to a decomposition of each perturbation. 

As usual, we consider that the perturbation due to~$\Jd$ can be split off in a secular part~$\mca{H}_{\Jd,sec}$ and periodic terms~$\mca{H}_{\Jd,per}$ \citep{Brouwer_1959aa}
\begin{subequations}
  \label{eq:HJ2_per_Br}
  \begin{align}
    \label{eq:HJ2_tot}
    & \mca{H}_{\Jd} 
    = {} \mca{H}_{\Jd,sec} + \mca{H}_{\Jd,per}
    \\
    \label{eq:HJ2_sec}
    & \mca{H}_{\Jd,sec} 
    = 8\gamma_{2}\omega_{0}\frac{\eta^3}{L}\,Y_{0}^{-1,0} (e)\,F_{2,0,1}(I) 
    \\
    \label{eq:HJ2_per}
    \begin{split}
      & \mca{H}_{\Jd,per}
      = 8\gamma_{2}\omega_{0}\frac{\eta^4}{L} \sum_{p=0}^{2} \sum_{q=-1}^{1} 
      \left[ \eta \fracp*{a}{r}^{2} - \delta_{0}^{q} \delta_{2}^{2p} \right]
        Y_{q}^{-1,0} (e)\,F_{2,0,p}(I)
        \\ & \hspace{4em} \times
        \cos \left[ (q+2-2p)\nu + (2-2p)\,\omega \right] 
      \end{split}
  \end{align}
\end{subequations}
with~$\omega_{0}$ the mean motion
\begin{equation}
  \label{eq:def_w0}
  \omega_{0} = \frac{\mu^{2}}{L^{3}}
\end{equation}
and
\begin{equation}
	\label{eq:def_gamma_2_EO}
	\gamma_{2} 
	= {-} \frac{\Jd}{8\,\eta^4} \fracp*{R_{\varoplus}}{a}^2
	= {-} \frac{\Jd}{8} \fracp*{{\mu_{\varoplus}} R_{\varoplus}}{G^2}^{2} 
\end{equation}

Concerning the third-body perturbation, we rewrite~$\mca{H}_{3b}$ in order to isolate the secular~$\mca{H}_{3b,sec}$, the long-periodic~$\mca{H}_{3b,lp}$ and the short-periodic~$\mca{H}_{3b,sp}$ terms%
\begin{equation}
\label{eq:2}
\mca{H}_{3b} = {}
\mca{H}_{3b} (\vecbf{y},\vecbf{y}^{\prime},\vecbf{Y},\vecbf{Y}^{\prime}) = {} 
{} \mca{H}_{3b,sec} + \mca{H}_{3b,sp} + \mca{H}_{3b,lp} \;.
\end{equation}
We define the secular part such that it does not contain any term depending of any angular variables%
%
  \label{eq:134}
  \begin{align}
  \mca{H}_{3b,sec} 
  & = {} \mca{H}_{3b,sec} (\_,\_,\vecbf{Y},\vecbf{Y}^{\prime}) 
  = {} \lim_{T\to\infty}\frac{1}{T} \int_{0}^{T} \mca{H}_{3b} \, \diff{t} 
  = {} \mean*{\mca{H}_{3b,sec}}_{l,l',g,g',h,h'}
  \end{align}
%

%
%
Knowing that~$l$ and~$E$ are connected by~\eqref{eq:20}, we introduce the intermediate function~$\mean*{\mca{H}_{3b}}_{l}$ :%
\begin{align}
\label{eq:132b}
\mean*{\mca{H}_{3b}}_{l}  = {}
\frac{1}{2\pi}\int_{0}^{2\pi} \frac{r}{a} \mca{H}_{3b} \diff{E} \;.
\end{align}
This step was anticipated in the development of the third-body disturbing function. The factor~$a/r$ kept in~$\mca{H}_{3b}$ (see Eq. \eqref{eq:R_moon_EO_exp_Fnmp} and \eqref{eq:R_sun_EO_exp_Fnmp}) is used to offset~$r/a$ in~\eqref{eq:132b} and therefore, we can integrate with respect to~$E$ a function that depends explicitly on~$E$.

Hence, the secular terms are given by 
\begin{align}
\label{eq:134b}
\mca{H}_{3b,sec} 
& = \frac{1}{(2\pi)^{5}} {\int_{0}^{2\pi} \int_{0}^{2\pi} \cdots\!\int_{0}^{2\pi}} 
\mean*{\mca{H}_{3b}}_{l} \diff{l'} 
\diff{g} \diff{g'} 
\diff{h} \diff{h'} \;,
\end{align}
the long-periodic terms, which correspond to the slow angular variables, are obtained by removing the secular terms in~$\mean*{\mca{H}_{3b}}_{l}$:%
\begin{equation}
\label{eq:5}
\mca{H}_{3b,lp} = {}
\mean*{\mca{H}_{3b}}_{l} - \mca{H}_{3b,sec} \;,
\end{equation}
and the short-period terms are computed by eliminating in~$\mca{H}_{3b}$ all terms that do not depend on the fast variable~$l$ through~$E$:%
\begin{equation}
\label{eq:4}
\mca{H}_{3b,sp} = {}
\mca{H}_{3b} - \mean*{\mca{H}_{3b}}_{l} \;.
\end{equation}
In practice, the splitting of~$\mca{H}_{3b}$ is equivalent to an appropriate sorting of the indices in the development of the third-body disturbing function. Results for Moon and Sun are established in Section~\ref{sec:Lie_Transforms_principle}.

\subsection{Perturbations classification}
\label{ssec:Pert_class}

Assume that the initial Hamiltonian can be sorted as follows:%
\begin{equation}
  \label{eq:6a}
  \mca{H} = \mca{H}_{0} + \mca{H}_{1} + \frac{1}{2} \mca{H}_{2} + \gordre{3}  
\end{equation}
with~$\mca{H}_{0}$ the keplerian part. As usual, we put in the perturbing part~$\mca{H}_{1}$ the secular variations and the periodic terms due to~$\Jd$ in order to reuse results from~\citet{Brouwer_1959aa}. Concerning the third-body perturbations, we have chosen to put in the~$\mca{H}_{1}$ their secular part and in~$\mca{H}_{2}$ their periodic contribution, improving the degree of accuracy of the theory. Hence,
\begin{subequations}
  \begin{align}
    \label{eq:Hierarchie_H}
    \mca{H}_{0} 
    & = \mca{H}_{0}(L) 
    = \mca{H}_{Kep}
    \\
    \mca{H}_{1} 
    & = \mca{H}_{1}(l,g,\vecbf{Y},\vecbf{Y}^{\prime}) 
    = \mca{H}_{\Jd,sec} + \mca{H}_{3b,sec} + \mca{H}_{\Jd,per} 
    \\
    \mca{H}_{2} 
    & = \mca{H}_{2}(\vecbf{y},\vecbf{y}^{\prime},\vecbf{Y},\vecbf{Y}^{\prime}) 
    = 2 \mca{H}_{3b,per} = 2 \mca{H}_{3b,sp} + 2 \mca{H}_{3b,lp}
  \end{align}
\end{subequations}
%

\subsection{Elimination of the short period terms} 
\label{ssec:Rm_sp}

In order to remove the fast variable~$l$ from the Hamiltonian~$\mca{H}$, we shall apply up to the order 2 a change of variables that transforms~$\mca{H}$ to a new one~$\mca{K}$ through a generating function~$\mca{V}$:%
\begin{equation}
  \label{eq:TC_H2Ko2_a}
  \begin{array}{ccc}
    (\vecbf{y},\vecbf{y}^{\prime},\vecbf{Y},\vecbf{Y}^{\prime})
    & \xlongrightarrow{\mca{V}}
    & (\vecbf{y}^{\star},\vecbf{y}^{\prime},\vecbf{Y}^{\star},\vecbf{Y}^{\prime})
    \\
    \mca{H}(\vecbf{y},\vecbf{y}^{\prime},\vecbf{Y},\vecbf{Y}^{\prime})
    & \longrightarrow  
    & \mca{K}(\_,g^{\star},h^{\star},\vecbf{y}^{\prime},\vecbf{Y}^{\star},\vecbf{Y}^{\prime})
  \end{array}
\end{equation}
We then assume that~$\mca{K}$ and~$\mca{V}$ can be expanded as a series of the form
\begin{subequations}
  \begin{align}
    \label{eq:TC_H2Ko2_b}
    \mca{K} (\_,g,h,\vecbf{y}^{\prime},\vecbf{Y},\vecbf{Y}^{\prime})
    & = \mca{K}_{0} +\mca{K}_{1} + \frac{1}{2}\mca{K}_{2} + \gordre{3}
    \\
    \mca{V} (l,g,h,\vecbf{y}^{\prime},\vecbf{Y},\vecbf{Y}^{\prime})
    & = \mca{V}_{1} + \frac{1}{2}\mca{V}_{2} + \gordre{3}
  \end{align}
\end{subequations}
%
%
%
%
Knowing that~$\mca{H}$ is time-dependent, we shall use the time-dependent Lie Transfrom \cite{Deprit_1969aa} to find the determining functions~$\mca{V}_{1}$ and~$\mca{V}_{2}$. 

\paragraph{\textbf{Order 0}}

The Lie's Triangle is initialized with the identity transformation%
\begin{equation}
  \label{eq:EHo0_rm_sp}
  \mca{K}_{0} = {} \mca{H}_{0}
\end{equation}

\paragraph{\textbf{Order 1}}

The first order homological equation is given by%
\begin{subequations}
  \begin{align}
    \label{eq:K1_def_aa}
    \mca{K}_{1} & = {}
    \mca{H}_{1} + \cpoiss{\mca{H}_{0}}{\mca{V}_{1}} - \dpartial{\mca{V}_{1}}{t}
    \\
    \label{eq:K1_def_ab}
    & = {}
    \mca{H}_{\Jd,sec} + \mca{H}_{3b,sec} + \mca{H}_{\Jd,per} 
    - \omega_{0} \dpartial{\mca{V}_{1}}{l}
    - \dpartial{\mca{V}_{1}}{t}
  \end{align}
\end{subequations}
where~$\cpoiss{\alpha}{\beta}$ is the Poisson brackets defined by%
  \begin{equation}
  \label{eq:def_cpoiss}
  \cpoiss{\alpha}{\beta}_{y,Y} = {} \sum_{j=1}^{3} 
  \left(
  \dpartial{\alpha}{y_j} \dpartial{\beta}{Y_j}  - \dpartial{\alpha}{Y_j} \dpartial{\beta}{y_j}  
  \right)
  =  {-} \cpoiss{\beta}{\alpha}_{y,Y} \;,
  \end{equation}
We choose~$\mca{K}_{1}$ such that it does not depend on any angle variables:%
\begin{equation}
  \label{eq:K1_def_b}
  \mca{K}_{1} 
  = \mean{\mca{H}_{1}}_{l} 
  = \mca{H}_{\Jd,sec} + \mca{H}_{3b,sec} 
\end{equation}
Moreover, since~$\mca{H}_{\Jd,per}$ is not explicitly time-dependent, the PDE \eqref{eq:K1_def_ab} reduces to the classical equation %
\begin{equation}
  \label{eq:EDP_V1Br}
  \omega_{0} \dpartial{\mca{V}_{1}}{l}
  = \mca{H}_{\Jd,per} 
\end{equation}
which gives the first order determining function of the short-periodic terms due to~$\Jd$. Denoted~$\mca{V}_{1,\Jd}$, this corresponds to the solution established by \citet[][]{Brouwer_1959aa}%
\begin{equation}
  \label{eq:V1Br_id}
  \begin{split}
    \mca{V}_{1,\Jd}
    & = \gamma_{2} G \left[
      2 \left( -1 + 3 c^{2} \right)
      \left( \phi + e\sin\nu \right)
    \right. \\ & \left. \hspace{2em} 
      + s^2 \left(
        3 \sin (2\nu + 2g) + 3 e \sin (\nu + 2g) + \sin (3\nu + 2g)
      \right)
    \right]
  \end{split}
\end{equation}
with~$\phi = \nu - l$ the equation of the center, $c = \cos{I}$, $s = \sin{I}$.
%

\paragraph{\textbf{Order 2}}

The second order of the time-dependent Lie Transfom \citep{Deprit_1969aa} is given by%
\begin{subequations}
  \begin{align}
    \label{eq:K2_def_a}
    \mca{K}_{2} & = {}
    \mca{H}_{2} + \cpoiss{\mca{H}_{1}+\mca{K}_{1}}{\mca{V}_{1}} + \cpoiss{\mca{H}_{0}}{\mca{V}_{2}} - \dpartial{\mca{V}_{2}}{t}
    \\
    \begin{split}
      & = {}
      2 \mca{H}_{3b,sp} + 2\mca{H}_{3b,lp}
      + \cpoiss{2\mca{H}_{\Jd,sec} + 2\mca{H}_{3b,sec} + \mca{H}_{\Jd,per}}{\mca{V}_{1,\Jd}}
      \\
      & \hspace{1em} - \omega_{0} \dpartial{\mca{V}_{2}}{l} - \dpartial{\mca{V}_{2}}{t}
    \end{split}
  \end{align}
\end{subequations}
and we choose~$\mca{K}_{2}$ independent of~$l$%
\begin{subequations}
  \begin{align}
    \label{eq:K2_def_b}
    \mca{K}_{2} 
    & = \mean*{2\mca{H}_{3b,sp} + 2\mca{H}_{3b,lp}
      + \cpoiss{2\mca{H}_{\Jd,sec} + 2\mca{H}_{3b,sec} + \mca{H}_{\Jd,per}}{\mca{V}_{1,\Jd}}}_{l} 
    \\
    & =  2 \mca{H}_{3b,lp} 
    + \mean*{\cpoiss{2\mca{H}_{\Jd,sec} + 2\mca{H}_{3b,sec} + \mca{H}_{\Jd,per}}{\mca{V}_{1,\Jd}}}_{l}
  \end{align}
\end{subequations}
The term~$\cpoiss{2\mca{H}_{\Jd,sec} + \mca{H}_{\Jd,per}}{\mca{V}_{1,\Jd}}$ is the same as those  involved in \citet{Brouwer_1959aa} or \citet{Kozai_1962aa} when eliminating the short period at the second order. We set,%
\begin{subequations}
  \begin{align}
    \label{eq:V2Brlp_id}
    & \mca{K}_{2,\Jd,lp}(g,L,G,H)
    = \mean*{\cpoiss{2\mca{H}_{\Jd,sec} + \mca{H}_{\Jd,per}}{\mca{V}_{1,\Jd}}}_{l}
    \\
    \begin{split}
      & \hspace{2em} = {3} \omega_{0} \gamma_{2}^{2} G \left[
        2 c^2 \left( 4 - 15 c^2 \right) 
        - 4 \eta \left( 1 - 3 c^2 \right)^2
        + e^2 \left( 5 - 18 c^2 - 5 c^4 \right)
      \right. \\ & \left. \hspace{3em}
        + e^2 s^2  \left( 28 - 30 s^2 \right) \cos 2 g
      \right]
    \end{split}
  \end{align}
\end{subequations}
Furthermore, as~$\mca{H}_{3b,sec}$ is independent of~$l$, we have%
\begin{equation}
  \label{eq:1}
  \mean*{\cpoiss{2\mca{H}_{3b,sec}}{\mca{V}_{1,\Jd}}}_{l}
  = \cpoiss{2\mca{H}_{3b,sec}}{\mean*{\mca{V}_{1,\Jd}}_{l}}
\end{equation}
Although~$\lpartial{\mca{V}_{1,\Jd}}{l}$ is a purely short periodic term, the generating function~$\mca{V}_{1,\Jd}$ used by Brouwer is not. Indeed, contrary to those chosen by \citet{Metris_1991aa}, its average with respect to~$l$ is not null and depends on long periodic terms through the angle variable~$g$ (see Eq. \ref{eq:16} in Appendix~\ref{anx:J2_astuces}):%
\begin{equation}
  \label{eq:mean_V1_wrt_l}
  \mean*{\mca{V}_{1,\Jd}}_{l}
  = {-} \gamma_{2}\,s\,G\,\frac{(1-\eta)(1+2\eta)}{1+\eta}  \sin 2 g
\end{equation}
Then, the contribution of \eqref{eq:mean_V1_wrt_l} in \eqref{eq:1} yields to a coupling term between~$\Jd$ and the third-body:%
\begin{subequations}
  \label{eq:K2coup_def}
  \begin{align}
    \mca{K}_{2,coup}(g,\vecbf{Y},\vecbf{Y}^{\prime})
    & = {-}2 \dpartial{\mca{H}_{3b,sec}}{G}\dpartial{\mean{\mca{V}_{1,\Jd}}_{l}}{g}
    \\
    & = 4 \gamma_{2}\omega_{g}^{\prime}\,s\,G\,\frac{(1-\eta)(1+2\eta)}{1+\eta} \cos 2 g
  \end{align}
\end{subequations}
with~$\omega_{g}^{\prime} = \lpartial{\mca{H}_{3b,sec}}{G}$ the secular effect due to the third-body
on the argument of the perigee~$g$; and finally, we get
\begin{equation}
  \label{eq:K2_def_c}
  \mca{K}_{2}
  = \mca{K}_{2,\Jd,lp}(g,\vecbf{Y}) + 2 \mca{H}_{3b,lp}(g,h,\vecbf{y}^{\prime},\vecbf{Y},\vecbf{Y}^{\prime}) + \mca{K}_{2, coup}(g,\vecbf{Y},\vecbf{Y}^{\prime})
\end{equation}

The homological equation \eqref{eq:K2_def_a} involves the~$t$-partial derivative. To absorb the time-dependence due to the external body motion into the Poisson bracket, we have assumed in Section~\ref{ssec:H_formalism} that the angles~$\vecbf{y}^{\prime}$ related to the third-body vary linearly with time and the momenta~$\vecbf{Y}^{\prime}$ are constants (which is a good approximation). In this way, we have%
\begin{equation}
  \label{eq:dW1sdt}
  \dpartial{\mca{V}_{2}}{t} 
  = {} \omega_{l'} \dpartial{\mca{V}_{2}}{l'} 
  + \omega_{g'} \dpartial{\mca{V}_{2}}{g'} 
  + \omega_{h'} \dpartial{\mca{V}_{2}}{h'}
  = {} \sum_{j=1}^{3} \omega_{j} \dpartial{\mca{V}_{2}}{y_{j}^{\prime}} 
\end{equation}
with~$\omega_{j} = \left\{ \omega_{l'} = \dot{l}_{}^{\prime}, \omega_{g'} = \dot{g}_{}^{\prime}, \omega_{h'} = \dot{h}_{}^{\prime}\right\}$ assimilated to constant pulsations.

It results that the remaining short periods to be absorbed by~$\mca{V}_{2}$ to satisfy the following PDE%
\begin{align}
  \label{eq:EDP_ordre2}
  \begin{split}
    & \omega_{0} \dpartial{\mca{V}_{2}}{l}
    + \sum_{j=1}^{3} \omega_{j} \dpartial{\mca{V}_{2}}{y_{j}^{\prime}} 
    = 
    2 \mca{H}_{3b,sp}
    + \cpoiss{2\mca{H}_{\Jd,sec} + 2\mca{H}_{3b,sec} + \mca{H}_{\Jd,per}}{\mca{V}_{1,\Jd}}
    \\
    & \hspace{1em} 
    - \mean*{\cpoiss{2\mca{H}_{\Jd,sec} + \mca{H}_{\Jd,per}}{\mca{V}_{1,\Jd}}}_{l}
    \;.
  \end{split}
\end{align}
\\
The two Poisson brackets contain short periodic terms in~$\Jd^2$, neglected in \citet{Brouwer_1959aa} but not in \citet{Kozai_1962aa}, and short periodic terms derived from the coupling between~$\Jd$ and the third-body. As their contribution is small compared to the first order in~$\Jd$, we can neglect them. 
So, by keeping only the direct effects due to the third body, the PDE \eqref{eq:EDP_ordre2} reduces to%
\begin{align}
  \label{eq:EDP_ordre2_reduc}
  & \omega_{0} \dpartial{\mca{V}_{2}}{l}
  + \sum_{j=1}^{3} \omega_{j} \dpartial{\mca{V}_{2}}{y_{j}^{\prime}} 
  = 
  2 \mca{H}_{3b,sp}
  \;.
\end{align}
Since~$\mca{H}_{3b,sp}$ depends explicitly of~$E$, this PDE can be rewritten as%
%
%
%
%
\begin{equation}
  \label{eq:89}
  \dpartial{\mca{V}_{2}^{}}{E} 
  + \sum_{j=1}^{3} \frac{r}{a} \beta_{j} \dpartial{\mca{V}_{2}}{y_{j}^{\prime}} 
  = {} \frac{2}{\omega_{0}}\frac{r}{a} \mca{H}_{3b,sp} 
  \;.
\end{equation}
The small parameters~$\beta_{j} = \omega_{j}/\omega_{0}$ correspond to the ratio between the slow pulsations normalized by the fast pulsation. Since the fastest long-period is~$2\pi/\omega_{l'}$ (about~$28$ days for the Moon) and supposing that the satellite orbital period for a highly elliptic orbit can reach 1--2 days, $\beta_{j}$ can not exceed~${1/15}$. 

We note that we have in factor of~$\mca{H}_{3b,sp}$ the ratio~$a/r$. This term will simplify due to the fact that we have anticipated this factor in the development of the disturbing functions \eqref{eq:R_moon_EO_exp_Fnmp} and \eqref{eq:R_sun_EO_exp_Fnmp}.

Then, we solve \eqref{eq:89} by means of a recursive process. Given that~$\beta_{j} \ll 1$, we can assume that~$\mca{V}_{2}$ can be expandable in power series of the quantity~$\beta_{j}$:%
\begin{equation}
  \label{eq:Taylor_W1sp}
  \mca{V}_{2}
  = {} \mca{V}_{2}^{(0)} + \sum_{\sigma \geq 1}^{} \mca{V}_{2}^{(\sigma)} \;.
\end{equation}
In practice, a very small number of iterations are required and the question of the theoretical convergence of this series will not be discussed. Inserting this series in \eqref{eq:89}, the generating function~$\mca{V}_{2}$ can be recursively determined by using the relations%
\begin{subequations}
  \label{eq:91}
  \begin{empheq}[left=\empheqlbrace]{alignat=1}
    \label{eq:91a} 
    & \dpartial{\mca{V}_{2}^{(0)}}{E} = {} 
    \frac{2}{\omega_{0}} \frac{r}{a} \mca{H}_{3b,sp} \;,
    \\
    \label{eq:91b} 
    & \dpartial{\mca{V}_{2}^{(\sigma+1)}}{E} = {} 
    {-} \sum_{j=1}^{3} \frac{r}{a} \beta_j \dpartial{\mca{V}_{2}^{(\sigma)}}{y_{j}^{\prime}} \;, \quad \sigma \geq 0
    \;.
  \end{empheq}
\end{subequations}
The order 0 is considered as the initial value and the order~$(\sigma+1)$ as a correction of the solution of order~$\sigma$.
We impose also that the mean value of the generator~$\mca{V}_{2}^{(\sigma)}$ over the mean anomaly~$l$ is zero:~$\mean*{\mca{V}_{2}}_{l} = 0$. This can be realized by adding a constant~$C^{(\sigma)}$ independent of the eccentric anomaly.

\subsection{Elimination of the long period terms} 
\label{ssec:Rm_lp}

To make the new dynamical system~$\mca{K}$ integrable, we shall now remove all the long-period perturbations. Starting from the following perturbations classification%
\begin{subequations}
  \begin{align}
    \label{eq:Hierarchie_K}
    & \mca{K}_{0} (L) 
    = \mca{H}_{Kep}
    \\
    &\mca{K}_{1} (\vecbf{Y},\vecbf{Y}^{\prime}) 
    = \mca{H}_{\Jd,sec} + \mca{H}_{3b,sec}
    \\
    & \mca{K}_{2} (g,h,\vecbf{y}^{\prime},\vecbf{Y},\vecbf{Y}^{\prime})
    = \mca{K}_{2,\Jd,lp} + 2 \mca{H}_{3b,lp} + \mca{K}_{2, coup}
  \end{align}
\end{subequations}
we shall make another change of canonical coordinates~$(\vecbf{y}^{\star},\vecbf{Y}^{\star}) \to (\vecbf{y}^{\dstar},\vecbf{Y}^{\dstar})$ such that the transformed Hamiltonian~$\mca{M}$ is independent of any angle%
\begin{equation}
  \label{eq:TC_K2Mo2_a}
  \begin{array}{ccc}
    (\vecbf{y}^{\star},\vecbf{y}^{\prime},\vecbf{Y}^{\star},\vecbf{Y}^{\prime})
    & \xlongrightarrow{\mca{W}}
    & (\vecbf{y}^{\dstar},\vecbf{y}^{\prime},\vecbf{Y}^{\dstar},\vecbf{Y}^{\prime})
    \\
    \mca{K}(\_,g^{\star},h^{\star},\vecbf{y}^{\prime},\vecbf{Y}^{\star},\vecbf{Y}^{\prime})
    & \longrightarrow  
    & \mca{M}(\_,\_,\_,\_,\vecbf{Y}^{\dstar},\vecbf{Y}^{\prime})
  \end{array}
\end{equation}
with~$\mca{W}$ the generating function related to this mapping.

Similarly to the previous Section~\ref{ssec:Rm_lp}, we assume that~$\mca{M}$ and~$\mca{W}$ can be expanded as a series:%
\begin{subequations}
  \label{eq:TC_K2Mo2}
  \begin{align}
    \label{eq:TC_K2Mo2_M}
    \mca{M} (\_,\_,\_, \_, \vecbf{Y},\vecbf{Y}^{\prime})
    & = \mca{M}_{0} +\mca{M}_{1} + \frac{1}{2}\mca{M}_{2} + \gordre{3}
    \\
    \label{eq:TC_K2Mo2_K}
    \mca{W} (\_,g,h,\vecbf{y}^{\prime},\vecbf{Y},\vecbf{Y}^{\prime})
    & = \mca{W}_{1} + \frac{1}{2}\mca{W}_{2} + \gordre{3}
  \end{align}
\end{subequations}
%
%
and the new variables satisfy
\begin{align}
  \label{eq:Eq_Motion_Final}
  \derive{\vecbf{y}^{\dstar}}{t} = {}\dpartial{\mca{M}}{\vecbf{Y}^{\dstar}} = 0
  \;, \quad
  \derive{\vecbf{Y}^{\dstar}}{t} = {-}\dpartial{\mca{M}}{\vecbf{y}^{\dstar}} = \textrm{Cst}
\end{align}
We apply now the Lie-Deprit algorithm \citep{Deprit_1969aa} as canonical perturbation method, and solve the chain of the homological equations up to the order 2.

\paragraph{\textbf{Order 0}} At the order~0, we define
\begin{equation}
  \label{eq:Eo0_rm_lp}
  \mca{M}_{0} = {} \mca{K}_{0}
\end{equation}

\paragraph{\textbf{Order 1}}
The determining equation at order~1 is given by
\begin{subequations}
  \begin{align}
    \label{eq:M1_def_a}
    \mca{M}_{1} & = {}
    \mca{K}_{1} 
    - \omega_{0} \dpartial{\mca{W}_{1}}{l}
    - \sum_{j=1}^{3} \omega_{j} \dpartial{\mca{W}_{1}}{y_{j}^{\prime}} 
  \end{align}
\end{subequations}
and we choose%
\begin{equation}
  \label{eq:M1_def_b}
  \mca{M}_{1} 
  = \mca{K}_{1} 
  = \mca{H}_{\Jd,sec} + \mca{H}_{3b,sec}
  \;.
\end{equation}
It results that~$\mca{W}_{1}$ is null up to an arbitrary function independent of~$(l, l',g',h')$, denoted~$w_{1}$, and determined at the next order:%
\begin{equation}
  \label{eq:W1_def_a}
  \mca{W}_{1}
  = 0 + w_{1}(g,h)
\end{equation}

\paragraph{\textbf{Order 2}}

Thus, we have%
\begin{align}
  \label{eq:M2_def_a}
  \mca{M}_{2} & = {}
  \mca{K}_{2} + \cpoiss{\mca{K}_{1}+\mca{M}_{1}}{\mca{W}_{1}} 
  + \cpoiss{\mca{K}_{0}}{\mca{W}_{2}} - \dpartial{\mca{W}_{2}}{t}
\end{align}
then substituting the equations \eqref{eq:Hierarchie_K}, \eqref{eq:W1_def_a} and \eqref{eq:dW1sdt}, we get
\begin{align}
  \label{eq:M2_def_ab}
  \mca{M}_{2} & = {}
  \mca{K}_{2,\Jd,lp} + 2 \mca{H}_{3b,lp} + \mca{K}_{2, coup}
  - \varpi_{g} \dpartial{w_{1}}{g}
  - \varpi_{h} \dpartial{w_{1}}{h}
  - \sum_{j=1}^{3} \omega_{j} \dpartial{\mca{W}_{2}}{y_{j}^{\prime}} 
\end{align}
where
\begin{subequations}
  \begin{align}
    \label{eq:pulsation_coup}
    \varpi_{g}
    &={2} \dpartial{}{G} \left( \mca{H}_{\Jd,sec} + \mca{H}_{3b,sec} \right)
    ={2} \left( \omega_{g,\Jd} + \omega_{g,3b} \right)
    \\
    \varpi_{h}
    &={2} \dpartial{}{H} \left( \mca{H}_{\Jd,sec} + \mca{H}_{3b,sec} \right)
    ={2} \left( \omega_{h,\Jd} + \omega_{h,3b} \right)
  \end{align}
\end{subequations}
Now, let's select for~$\mca{M}_{2}$ the terms independent of any angular variables
\begin{subequations}
  \begin{align}
    \label{eq:M2_def_b}
    \mca{M}_{2}
    & = \mca{M}_{2,\Jd,sec}
    \\
    & = \mean*{\mca{K}_{2,\Jd,lp} + 2 \mca{H}_{3b,lp} + \mca{K}_{2, coup}}_{g,h,l',g',h'}
    \\
    & = {3} \omega_{0} \gamma_{2}^{2} G \left[
      2 c^2 \left( 4 - 15 c^2 \right) 
      - 4 \eta \left( 1 - 3 c^2 \right)^2
      + e^2 \left( 5 - 18 c^2 - 5 c^4 \right)
    \right]
  \end{align}
\end{subequations}
and make appear the long-period terms, 
%
  \begin{align}
    \label{eq:9}
    \mca{M}_{2,\Jd, lp}
    & = {} \mca{K}_{2,\Jd, lp} - \mca{M}_{2,\Jd, sec}
    = {3} \omega_{0} \gamma_{2}^{2} G e^2 s^2  \left( 28 - 30 s^2 \right) \cos 2 g
  \end{align}
%
It turns out that the PDE \eqref{eq:M2_def_ab} reads%
\begin{align}
  \label{eq:EDP_lp_ordre2}
  & \varpi_{g} \dpartial{w_{1}}{g}
  + \varpi_{h} \dpartial{w_{1}}{h}
  + \sum_{j=1}^{3} \omega_{j} \dpartial{\mca{W}_{2}}{y_{j}^{\prime}} 
  = {} \mca{M}_{2,\Jd,lp} + 2 \mca{H}_{3b,lp} + \mca{K}_{2, coup}
\end{align}
This can be solved by using the principle of superposition and the separation of variables. By isolating in~$\mca{H}_{3b,lp}$ the terms that depend on the angular variables of the disturbing body orbit~$(l',g',h')$ from those that do not depend, respectively denoted~$\mca{H}_{3b,lp 2}$ and~$\mca{H}_{3b,lp 1}$, we get%
\begin{subequations}
  \begin{empheq}[left=\empheqlbrace]{alignat=1}
    \label{eq:EDP_lp_w1} 
    & \varpi_{g} \dpartial{w_{1}}{g}
    + \varpi_{h} \dpartial{w_{1}}{h}
    ={} \mca{M}_{2,\Jd,lp} + 2 \mca{H}_{3b,lp1} + \mca{K}_{2, coup}
    \\
    \label{eq:EDP_lp_W2} 
    &\sum_{j=1}^{3} \omega_{j} \dpartial{\mca{W}_{2}}{y_{j}^{\prime}} 
    ={} 2 \mca{H}_{3b,lp2}
  \end{empheq}
\end{subequations}
Since the right-hand-side members contain trigonometric terms that are explicitly dependent of the variables of differentiation involved in the left-hand-side members, both generating functions can be easily determined. Thus, the generator~$w_{1}$ will contain the long-period part due to the~$\Jd$ effect (same expression as Brouwer) noted~$w_{1,\Jd}$, the long-period part of the third-body disturbing function independent of~$(l',g',h')$ noted~$w_{1,3b}$, and the coupling terms~$w_{coup}$; $\mca{W}_{2}$ will contain the long-periodic terms involving at least one angular variable related to the disturbing body orbit :~
\begin{equation}
  \label{eq:w1_def}
  w_{1} = w_{1,\Jd} + w_{1,3b} + w_{coup}
\end{equation}
According to \eqref{eq:K2coup_def} and \eqref{eq:M2_def_b}, we have
\begin{align}
  \label{eq:w1_br}
  & w_{1,\Jd} = {} 3 \frac{\omega_{0}}{\varpi_{g}}  \gamma_{2}^{2} e^2 G s^2 \left( 14-15 s^{2} \right)
  \sin 2g
  \\
  & w_{coup} = 2 \gamma_{2}\,s\,G\,\frac{(1-\eta)(1+2\eta)}{1+\eta} \frac{\omega_{g,3b}}{\varpi_{g}} \sin 2 g
\end{align}
The derivation of $w_{1,3b}$ will be discussed in the next section.



\section{Determination of the generating functions related to the Moon and Sun}
\label{sec:Lie_Transforms_principle}

The purpose of this section is to determine the generators eliminating the short-period terms~$\mca{V}_{2}$ and the long-period terms~$w_{1,3b}$ and $\mca{W}_{2}$ induced by the Moon and the Sun. The elimination of the periodic terms is carried out by applying the scheme exposed in the previous section.


\subsection{Lunar perturbations}
\label{ssec:lunar_pert_apps}

Let's adopt the compact notation
\begin{equation}
  \label{eq:sum_moon_exp}
  \sum_{}^{\bullet} = {}
  \sum_{n>=2}^{}
  \sum_{m=-n}^{n} 
  \sum_{m'=-n}^{n}
  \sum_{p=0}^{n}
  \sum_{p'=0}^{n}
  \sum_{q=-(n+1)}^{n+1}
  \sum_{q'=-\infty}^{\infty}
\end{equation}
and consider the perturbation of the Moon given in \eqref{eq:R_moon_EO_exp_Fnmp}%
\begin{equation}
  \label{eq:H_gen_moon}
  \mca{H}_{3b} 
  = {-} \mca{R}_{\leftmoon} 
  = {-} \sum_{}^{\bullet}
  \frac{a}{r} \mcat{A}_{n,m,m',p,p',q,q'} \exp \ci \Theta_{n,m,m',p,p',q,q'}^{-}
  \;.
\end{equation}
For ease of notation, we will use the dots "$\ldots$" to denote the indices $\{m,m',p,p'\}$. 

The intermediate function~\eqref{eq:132b} requires that~$\mca{H}_{3b}~$ satisfies~$q=0$:%
\begin{equation}
  \label{eq:H_inter_moon}
  \mean*{\mca{H}_{3b}}_{l} = {}
  {-} \sum_{q=0}^{\bullet}
  \mcat{A}_{n,\ldots,0,q'} \exp \ci \Theta_{n,\ldots,0,q'}^{-} \;.
\end{equation}
and the secular part is determined by choosing the indices combination that vanishes the phase~$\Theta_{n,\ldots,q,q'}$:%
\begin{align}
  \label{eq:132}
  \begin{cases}
    n=2p , \quad p' = p , \quad \forall p\geq 1
    \\
    m = m'= q = q'=0 
  \end{cases}
\end{align}
thus,%
\begin{align}
  \label{eq:H_split_moon_sec}
  \mca{H}_{3b,sec} 
  = {-} \sum_{p \geq 1}^{\bullet} \,\mcat{A}_{2p,0,0,p,p,0,0} 
  \;.
\end{align}


\subsubsection{Short-periodic generating function}
\label{sec:W_sp_derivees}

Starting from~\eqref{eq:4}, deriving short-periodic terms from~$\mca{H}_{3b}$ implies to satisfy the condition~$q \neq 0$:%
\begin{align}
  \label{eq:133b}
  \mca{H}_{3b,sp}
  & = {-} \sum_{}^{\bullet}
  \left( \frac{a}{r} -\delta^q_{0} \right) 
  \mcat{A}_{n,\ldots,q,q'} \exp \ci \Theta_{n,\ldots,q,q'}^{-}
  \;.
\end{align}
%
The generating function~$\mca{V}$ can be represented in series and determined by solving the iterative scheme formulated in~\eqref{eq:91}. 
We prove in Appendix~\ref{anx:proof_rec} that the solution at the order~$\sigma \ge 0$ can be put in the form%
\begin{align}
  \label{eq:103_a_bis}
  \mca{V}^{(\sigma)} 
  & = {-} \frac{(-1)^{\sigma}}{\ci} \sum_{n,\ldots,q,q'}^{\bullet}
  \mcat{A}_{n,\ldots,q,q'}^{(\sigma)}
  \sum_{\mathclap{s=-(\sigma+1)}}^{\sigma+1} \zeta_{q,s}^{(\sigma)}(e) \exp\ci \Theta_{n,\ldots,q+s,q'}^{-}
  \;,
\end{align}
or in trigonometric form (see Appendix~\ref{sanx:sp_det_func})%
\begin{align}
  \label{eq:Gen_CP_moon_trig}
  \begin{split}
    \mca{V}_{}^{(\sigma)} 
    & = {-} (-1)^{\sigma}
    \sideset{}{}\sum_{}^{\circ}
    \Delta_{0}^{m,m'} \frac{\mca{A}_{n,\ldots,q,q'}}{(q+\delta_{0}^{q}) \omega_{0}} 
    \sum_{s=-(\sigma+1)}^{\sigma+1} \zeta_{q,s}^{(\sigma)}(e)
    \\
    {} & \hspace{2em} \times 
    \left[ 
      \overline\varepsilon_{}^{\,-,\sigma} U_{n,m,m'}(\epsilon) 
      \sin \Theta_{}^{-}
      +  (-1)^{n-m'} \overline\varepsilon_{}^{\,+,\sigma} U_{n,m,-m'}(\epsilon)
      \sin \Theta_{}^{+}
    \right]
  \end{split}
\end{align}
with%
\begin{subequations}
  \begin{alignat}{3}
    & \Theta_{}^{\pm}
    = {} \Psi_{n,m,p,q+s} \pm \Psi_{n,m',p',q'}^{\prime} 
    \;,
    \\
    & \overline\varepsilon_{}^{\pm}
    = {} (n-2p) \overline\omega_{g} + m \overline\omega_{h} 
    \pm \left[ q' \overline\omega_{l'} + (n-2p') \overline\omega_{g'} + m' \overline\omega_{h'} \right]
    \;.
  \end{alignat}
\end{subequations}
The summations designated by~$\sum\limits_{}^{\circ}$ are similar to~$\sum\limits_{}^{\bullet}$ except that the indexes~$m$ and~$m'$ run from~$0$ to~$n$ instead of~$-n$ to~$n$. 

Initial values for the functions~$\mcat{A}_{n,\ldots,q,q'}^{(0)}~$ is%
  \begin{equation}
    \label{eq:def_A0}
    \mcat{A}_{n,\ldots,q,q'}^{(0)} 
  =\frac{\mcat{A}_{n,\ldots,q,q'}}{(q + \delta_{0}^{q}) \omega_{0}}
  , \quad \forall q
  \end{equation}
and for~$\zeta_{q,s}^{(0)}$:%
\begin{subequations}
  \label{eq:def_zeta_o0}
  \begin{empheq}[left=\empheqlbrace]{alignat=3}
    & \zeta_{q,0}^{(0)}=1,
    \quad
    & \zeta_{q,-1}^{(0)}=\delta^q_{1} \frac{e}{2},
    \quad
    & \zeta_{q,1}^{(0)}=\delta^{q}_{-1} \frac{e}{2}
    & , \quad \mbox{ if } q\neq 0 
    \\
    & \zeta_{0,0}^{(0)}=0,
    \quad
    & \zeta_{0,-1}^{(0)}=-\frac{e}{2},
    \quad
    & \zeta_{0,1}^{(0)}= \frac{e}{2}
    & , \quad \mbox{ if } q=0 
  \end{empheq}
\end{subequations}
\\
The next order is determined recursively by using the relations:%
\begin{align}
  \label{eq:def_As_wrt_A0}
  & \mcat{A}_{n,\ldots,q,q'}^{(\sigma+1)} 
  = \overline\varepsilon_{n,\ldots,q'}^{\sigma+1} \mcat{A}_{n,\ldots,q,q'}^{(0)}
\end{align}
and 
\begin{small}
  \begin{subequations}
  \label{eq:def_zeta_os}
  \begin{empheq}[left= {\zeta_{q,s}^{(\sigma+1)} = \empheqlbrace}]{alignat=4}
    & {}\frac{1}{(q+s)}\left( 
      \zeta_{q,s}^{(\sigma)} -\frac{e}{2}\zeta_{q,s-1}^{(\sigma)} -\frac{e}{2} \zeta_{q,s+1}^{(\sigma)} 
    \right)
    & , \quad \mbox{ if } s \neq -q
    \\
    & \frac{e}{2} \left(-\zeta_{q,-q-1}^{(\sigma)} +\zeta_{q,-q+1}^{(\sigma)} + 
      \frac{e}{2}\zeta_{q,-q-2}^{(\sigma)}-\frac{e}{2}\zeta_{q,-q+2}^{(\sigma)}\right)
    & , \quad \mbox{ if } s = -q 
  \end{empheq}
\end{subequations}
\end{small}
\\
We can show by induction that the elements~$\zeta_{q,s}^{(\sigma)}$ verify the property:%
\begin{equation}
  \label{eq:8}
  \zeta_{-q,-s}^{(\sigma)} = (1 - 2\,\delta_{0}^{q}) (-1)^{\sigma} \zeta_{q,s}^{(\sigma)}
  \;.
\end{equation}
Remark that the~$\zeta$-elements are chosen such that~$\lpartial{\mca{V}^{(\sigma)}}{E}$ contains no terms independent of~$E$, so~$\mean{\mca{V}_{}^{(\sigma)}}_{l} = 0$ .
In practice, the corrections~$\sigma > 0~$ only permit to improve the initial solution by about a few meters.

\subsubsection{Long-periodic generating function}
\label{sec:W_lp_derivees}

To determine~$w_{1,3b}$ and~$\mca{W}_{2}$ in \eqref{eq:EDP_lp_W2}, we shall isolate all the long-period perturbations related to~$\mca{H}_{3b,lp}$ from the  PDE~\eqref{eq:EDP_lp_ordre2}%
\begin{align}
  \label{eq:EDP_lp_ordre2_H3c}
  & \varpi_{g} \dpartial{\mcat{W}_{2,3b}}{g}
  + \varpi_{h} \dpartial{\mcat{W}_{2,3b}}{h}
  + \sum_{j=1}^{3} \omega_{j} \dpartial{\mcat{W}_{2,3b}}{y_{j}^{\prime}} 
  = {} 2 \mca{H}_{3b,lp}
\end{align}
such that
\begin{equation}
  \label{eq:W2t_def}
  \mcat{W}_{2,3b} = \mca{W}_{2,3b} + w_{1,3b}
\end{equation}
According to~\eqref{eq:5}, the long-periodic part of~$\mca{H}_{3b}$ corresponds to the terms that satisfy~$q = 0$ and do not simultaneously satisfy the condition~\eqref{eq:132}:%
  \begin{align}
    \mca{H}_{3b,lp} 
    & = {-} \sum_{q=0}^{\bullet}
    \mcat{A}_{n,\ldots,0,q'} \exp \ci \Theta_{n,\ldots,0,q'}^{-}
    + 
    \sum_{p \geq 1}^{} \mcat{A}_{2p,0,0,p,p,0,0} 
    \nonumber
    \;,
    \\
    \label{eq:H3c_lp}
    & = {-} \sum_{}^{\bullet\bullet}
    \mcat{A}_{n,\ldots,0,q'} \exp \ci \Theta_{n,\ldots,0,q'}^{-}
    \;,
  \end{align}
with%
\begin{align}
  \label{eq:intro_somme_Moon_LP}
  \sum\limits_{}^{\bullet\bullet}
  = {} 
  \sum_{q=0}^{\bullet} 
  - \sum_{%
    \substack{%
      n=2p=2p' \\
      m=m'=0 \\
      q=q'=0}}^{\bullet} 
  = {}
  \sum_{%
    \substack{%
      p \geq 1, q = 0 \\
      (n-2p, p'-p) \neq (0, 0) \\ 
      (m, m', q') \neq (0, 0, 0)} }^{\bullet} 
\end{align}
Therefore, substituting~\eqref{eq:H3c_lp} in \eqref{eq:EDP_lp_ordre2_H3c} and solving the PDE, we get
%
\begin{equation}
  \label{eq:Ulp_exp}
  \mcat{W}_{2,3b} 
  = {-}\frac{2}{\ci} \sum^{\bullet\bullet}
  \frac{\mcat{A}_{n,\ldots,0,q'} } {\varepsilon_{n,\ldots,q'}}
  \exp\ci \Theta_{n,\ldots,0,q'}^{-}
  + C
  \;,
\end{equation}
with 
\begin{subequations}
  \begin{alignat}{2}
    & \Theta_{n,\ldots,0,q'}^{\pm}
    && = {} \Psi_{n,m,p,0} \pm \Psi_{n,m',p',q'}^{\prime} \;,
    \\
    & \varepsilon_{n,\ldots,q'}^{\pm}
    && = {} (n-2p) \varpi_{g} + m \varpi_{h} \pm 
    \left[ 
      q' \omega_{l'} + (n-2p') \omega_{g'} + m' \omega_{h'} 
    \right]
  \end{alignat}
\end{subequations}
and~$C$ an arbitrary function independent of~$l$. We take~$C = 0$. 

Converting~\eqref{eq:Ulp_exp} into trigonometric form for numerical computations (see Appendix~\ref{sanx:lp_det_func}), we find
\begin{align}
  \label{eq:Gen_LP_moon_trig}
  \begin{split}
    \mcat{W}_{2,3b} = {} & 
    -2
    \sum_{}^{\circ\circ}
    \Delta_{0}^{m,m'} \mca{A}_{n,\ldots,0,q'}
    \left[
      \frac{U_{n,m,m'}(\epsilon)}{\varepsilon^{-}}
      \sin \Theta_{n,\ldots,0,q'}^{-} 
    \right. \\  {} & \hspace{2em} \left. 
      +  (-1)^{n-m'} \frac{U_{n,m,-m'}(\epsilon)}{\varepsilon^{+}}
      \sin \Theta_{n,\ldots,0,q'}^{+}
    \right] \;,
  \end{split}
\end{align}
with
\begin{equation}
  \label{eq:Delta_def}
  \Delta_{0}^{m,m'} 
  = {} \frac{(2 - \delta_{0}^{m}) \, (2 - \delta_{0}^{m'})}{2} 
  \;.
\end{equation}
The summations designated by~$\sum\limits_{}^{\circ\circ}$ are similar to~$\sum\limits_{}^{\bullet\bullet}$ except that the indices~$m$ and~$m'$ run from~$0$ to~$n$ instead of~$-n$ to~$n$.

Finally, we deduce from \eqref{eq:Ulp_exp} (or \eqref{eq:Gen_LP_moon_trig}) and \eqref{eq:W2t_def}:%
\begin{subequations}
  \begin{align}
    \label{eq:6}
    & w_{1,3b} = \left. \mcat{W}_{2,3b} \right\vert_{m' = n-2p' = q' = 0}
    \;,
    \\
    & \mca{W}_{2,3b} = \mcat{W}_{2,3b} - w_{1,3b}
    \;.
  \end{align}
\end{subequations}

\subsection{Solar perturbations}
\label{ssec:solar_pert_apps}

Consider now the perturbations due to the Sun and let us define the symbols%
\begin{equation}
  \label{eq:sum_sun_exp}
  \sum_{}^{\bullet} = {}
  \sum_{n>=2}^{}
  \sum_{m=-n}^{n} 
  \sum_{p=0}^{n}
  \sum_{p'=0}^{n}
  \sum_{q=-(n+1)}^{n+1}
  \sum_{q'=-\infty}^{\infty} 
\end{equation}
and
\begin{align}
  \label{eq:intro_somme_sun_LP}
  \sum\limits_{}^{\bullet\bullet}
  = {} 
  \sum_{q=0}^{\bullet} 
  - \sum_{%
    \substack{%
      n=2p=2p' \\
      m=0 \\
      q=q'=0}}^{\bullet} 
  = {}
  \sum_{%
    \substack{%
      p \geq 1, q = 0 \\
      (n-2p, p'-p) \neq (0, 0) \\ 
      (m,  q') \neq (0, 0)} }^{\bullet} 
\end{align}
such that,
\begin{equation}
  \label{eq:H_gen_sun}
  \mca{H}_{\astrosun} 
  = {-} \mca{R}_{\astrosun} 
  = {-} \sum_{}^{\bullet}
  \frac{a}{r} \mcat{A}_{n,m,p,p',q,q'} \exp \ci \Theta_{n,m,p,p',q,q'} 
  \;.
\end{equation}
Proceeding as for the Moon case, expressions of the secular and the long and short-periodic part are respectively,%
\begin{subequations}
  \label{eq:H_split_sun}
  \begin{align}
    \label{eq:H_split_sun_sec}
    & \mca{H}_{\astrosun,sec} 
    = {-} \sum_{p \geq 1}^{\bullet} \,\mcat{A}_{2p,0,p,p,0,0} 
    \\
    \label{eq:H_split_sun_sp}
    & \mca{H}_{\astrosun,sp}
    = {-} \sum_{}^{\bullet}
    \left( \frac{a}{r} -\delta^q_{0} \right) 
    \mcat{A}_{n,\ldots,q,q'} \exp \ci \Theta_{n,\ldots,q,q'} 
    \\
    \label{eq:H_split_sun_lp}
    & \mca{H}_{\astrosun,lp} 
    = {-} \sum_{}^{\bullet\bullet}
    \mcat{A}_{n,\ldots,0,q'} \exp \ci \Theta_{n,\ldots,0,q'}
  \end{align}
\end{subequations}
The generating function eliminating the short-periodic terms at the order~$\sigma \ge 0$ reads%
\begin{subequations}
  \label{eq:Gen_SP_Sun_exp}
  \begin{alignat}{2}
    & \mca{V}^{(\sigma)} 
    = {}\sideset{}{}\sum_{n,m,p,p',q,q'}^{\bullet} \mca{V}^{(\sigma)}_{n,\ldots,q,q'} 
    \\
    \label{eq:103_sun_ci}
    & \mca{V}^{(\sigma)}_{n,\ldots,q,q'} 
    = {-} \frac{(-1)^{\sigma}}{\ci}
    \mcat{A}_{n,\ldots,q,q'}^{(\sigma)}
    \sum_{s=-(\sigma+1)}^{\sigma+1} \zeta_{q,s}^{(\sigma)} (e) 
    \exp\ci\Theta_{n,m,p,p',q+s,q'}
  \end{alignat}
\end{subequations}
or
\begin{align}
  \label{eq:Ur_sun_trigp_final}
  \mca{V}^{(\sigma)}_{n,\ldots,q,q'} 
  = {-} (-1)^{\sigma} (2 - \delta_{0}^{m}) \mcat{A}_{n,\ldots,q,q'}^{(\sigma)}
  \sum_{s = -(\sigma+1)}^{\sigma+1} \zeta_{q,s}^{(\sigma)}(e)
  \sin \Theta_{n,\ldots,q+s,q'}
  \;.
\end{align}
and the generating function eliminating the long-periodic terms can be written as%
\begin{equation}
  \label{eq:Ulp_sun_exp}
  \mcat{W}_{2,\astrosun} 
  = {-}\frac{2}{\ci} \sum^{\bullet\bullet}
  \frac{\mcat{A}_{n,\ldots,0,q'} } {\varepsilon_{n,\ldots,q'}}
  \exp\ci \Theta_{n,\ldots,0,q'}
  \;,
\end{equation}
or 
\begin{align}
  \label{eq:Gen_LP_sun_trig}
  \mcat{W}_{2,\astrosun} = {} & 
  -2
  \sum_{}^{\circ\circ}
  \left( 2 - \delta_{0}^{m} \right) 
  \frac{\mca{A}_{n,\ldots,0,q'}}{\varepsilon_{n,\ldots,q'} }
  \sin \Theta_{n,\ldots,0,q'} 
\end{align}
with 
\begin{align}
  \label{eq:vareps_sun}
  \varepsilon_{n,\ldots,q'} 
  & = {} (n-2p) \omega_{g} + m \omega_{h} - q' \omega_{l'} - (n-2p') \omega_{g'} 
\end{align}
%

\section{Complete solution of the motion equations}
\label{sec:solution_complete}

Suppose that the initial conditions~$\mca{E}_{i} = ( \vecbf{y}, \vecbf{Y} )$ (or equivalently~$(a,e,I,h,g,l)$) are known at the instant~$t_{0}$. 

\begin{figure}[!htb]
  \captionsetup{width=0.65\textwidth}  
  \centering
\includegraphics[clip=true, trim = 0 765 415 0, width=0.60\linewidth]{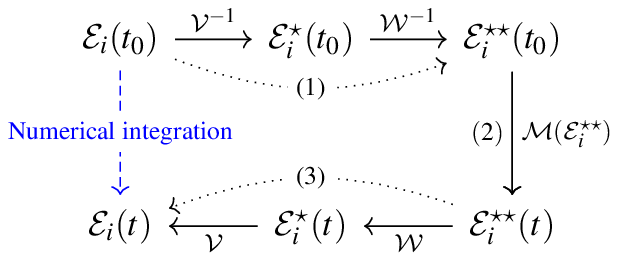}
\caption{Diagram of the change of variables between the osculating elements and the mean elements. Steps (1) to (3) correspond to the analytical propagation and the "Numerical integration" is assumed to be the reference solution.}
\label{fig:diag}
\end{figure}

We present in this section the procedure to determine the complete solution of the dynamical system~$\mca{H}$ at any instant~$t$. This is illustrated through the diagram in Figure~\ref{fig:diag} with%

\begin{enumerate}[label=(\arabic*)]
	\setlength{\itemsep}{0pt}
	\item The transformation of the initial osculating elements into mean elements with~$\mca{V}^{-1}$ and~$\mca{W}^{-1}$;
	\item The propagation of the mean elements at any time~$t$ thanks to the normalized Hamiltonian, such as the action variables are constant and the angular variables are linear with time;
	\item The transformation of the mean elements into osculating elements with~$\mca{W}$ and~$\mca{V}$.
\end{enumerate}

\subsection{Transformation of the initial elements}
\label{ssec:TC_inverse}

The new set of variables~$\mca{E}_{i}^{\star}$ can be expressed from the old variables~$\mca{E}_{i}^{}$ through the determining function~$\mca{V}$, eliminating the short-period variations, and by means of the Lie series~\citep{Deprit_1969aa}%
\begin{equation}
  \label{eq:TCinv_H}
  \mca{E}_{i}^{\star}
  = {} \mca{E}_{i}^{} - \sum_{k > 0}^{} \frac{1}{n!}
    \Lambda_{\mca{V}}^{k} \mca{E}_{i} 
\end{equation}
where~$\Lambda_{\mca{V}} \mca{E}_{i}$ denotes the Lie derivative%
\begin{equation}
\label{eq:def_Delta_Ei}
\Lambda_{\mca{V}} \mca{E}_{i}
= {} \cpoiss{\mca{E}_{i}^{}}{\mca{V}}  
= {} \sum_{j=1}^{6} \cpoiss{ \mca{E}_{i}^{} }{ \mca{E}_{j}^{} } \dpartial{\mca{V}}{\mca{E}_{j}^{}}
\end{equation}
and~$\Lambda_{\mca{V}}^{k} = \Lambda_{\mca{V}} \left( \Lambda_{\mca{V}}^{k-1} \right)$ the~$k$-th derivative.
\\
Up to the order~$k=2$, \eqref{eq:TCinv_H} writes
\begin{align}
  \label{eq:TCinv_H2Ko2_bis}
  \mca{E}_{i}^{\star} 
  = {} \mca{E}_{i}
  - \cpoiss{ \mca{E}_{i} }{\mca{V}_{1}} 
  - \frac{1}{2} \left( 
    \cpoiss{ \mca{E}_{i} }{\mca{V}_{2}} 
    - \cpoiss{ \cpoiss{\mca{E}_{i}}{\mca{V}_{1}} }{\mca{V}_{1}} 
  \right) + \gordre{3}
  \;.
\end{align}
\textcolor{black}{Some numerical tests permitted us to deduce that periodic terms in~$\Jd^2$ can be neglected in the theory without  significant loss of accuracy.} It results,%
%
%
\begin{align}
  \label{eq:TCinv_H2Ko2-red}
  \mca{E}_{i}^{\star} 
  = {} \mca{E}_{i}
  - \cpoiss{ \mca{E}_{i} }{\mca{V}_{1}} 
  - \frac{1}{2} 
  \cpoiss{ \mca{E}_{i} }{\mca{V}_{2}}
  \;.
\end{align}
Applying the inverse transformation~\eqref{eq:TCinv_H2Ko2-red} to the initial osculating elements \linebreak[4]$\mca{E}_{i}^{} = \mca{E}_{i}^{}(t_{0})$, we get%
    \begin{align}
      \begin{split}
        \mca{E}_{i}^{\star}(t_{0})
        & = {} \mca{E}_{i}^{}
        - \Lambda_{\mca{V}_{1,\Jd}} \mca{E}_{i} 
        - \frac{1}{2} \left( 
          \Lambda_{\mca{V}_{2,\leftmoon}} \mca{E}_{i} 
          + \Lambda_{\mca{V}_{2,\astrosun}} \mca{E}_{i} 
        \right)
      \end{split}
    \end{align}
with~$\mca{V} = \mca{V} \left( \mca{E}_{i}(t_{0}) \right)$.

In the same way, we can now remove the long-period variations in~$\mca{E}_{i}^{\star}$ with the generating function~$\mca{W}$, providing the mean elements~$\mca{E}_{i}^{\dstar}$ used for the secular solution%
\begin{align}
      \begin{split}
        \mca{E}_{i}^{\dstar}(t_{0})
        & = {} \mca{E}_{i}^{\star}
        - \Lambda_{\mca{W}_{1,\Jd}} \mca{E}_{i}^{\star}
        - \Lambda_{w_{coup}} \mca{E}_{i}^{\star}
        - \Lambda_{w_{1,\leftmoon}} \mca{E}_{i}^{\star} 
        - \Lambda_{w_{1,\astrosun}} \mca{E}_{i}^{\star} 
        \\ & \hspace{1em}
        - \frac{1}{2} \left( 
          \Lambda_{\mca{W}_{2,\leftmoon}} \mca{E}_{i}^{\star} 
          + \Lambda_{\mca{W}_{2,\astrosun}} \mca{E}_{i}^{\star} 
        \right)
      \end{split}
    \end{align}
with~$\mca{W} = \mca{W} \left( \mca{E}_{i}^{\star}(t_{0}) \right)$ and we verify that~$\Lambda_{\mca{W}_{}} a^{\star} = 0$.

If we consider that~$\mca{E}_{i}$ are keplerian elements then, for any function \linebreak[4]$f=f(\mca{E}_{i}) \in \R$, the derivative~\eqref{eq:def_Delta_Ei} transforms into%
\begin{subequations}
  \allowdisplaybreaks[3]
  \label{eq:dEO_new_TF_o1}
  \begin{empheq}[left=\empheqlbrace]{alignat=3}
    & \Lambda_{f} {a}  
    && = -J_{a,L} \dpartial{f}{l} 
    \\
    & \Lambda_{f} {e }  
    && =  -J_{e,L} \dpartial{f}{l} - J_{e,G} \dpartial{f}{g}
    \\
    & \Lambda_{f} {I }  
    && = -J_{I,G} \dpartial{f}{g} - J_{I,H} \dpartial{f}{h}
    \\
    & \Lambda_{f} {h }  
    && = J_{I,H} \dpartial{f}{I}
    \\
    & \Lambda_{f} {g }  
    && = {} 
    J_{e,G} \dpartial{f}{e} +  J_{I,G} \dpartial{f}{I}
    \\
    & \Lambda_{f} {l }  
    && = {} 
    J_{a,L} \dpartial{f}{a} +  J_{e,L} \dpartial{f}{e}
  \end{empheq}
\end{subequations}
with the Jacobian matrix~$J_{i,j} = \lpartial{x_{i}}{Y_{j}}$ defined in~\eqref{eq:dJsdx}.

For each perturbation, the associated derivatives of~$\mca{V}$ and~$\mca{\mca{W}}$ with respect to keplerian elements are established in Appendix~\ref{anx:proof_rec}.

\subsection{The secular solution}
\label{ssec:secular solution}

The secular solution of the system \eqref{eq:Ham_sys} derives from the normalized Hamiltonian~\eqref{eq:TC_K2Mo2_M}
\begin{small}
\begin{subequations}
  \begin{align}
    \mca{M} 
    & = \mca{M}_{0} +\mca{M}_{1} + \frac{1}{2}\mca{M}_{2} + \gordre{3}
    \\
    \mca{M}_{0} 
    & = \mca{H}_{Kep}
    = {-}\frac{\mu_{\varoplus}}{2\,L^2}
    \\
    \mca{M}_{1} 
    &  = \mca{H}_{\Jd,sec} + \mca{H}_{\leftmoon,sec} + \mca{H}_{\astrosun,sec}
    \nonumber
    \\
    & = {-} 2 \gamma_{2} G \left( 1 - 3c\right)
    - \sum_{p \geq 1}^{\bullet} \left( \mcat{A}_{\leftmoon\,2p,0,0,p,p,0,0}  +  \mcat{A}_{\astrosun\,2p,0,p,p,0,0} \right)
    \\
    \mca{M}_{2} 
    & = \frac{3}{2} \omega_{0} \gamma_{2}^{2} G \left[
      2 c^2 \left( 4 - 15 c^2 \right) 
      - 4 \eta \left( 1 - 3 c^2 \right)^2
      + e^2 \left( 5 - 18 c^2 - 5 c^4 \right)
    \right]
  \end{align}
\end{subequations}
\end{small}
\\
Knowing that for any set of variable~$\mca{E}_{i}$ (canonical or not)
\begin{equation}
\label{eq:def_pulsations}
\derive{\mca{E}_{i}}{t} 
= {} \Lambda_{\mca{M}} \mca{E}_{i}
= {} \sum_{j=1}^{6} \cpoiss{ \mca{E}_{i}^{} }{ \mca{E}_{j}^{} } \dpartial{\mca{M}_{}}{\mca{E}_{j}^{}}
\end{equation}
the solution of the equations of motion \eqref{eq:Eq_Motion_Final} expressed in keplerian elements is%
%
\begin{subequations}
  \label{eq:EO_mean_old}
  \begin{empheq}[left=\empheqlbrace]{alignat=3}
  & {a}_{}^{\dprime} (t) 
  && = {} {a}_{0}^{\dprime} 
  \\
  & {e}_{}^{\dprime} (t) 
  && = {} {e}_{0}^{\dprime} 
  \\
  & {I}_{}^{\dprime} (t) 
  && = {} {I}_{0}^{\dprime} 
  \\
  & {h}_{}^{\dprime} (t) 
  && = {} {h}_{0}^{\dprime}  
  + \left( 
    \omega_{h, \Jd}^{\dprime} 
    + \omega_{h, \Jd^2}^{\dprime}
    + \omega_{h, \leftmoon}^{\dprime}
    + \omega_{h, \astrosun}^{\dprime} \right) \Delta t
  \\
  & {g}_{}^{\dprime} (t) 
  && = {} {g}_{0}^{\dprime}  
  + \left( 
    \omega_{g, \Jd}^{\dprime}
    + \omega_{g, \Jd^2}^{\dprime}
    + \omega_{g, \leftmoon}^{\dprime}
    + \omega_{g, \astrosun}^{\dprime} \right) \Delta t
  \\
  & {l}_{}^{\dprime} (t) 
  && = {} {l}_{0}^{\dprime} 
  + \left(
    \omega_{0}^{} 
    + \omega_{l, \Jd}^{\dprime} 
    + \omega_{l, \Jd^2}^{\dprime}
    + \omega_{l, \leftmoon}^{\dprime}
    + \omega_{l, \astrosun}^{\dprime} \right) \Delta t
\end{empheq}
\end{subequations}
with~$\Delta t = t - t_{0}$, $\omega_{0}^{}$ the mean motion and~$\omega_{h,[\cdot]}, \omega_{g,[\cdot]}, \omega_{l,[\cdot]}$ the secular variations related to each perturbative term of the analytical theory:~$\Jd$, $\Jd^2$, $Moon$ and~$Sun$. Their expression are given below. Note that, as far as we know, it's the first time that a compact and general relation to compute the secular terms at any degree is proposed for the Moon and Sun.

\paragraph{\pucedingk~$\Jd$ effect}
~~ \\
Given that the normalized hamiltonians~\eqref{eq:HJ2_sec} for~$\Jd$ and~\eqref{eq:M2_def_b} for~$\Jd^{2}$ are similar to \citet{Brouwer_1959aa}, the secular variations are given, respectively, as%
\begin{subequations}
   \label{eq:ES_J2}
  \begin{alignat}{5}
    &\omega_{l,\Jd} 
    && = {} 6 \omega_{0} \gamma_{2} \, \eta \left( 1-3\cos^{2}I \right) \;,
    \\
    &\omega_{g,\Jd} 
    && = {} 6 \, \gamma_{2} \omega_{0} \left( 1-5\cos^{2}I \right) \;,
    \\
    &\omega_{h,\Jd} 
    && = {} 12 \, \gamma_{2} \omega_{0} \cos I \;,
  \end{alignat}
\end{subequations}
%
and
\begin{footnotesize}
\begin{subequations}
  \allowdisplaybreaks[3]
  \label{eq:ES_J2c}
  \begin{align}
  \label{eq:def_wl_J2c}
  \omega_{l, \Jd^{2}}
  & = {} \frac{3}{2} \omega_{0} \gamma_{2}^{2} \eta 
  \left[
  10 \left( 1 - 6 c^{2} + 13 c^{4 }\right)
  + 16 \eta  \left( 1 - 3 c^{2} \right)^{2}
  - 5 e^{2} \left( 5 - 18 c^{2} + 5 c^{4 }\right)
  \right] 
  \;,
  \\
  \begin{split}
  \label{eq:def_wg_J2c}
  \omega_{g, \Jd^{2}}
    & = {} \frac{3}{2} \omega_{0} \gamma_{2}^{2} \left[ - 2 \left( 1 - 5 c^{2} \right) \left( 5 + 43 c^{2} \right) + 24
      \eta  \left( 1 - 3 c^{2} \right) \left( 1 - 5 c^{2} \right) \right. \\ & \left. \hspace{1em} - e^{2} \left(
        25 - 126 c^{2} + 45 c^{4}\right) \right] \;,
  \end{split}
  \\
  \label{eq:def_wh_J2c}
  \omega_{h, \Jd^{2}}
  & = {} \frac{3}{2} \omega_{0} \gamma_{2}^{2} c
  \left[
  4 \left( 1 - 10 c^{2} \right) 
  + 12 \eta \left( 1 - 3 c^{2} \right)
  - e^{2} \left( 9 - 5 c^{2} \right)
  \right] 
  \;.
  \end{align}
\end{subequations}
\end{footnotesize}

\paragraph{\pucedingk Moon and Sun perturbations}
~~ \\
Consider that the secular part of the lunar perturbations~\eqref{eq:H_split_moon_sec} and the solar perturbations ~\eqref{eq:H_split_sun_sec} can be written%
\begin{equation}
  \label{eq:21}
  \mca{H}_{3b,sec}
  = {-} \sum_{p \ge 1}^{}
  \mca{B}_{p}  F_{2p,0,p} (I) Z_{0}^{2p+1,0}(e) 
\end{equation}
with
\begin{equation}
  \label{eq:11}
  \mca{B}_{p} = 
  \frac{\mu_{}^{\prime}}{a_{}^{\prime}} \fracp*{a}{a_{}^{\prime}}^{2p} 
  F_{2p,0,p} (I_{}^{\prime}) X_{0}^{-(2p+1),0} (e_{}^{\prime})
  \times \left[ 
    \begin{array}{l}
      U_{2p,0,0} (\epsilon) \\ 1
    \end{array}
  \right]_{3b = \astrosun}^{3b = \leftmoon}
  \;,
\end{equation}
then, we have:%
\begin{footnotesize}
\begin{subequations}
  \label{eq:dKmoonsdx}
    \begin{align}
        \omega_{l,3b}
        & {} = {-} \sum_{p \geq 1}^{} 
        \mca{B}_{p} {F}_{2p,0,p}(I)
        \left(
          \frac{2p}{a} J_{a,L} {Z}_{0}^{2p+1,0}(e)
          + J_{e,L} \dpartial{{Z}_{0}^{2p+1,0}(e)}{e}
        \right)
    \\
        \omega_{g,3b}
        & {} = {-} \sum_{p \geq 1}^{} 
        \mca{B}_{p}
        \left(
         J_{e,G} {F}_{2p,0,p}(I) \dpartial{{Z}_{0}^{2p+1,0}(e)}{e}
         + J_{I,G} \dpartial{{F}_{2p,0,p}(I)}{I} {Z}_{0}^{2p+1,0}(e)
        \right)
    \\
    \omega_{h,3b}
    & {} = {-} \sum_{p \geq 1}^{} 
        J_{I,G} \mca{B}_{p} \dpartial{{F}_{2p,0,p}(I)}{I} {Z}_{0}^{2p+1,0}(e)
    \end{align}
\end{subequations}
\end{footnotesize}
\\
For~$n = 2$ (or~$p = 1$), we find for the Moon case%
\begin{small}
\begin{subequations}
  \allowdisplaybreaks[3]
    \label{eq:ES_moon}
    \begin{align}
    \omega_{l, \leftmoon}
    & = \frac{\mu_{\leftmoon}}{32} 
    \frac{10 - 3 \eta^{2}}{\omega_{0} \, a_{\leftmoon}^{2} \eta_{\leftmoon}^{3}} 
    \left( 1-3\cos^2 I \right) 
    \left( 1-3\cos^2 \epsilon \right) 
    \left( 1-3\cos^2 I_{\leftmoon} \right) 
    \;,
    \\
    \omega_{g, \leftmoon}
    &  = {-} \frac{3}{32} \mu_{\leftmoon} 
    \frac{\eta^{2}  - 5\cos^2 I}{\omega_{0} \eta \, a_{\leftmoon}^{2} \eta_{\leftmoon}^{3}} 
    \left( 1-3\cos^2 \epsilon \right) 
    \left( 1-3\cos^2 I_{\leftmoon} \right)  
    \;,
    \\
    \omega_{h, \leftmoon}
    &  =  {-} \frac{3}{32} \mu_{\leftmoon}
    \frac{5 - 3 \eta^{ 2}}{\omega_{0} \eta \, a_{\leftmoon}^{2} \eta_{\leftmoon}^{3}} 
    \cos I 
    \left( 1-3\cos^2 \epsilon \right) 
    \left( 1-3\cos^2 I_{\leftmoon} \right) 
    \;.
    \end{align}
\end{subequations}
\end{small}
and for the Sun case
\begin{subequations}
  \allowdisplaybreaks[3]
  \label{eq:ES_sun}
  \begin{alignat}{3}
  \omega_{l, \astrosun}
  & =  {-}\frac{\mu_{\astrosun}}{16} \frac{10 - 3 \eta ^{2}}{\omega_{0} \, a_{\astrosun} \eta_{\astrosun}^{3}} 
  \left( 1-3\cos^2 I \right)  
  \left( 1-3\cos^2 I_{\astrosun} \right)  
  \;,
  \\
  \omega_{g, \astrosun}
  & = \frac{3}{16} \mu_{\astrosun} \frac{\eta^{2} - 5\cos^2 I}{\omega_{0} \eta \, a_{\astrosun}^{2} \eta_{\astrosun}^{3}} 
  \left( 1-3\cos^2 I_{\astrosun} \right)  
  \;,
  \\
  \omega_{h, \astrosun}
  & = \frac{3}{16} \mu_{\astrosun} \frac{5 - 3 \eta^{2}}{\omega_{0} \eta \, a_{\astrosun}^{2} \eta_{\astrosun}^{3}} 
  \cos I 
  \left( 1-3\cos^2 I_{\astrosun} \right)    
  \;.
  \end{alignat}
\end{subequations}

\subsection{Propagation of the elements}
\label{ssec:Propagation}

If the mean elements~$\mca{E}_{i}^{\dstar}$ are known, we can propagate the equation of motions at any instant~$t$. Beginning to add the long-periodic terms thanks to~$\mca{W}$, the new variables~$\mca{E}_{i}^{\star}$ can be expressed in Lie series \citep{Deprit_1969aa} as%
\begin{equation}
  \label{eq:17}
  \mca{E}_{i}^{\star}
  = {} \mca{E}_{i}^{\dstar} + \sum_{k > 0}^{} \frac{1}{n!} \left. 
    \Lambda_{\mca{V}}^{k} \mca{E}_{i} \right\rvert_{\mca{E}_{i}^{} = \mca{E}_{i}^{\dstar}} 
\end{equation}
By proceeding in the same way as in the inverse transformation case, if we consider a canonical transformation up to the order~2 and we discard the~$\Jd^2$ terms, we get
\begin{align}
      \begin{split}
        \mca{E}_{i}^{\star}(t)
        & = {} \mca{E}_{i}^{\dstar} 
        + \Lambda_{\mca{W}_{1,\Jd}} \mca{E}_{i}^{\dstar}
        + \Lambda_{w_{coup}} \mca{E}_{i}^{\dstar}
        + \Lambda_{w_{1,\leftmoon}} \mca{E}_{i}^{\dstar} 
        + \Lambda_{w_{1,\astrosun}} \mca{E}_{i}^{\dstar} 
        \\ & \hspace{1em}
        + \frac{1}{2} \left( 
          \Lambda_{\mca{W}_{2,\leftmoon}} \mca{E}_{i}^{\dstar} 
          + \Lambda_{\mca{W}_{2,\astrosun}} \mca{E}_{i}^{\dstar} 
        \right)
      \end{split}
    \end{align}
with~$\mca{W} = \mca{W} \left( \mca{E}_{i}^{\dstar}(t) \right)$.

Hence, add the short-period variations modeled by~$\mca{V} = \mca{V} \left( \mca{E}_{i}^{\star} (t) \right)$ to~$\mca{E}_{i}^{\star}$ gives the osculating elements~$\mca{E}_{i}^{}$, solution of the dynamical system~$\mca{H}$:%
    \begin{align}
      \begin{split}
        \mca{E}_{i}^{}(t)
        & = {} \mca{E}_{i}^{\star}
        + \Lambda_{\mca{V}_{1,\Jd}} \mca{E}_{i}^{\star} 
        + \frac{1}{2} \left( 
          \Lambda_{\mca{V}_{2,\leftmoon}} \mca{E}_{i}^{\star} 
          + \Lambda_{\mca{V}_{2,\astrosun}} \mca{E}_{i}^{\star} 
        \right)
      \end{split}
    \end{align}
All the derivatives with respect to keplerian elements involved in the Lie operator~\eqref{eq:dEO_new_TF_o1} are defined in Appendix~\ref{anx:proof_rec}.

\section{Numerical tests}
\label{sec:Num_Tests}

In this section, we present some numerical tests to show abilities of the theory. The complete analytical solution described in Section~\ref{sec:solution_complete} was implemented in Fortran~90 program APHEO (Analytical Propagator for Highly Elliptical Orbits). 

All the numerical tests have been realized with the object SYLDA, an Ariane~5 debris in Geostationary Transfer Orbit (GTO). The initial orbital elements are given in Table~\ref{fig:TLE}, with a semi-axis major $a=\num{24286.863}$~km, eccentricity $e=0.726$ and inclination $I=5.957$~\si{\degree}, perigee altitude $h_{p} = 267$~km and apogee altitude $h_{a} = \num{83555}$~km, and an orbital period of $T\approx 10.463$~h.

  \begin{verbbox}[\scriptsize]
    ARIANE 5 DEB [SYLDA]    
    1 40274U 14062D   14313.65939750  .00023668  00000-0  92879-2 0   135
    2 40274   5.9570 168.6919 7263810 197.5825 109.5543  2.29386099   532
  \end{verbbox}
\begin{table}[!ht]
  \centering
  \theverbbox
  \vspace{-10pt}
  \caption[toto]{Two-Line Elements of SYLDA \footnotemark. (NORAD Id:~40274)}
  \label{fig:TLE}
\end{table}

\footnotetext{Available on \url{http://celestrak.com/satcat}.}

In Table~\ref{tab:Eff_sec} we give the values of the secular effects on the satellite's angular variables $(l,g,h)$ induced by the $\Jd$ effect (Eq.~\ref{eq:ES_J2}--\ref{eq:ES_J2c}) and the luni-solar perturbations (Eq.~\ref{eq:dKmoonsdx} truncated at the degree 4), computed from the initial osculating elements.

\begin{table}[h!] 
  
  \sisetup{
    table-number-alignment = left,
    table-figures-integer = 1, 
    table-figures-decimal = 4,
    table-auto-round,
    table-sign-mantissa,
    table-sign-exponent,
    exponent-product =  \times, 
    output-exponent-marker 
  } 
  
  \captionsetup{width=0.95\textwidth}  
  \centering 
  
  \begin{tiny}
    \begin{tabular}{
        *{1}{R{1em}}|
        S[table-format = -2.5e1]
        *{1}{R{2em}}
        S[table-format = -4.5e1]
        *{1}{R{2em}}
        S[table-format = -4.5e1]
        *{1}{R{2em}}
        S[table-format = -4.5e1]
        *{1}{R{2em}}
      } 

      \toprule 
      
      {} 
      & {Keplerian} & {} 
      & $\Jd$ & {} 
      & {Sun}  & {} 
      & {Moon}  & {} 
      \\ 
      
      \cmidrule(r){1-1}
      \cmidrule(l){2-3}
      \cmidrule(l){4-5}
      \cmidrule(l){6-7}
      \cmidrule(l){8-9}
      
      \omega_{h} 
      & 0 & \textrm{rad/s}
      &  {-0.833774995391E-07} & \textrm{rad/s}
      &  {-0.352535863831E-09} & \textrm{rad/s}
      &  {-0.772650652420E-09} & \textrm{rad/s}
      \\
      T_{h}  
      & \textrm{Not defined} & \textrm{h}
      & -872.20 & \textrm{d}
      & -564.77 & \textrm{y}
      & -257.69 & \textrm{y}
      \\
      \omega_{g} 
      & 0 & \textrm{rad/s}
      & {0.165449887355E-06} & \textrm{rad/s}
      & {0.442584087739E-09} & \textrm{rad/s}
      & {0.969432099980E-09} & \textrm{rad/s}
      \\
      T_{g}   
      & \textrm{Not defined} & \textrm{h}
      &  {439.54} & \textrm{d}
      &  {449.86} & \textrm{y}
      &  {205.380} & \textrm{y}
      \\
      \omega_{l}  
      &  {0.166814278636E-03} & \textrm{rad/s}
      & {0.566636363022E-07} & \textrm{rad/s}
      & {-0.382764304828E-09} & \textrm{rad/s}
      & {-0.836496682109E-09} & \textrm{rad/s}
      \\
      T_{l}  
      & {10.4627} & \textrm{h}
      & {1283.4} & \textrm{d}
      & {-520.17} & \textrm{y}
      & {-238.02} & \textrm{y}
      \\
      \bottomrule 
    \end{tabular}
  \end{tiny}
  \caption{Values of the precession rate $\omega_{i}$ and their associated period $T_{i}$ on the satellite's angular variables $i=(l,g,h)$ induced by the effect of $\Jd$ and the luni-solar perturbations.}
  \label{tab:Eff_sec}
\end{table}

\subsection{Degree of accuracy}

In this part, we have sought to evaluate the degree of validity of our analytical model related to each external disturbing body, sketched in Figure~\ref{fig:diag}. As reference solution, we have integrated the motion equations defined in~\eqref{eq:Modele_dynamique} using a fixed step variational integrator at the order~6. It is based on a Runge-Kutta Nyström method, fully described in the thesis~\cite{Lion_2013ab}. This kind of integrators are well-adapted for high elliptical orbits and numerical propagation over long periods. For more details about the variational integrator, see e.g. \citet{Marsden_2001aa}, \citet{West_2004aa}, \citet{Farr_2007aa}, \citet{Farr_2009aa}. 

For both analytical and numerical propagations, we have assumed that the apparent motion for each disturbing body can be parametrized by a linear precessing model (see Section~\ref{ssec:Dyn_Model}). 
The Fourier series in multiple of the mean anomaly~\eqref{eq:dev_fourier_M} are expanded up to the order $Q=4$, which is quite enough for external bodies such as the Moon and Sun.


\paragraph{Perturbations related to the Sun}

\begin{figure}[H]
  \captionsetup{width=\textwidth}  
  \centering
  \subfloat[Analytical propagation.]{%
    \includegraphics[clip=true, trim = 0 0 0 0, width=0.95\linewidth]{%
      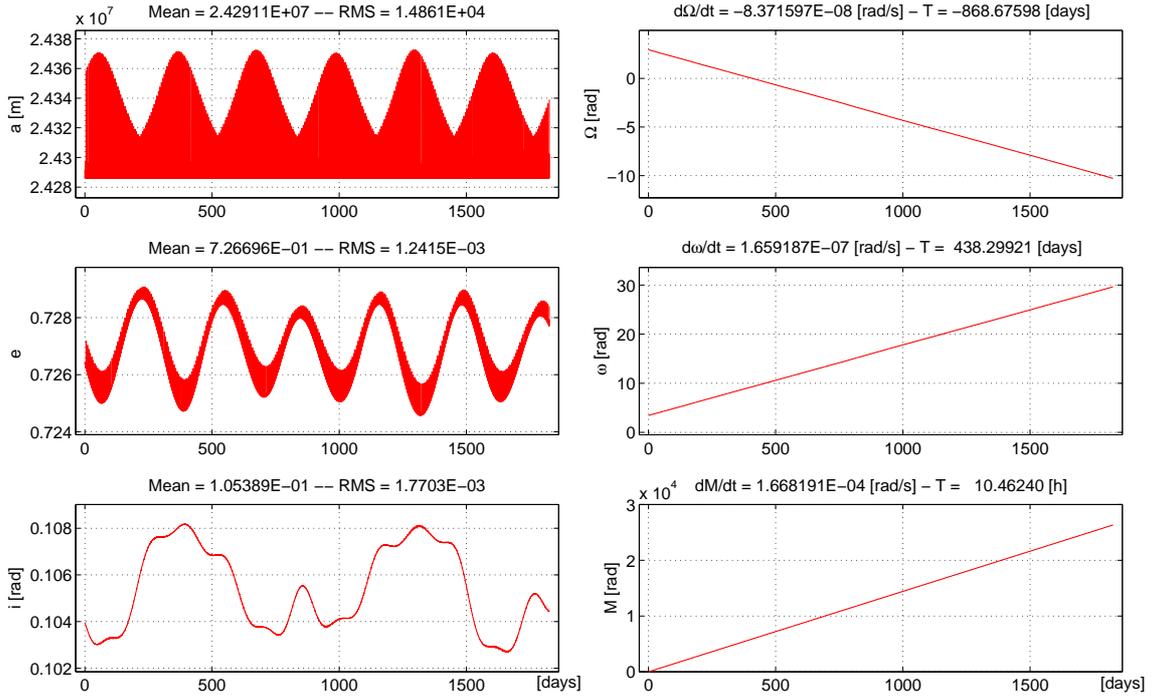}
    \label{fig:Propa_ana_J2o2_0S_3-0-4_3x2}}
  \\
  \subfloat[Comparison between the analytical solution and the numerical simulation.]{%
    \includegraphics[clip=true, trim = 0 0 0 0, width=0.95\linewidth]{%
      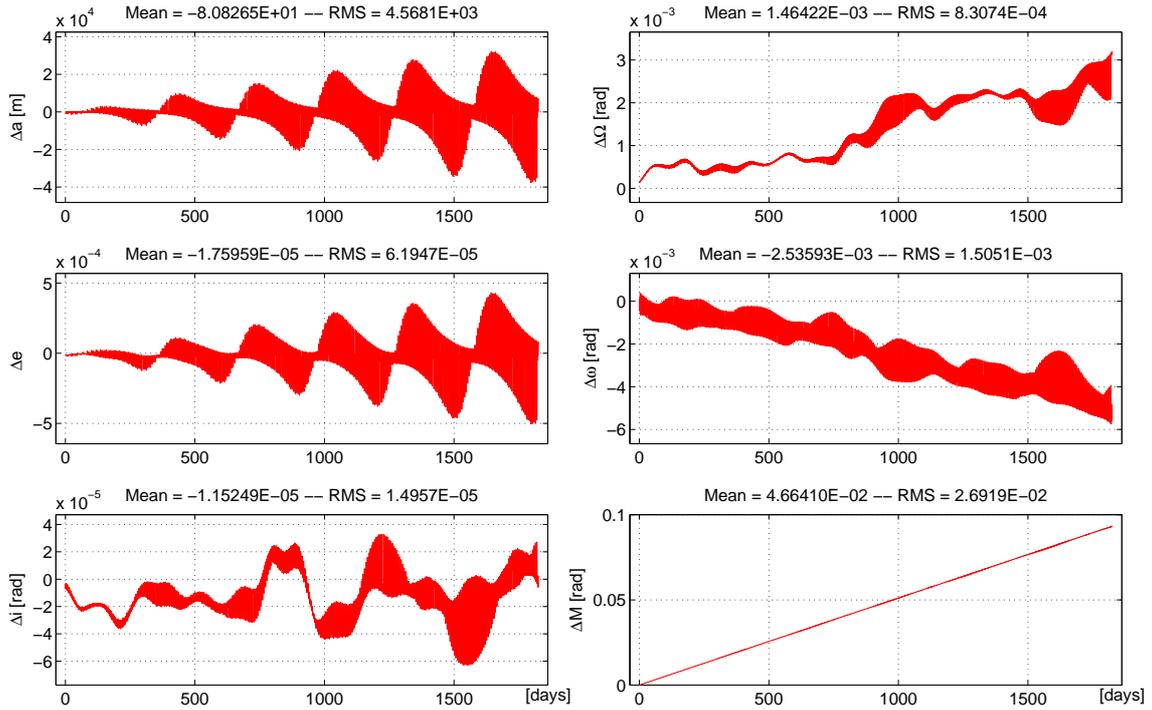}
    \label{fig:Erreurs_J2o2_S0_3-0-4_3x2}}
  \caption{Perturbations: $\Jd$ + Sun with settings $n = 3$, $\sigma = 0$.}
  \label{fig:J2o2_S0_3-0-4_3x2}
\end{figure}

Let us consider the perturbations of $\Jd$ and the Sun. 
Since the variations of $\mca{H}_{\astrosun}$ are proportional to~$(a/a_{\astrosun})^{n}$, it is enough to expand the series up to $n=3$, so~$\sim 4\times10^{-12}$. The parameter $\sigma$ is kept zero here because the short periodic corrections involved by the time dependence are very small. Indeed, we have for example for $a$ only few meters in RMS for the first order correction and a few centimeters beyond, to be compared to the $\sim 10$~km of the analytical solution plotted in Figure~\ref{fig:Propa_ana_J2o2_0S_3-0-4_3x2}. This permits us also to reduce considerably the time computation without loose in stability and accuracy.

In Figure~\ref{fig:Erreurs_J2o2_S0_3-0-4_3x2}, we show that the analytical model fits the numerical solution quite well. 
The main source of errors is the computation of the mean elements $\mca{E}_{i}^{\dstar}$ from the initial osculating elements~$\mca{E}_{i}$, which is truncated in our work at the order~1 in $\Jd$. If we apply the direct-inverse change of variables on the elements $\mca{E}_{i}(t_{0})$, which corresponds to steps~(1) and~(3) of the Figure~\ref{fig:diag}, the resulting new initial elements noted~ $\mcat{E}_{i}(t_{0})$ differ by a quantity that is not null. This is why the errors on the metric elements are not centered on zero. This yields a phase error increasing the amplitude of the error during the propagation as we can see clearly on $\Delta a$ or $\Delta e$. 
The problem is slighty different for the angular variables. The small remaining slopes result from the approximation of the secular effects due to $\Jd$:%
%
  \begin{enumerate}[label=\roman*)]
    \setlength{\itemsep}{0pt}
  \item We have used the Brouwer's expressions expanded up to $\Jd^2$, so we have not totally all the contribution of $\Jd$ compared to the numerical solution;
  \item The secular terms are evaluated from the mean elements at step (2).
  \end{enumerate}

\paragraph{Perturbations related to the Moon}

\begin{figure}[H]
   \captionsetup{width=0.55\textwidth}  
   \centering
        \includegraphics[clip=true, trim = 0 0 0 0, width=0.485\linewidth]{%
          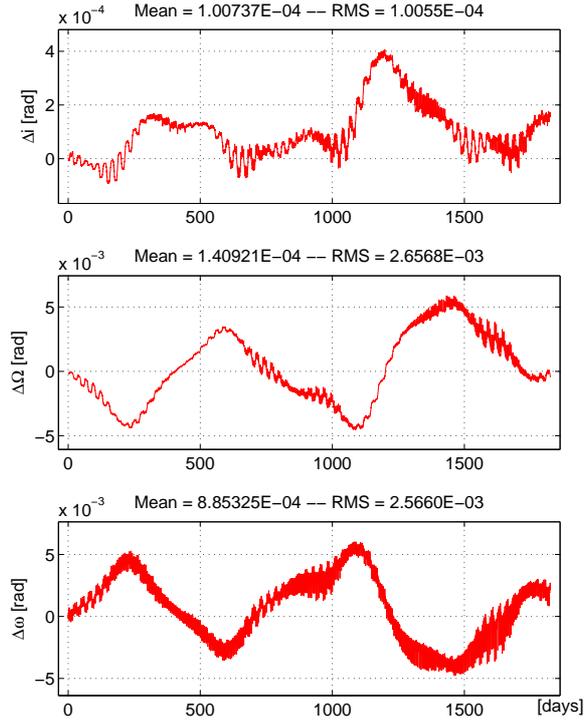}
      \caption{Perturbations: $\Jd$ + Moon with settings $n=~2$, $\sigma=0$. Comparison between the analytical solution and the numerical simulation.}
      \label{fig:Erreurs_J2o2_0M_2-0-4_ihg}
    \end{figure}

Similar tests have been done with the Moon in Figure~\ref{fig:J2o2_S0_3-0-4_3x2}. Because $a_{\leftmoon} \ll a_{\astrosun}$, it is necessary here to develop the disturbing function up to at least to $n=4$ to improve significantly the solution, see Figure~\ref{fig:Erreurs_J2o2_0M_2-0-4_ihg}. 
We remark that the modeling errors are more important than for the Sun, particularly on the long periodic part of $I$, $\omega$ and $\Omega$. This is not surprising since the motion of the Moon is both faster and more complicated than the motion of the Sun.


\begin{figure}[H]
  \captionsetup{width=\textwidth}  
  \centering
  \subfloat[Analytical propagation.]{%
    \includegraphics[clip=true, trim = 0 0 0 0, width=0.95\linewidth]{%
      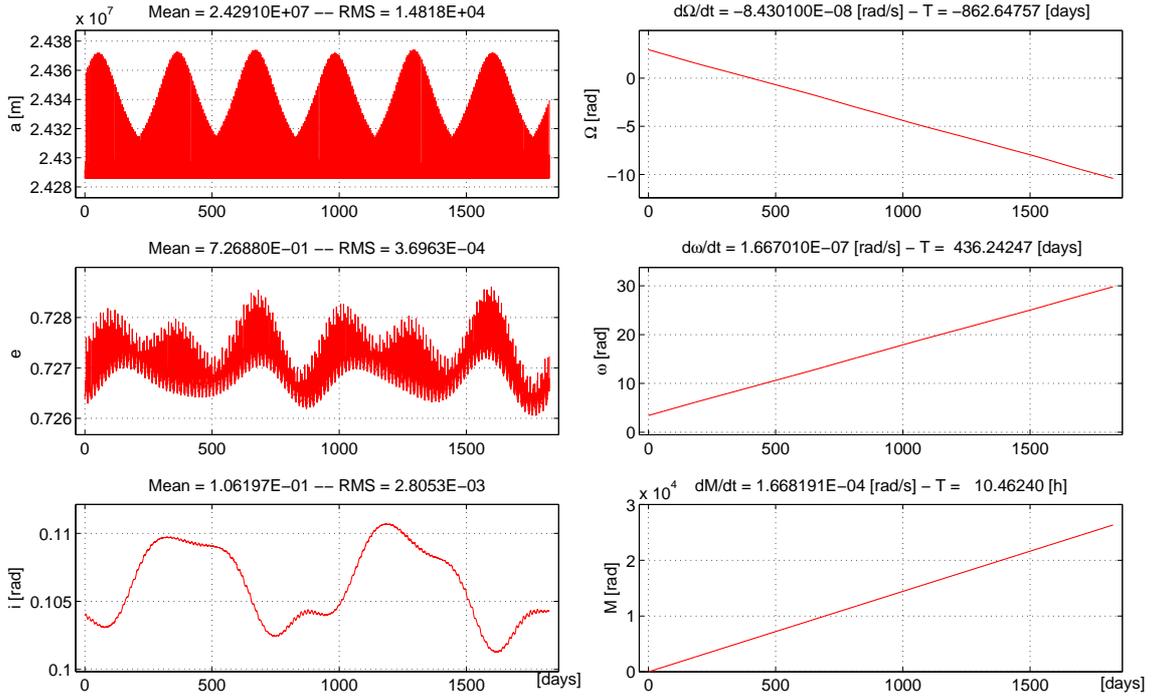}
    \label{fig:Propa_ana_J2o2_M0_4-0-4_3x2}}
  \\
  \subfloat[Comparison between the analytical solution and the numerical simulation.]{%
    \includegraphics[clip=true, trim = 0 0 0 0, width=0.95\linewidth]{%
      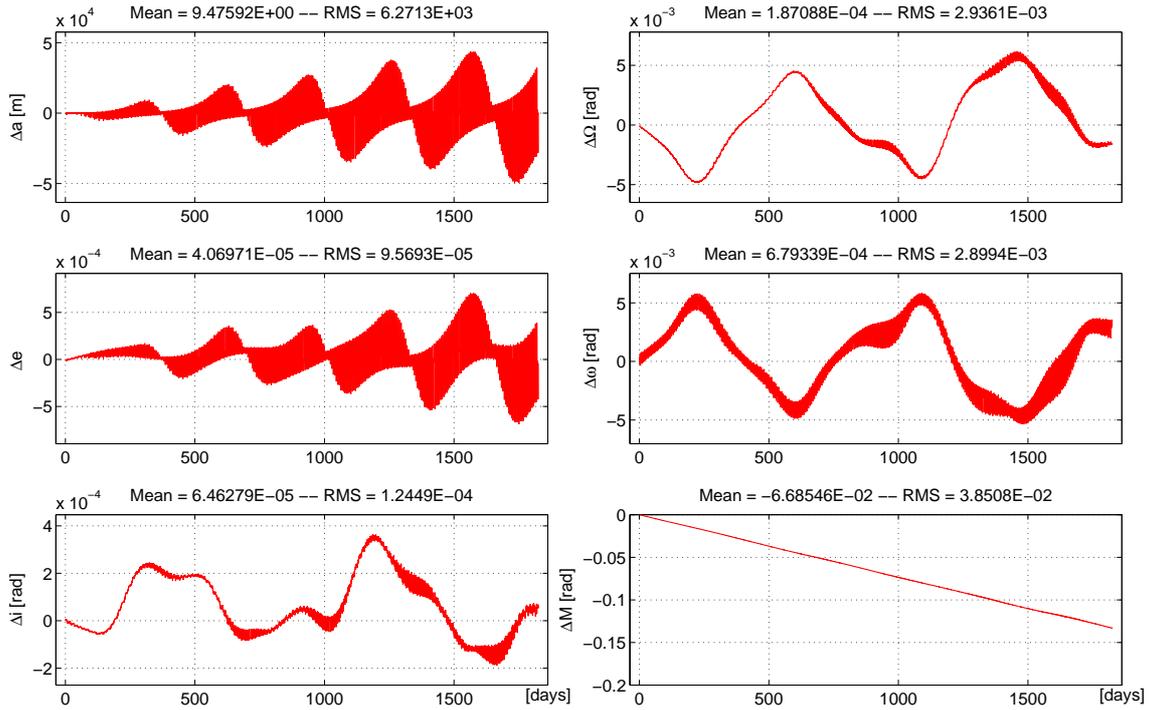}
    \label{fig:Erreurs_J2o2_0M_4-0-4_3x2}}
  \caption{Perturbations: $\Jd$ + Moon with settings $n =4$, $\sigma = 0$.}
  \label{fig:J2o2_0M_4-0-4_3x2}
\end{figure}

\subsection{Explicit time dependence}

\begin{figure}[H]
   \captionsetup{width=\textwidth}  
   \centering
     \subfloat[J2 + Sun with $n = 3$.]{%
        \includegraphics[clip=true, trim = 0 0 0 0, width=0.485\linewidth]{%
          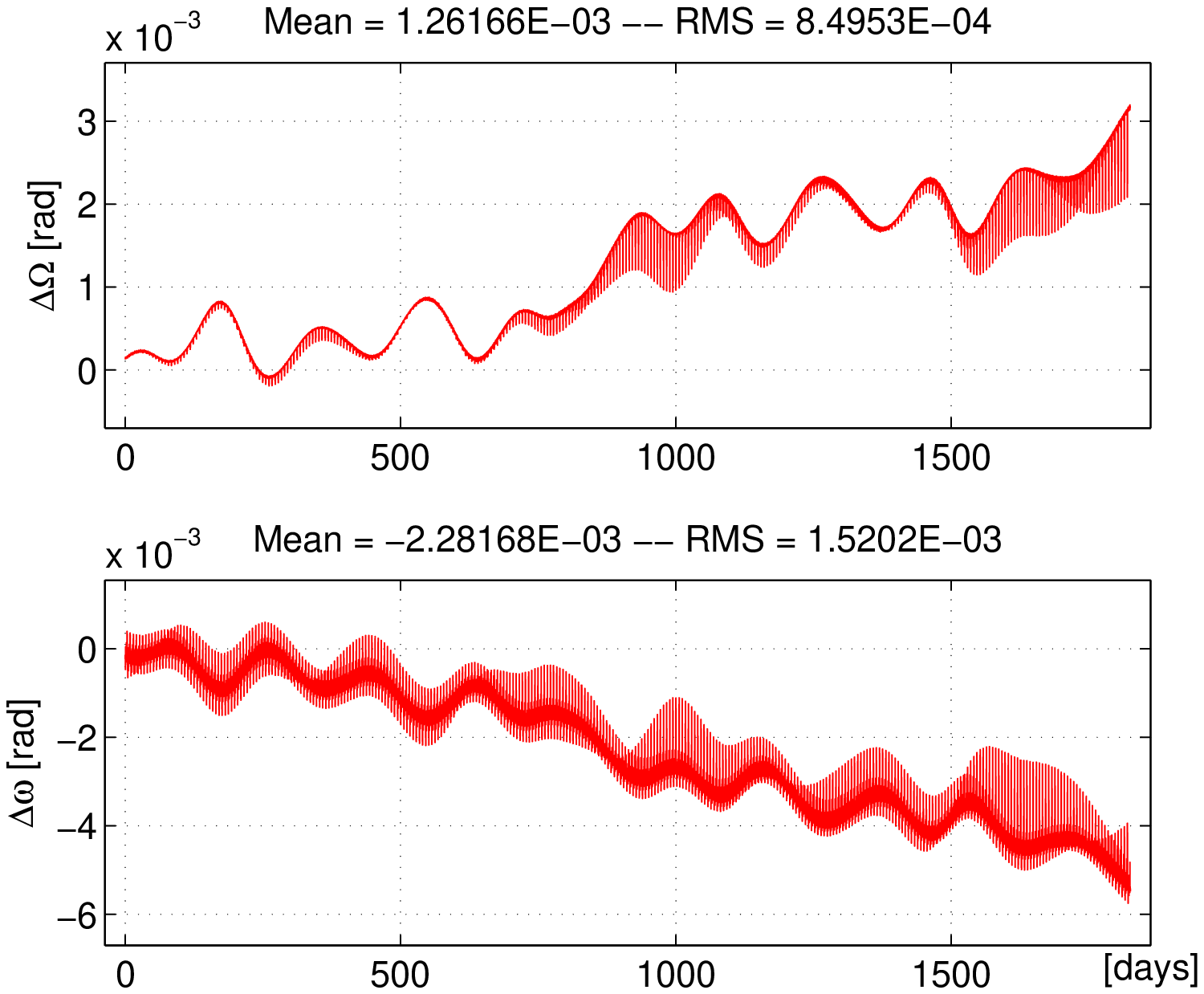}
        \label{fig:Erreurs_J2o2_S0_3-0-4_hg}}
      \hfill
      \subfloat[J2 + Moon with $n = 4$.]{%
        \includegraphics[clip=true, trim = 0 0 0 0, width=0.485\linewidth]{%
          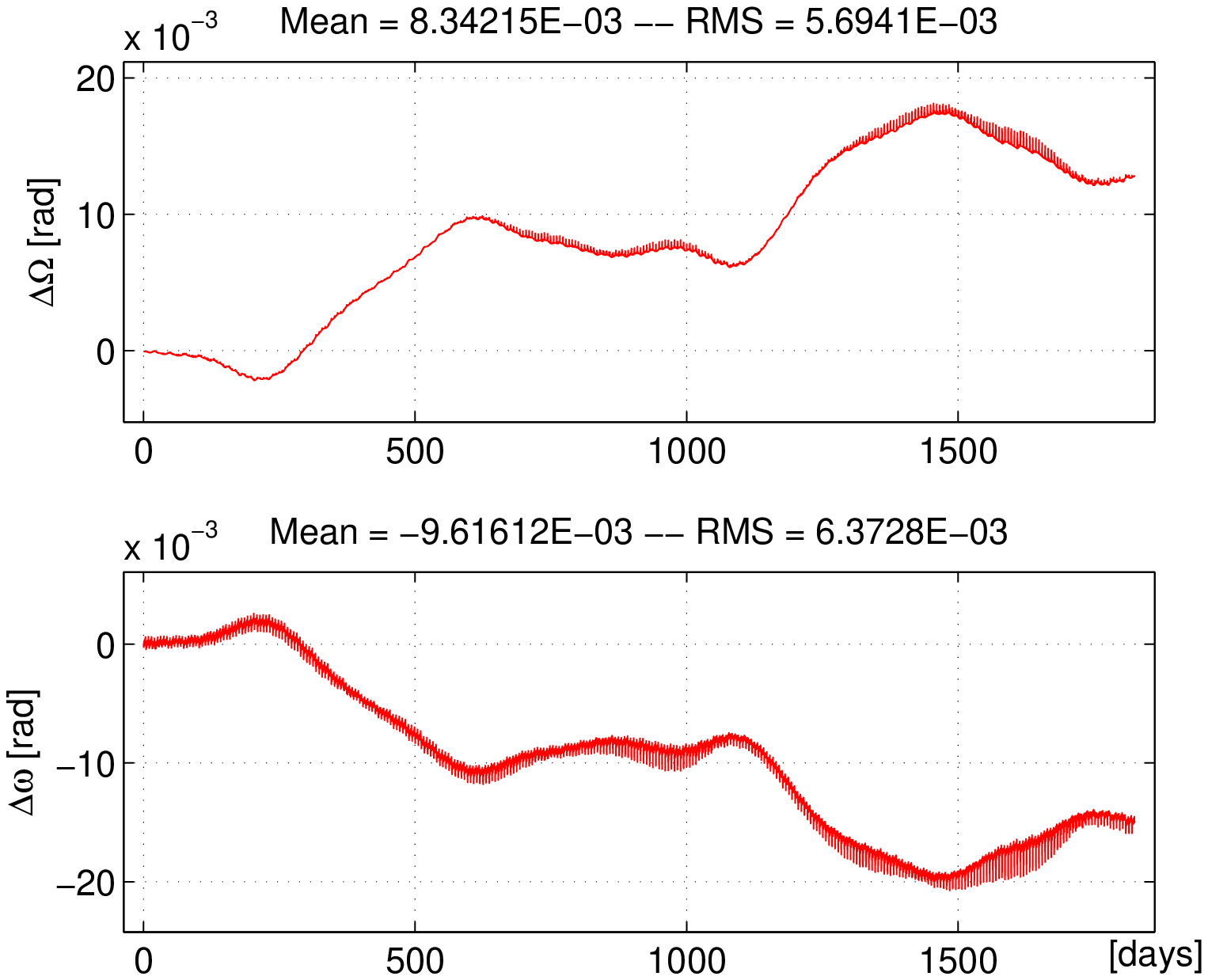}
        \label{fig:Erreurs_J2o2_0M_3-0-4_hg}}
      \caption{Comparison between the analytical solution with no time dependence and the numerical simulation.}
      \label{fig:Erreurs_J2o2_lunisol_3-0-4_hg_wt}
    \end{figure}

We have evaluated the contribution of the explicit time dependence due to the third body motion, modeled by the generating function~$\mca{W}_{2,3b}$ and the corrections~$\sigma>0$.
In Figure~\ref{fig:Erreurs_J2o2_lunisol_3-0-4_hg_wt}, we have performed similar tests than in the previously one, but with $\mca{W}_{2,3b} = 0$. By comparing the errors with the results in Figures~\ref{fig:Erreurs_J2o2_S0_3-0-4_3x2} and \ref{fig:Erreurs_J2o2_0M_4-0-4_3x2}, we can see that taking into account the time dependence permits to reduce the drift rate up to a factor of~3.



\FloatBarrier


\subsection{Inverse-direct change of variables}

Another way to evaluate the performance of our analytical propagator is to apply on a set of osculating elements an inverse transformation, then a direct transformation, and to verify that we find the identity.    

\begin{figure}[!htb]
  \captionsetup{width=0.9\textwidth}  
  \centering
  \includegraphics[clip=true, trim = 00 0 0 00, width=0.95\linewidth]{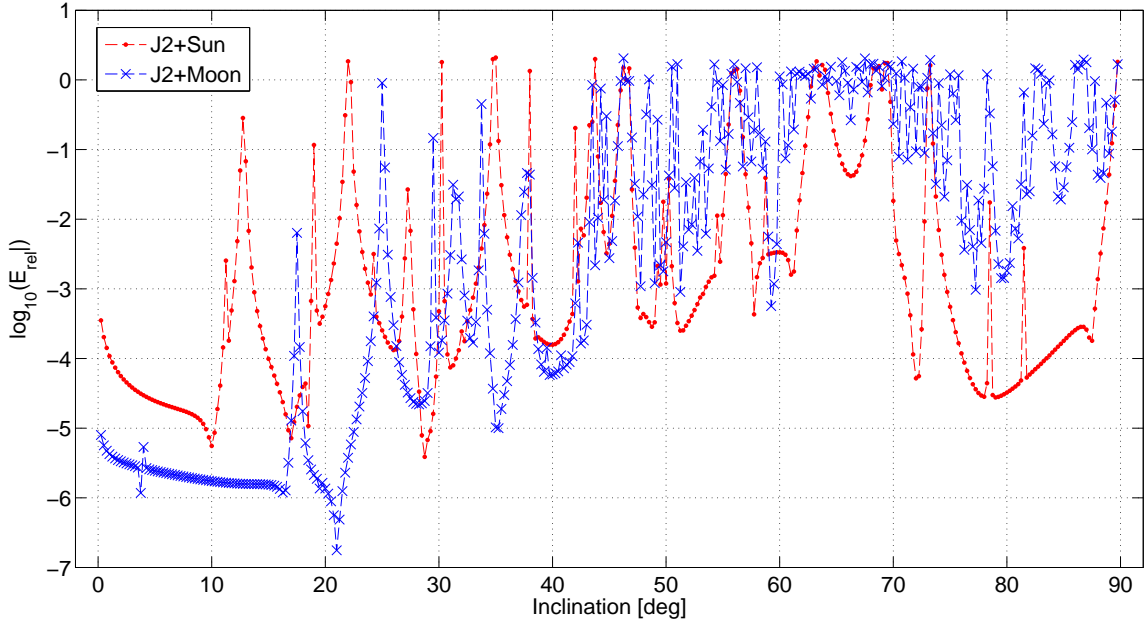}
  \caption{Relative errors occurring during the inverse-direct transformations as a function of the inclination. Settings $n = 3$, $\sigma = 0$.}
  \label{fig:Err_pos_J2o2_SM_merge_304e726}
\end{figure}

Figure~\ref{fig:Err_pos_J2o2_SM_merge_304e726} is a sample plot of the behavior of the relative errors in position due to the successive transformations of the initial osculating elements~$\mca{E}_{i}(t_{0})$ illustrated in Figure~\ref{fig:diag}, against inclination. Other parameters remain the same. For more clarity, results for the Sun and Moon have been computed separately and the relative error is defined by
\begin{equation}
  \label{eq:def_err_pos}
  E_{rel} = \frac{\norm{\vecbf{x}_{i} - \vecbf{x}_{f}}}{\norm{\vecbf{x}_{i}}}
\end{equation}
with $\vecbf{x}_{i}$ and $\vecbf{x}_{f}$ denoting respectively the rectangular coordinates before and after the transformation of the elements $\mca{E}_{i}$.




\FloatBarrier

As we can seen, the change of variables is very sensitive to the inclination. The peaks correspond to a resonant term, that the theory does not deal with. By collecting the resonant frequencies in APHEO satisfying the conditions:%
\begin{align}
  \label{eq:resonances_lunisol}
  & \varepsilon_{\leftmoon}^{\pm} 
  = {} (n-2p) \omega_{g,\Jd} + m \omega_{h,\Jd} \pm \left( q' \omega_{l_{\leftmoon}} + (n-2p') \omega_{g_{\leftmoon}} + m' \omega_{h_{\leftmoon}} \right) 
  \approx 0
  \\
  & \varepsilon_{\astrosun} 
  = {} (n-2p) \omega_{g,\Jd} + m \omega_{h,\Jd}  - \left(q' \omega_{l_{\astrosun}} + (n-2p') \omega_{g_{\astrosun}} \right) 
  \approx 0
\end{align}
we were able to identify the set of resonances given in~ \citep{Hughes_1980aa} up to $n=3$ for this test.



%
%
%

\section{Conclusions}
\label{sec:Ccl}

The construction of an analytical theory of the third-body perturbations in case of highly elliptical orbits is facing several difficulties. 
In term of the mean anomaly, the Fourier series converge slowly, whereas the disturbing function is time dependent.
Each of these difficulties can be solved separately with more-or-less classical methods. 
Concerning the first issue, it is already known that the Fourier series in multiple of eccentric anomaly are finite series. Their use in an analytical theory is less simple than classical series in multiple of the mean anomaly, but remains tractable. The time dependence is not a great difficulty, only a complication: after having introduced the appropriate (time linear) angular variables in the disturbing function, these variables must be taken into account in the PDE to solve during the construction of the theory. 

However, combining the two problems (expansion in terms of the eccentric anomaly and time dependence) in the same theory is a more serious issue. 
In particular, solving the PDE \eqref{eq:91} in order to express the short periodic terms generating function is not trivial. 
In this work we have proposed two ways:%
\begin{itemize}
\item using an appropriated development of the disturbing function involving the Fourier series with respect to the eccentric anomaly;
\item computing the solution of the PDE by means of an iterative process, which is equivalent to a development of a generator in power series of a small ratio of angular frequencies.
\end{itemize}
These allowed us to get a compact solution using special functions. The main advantage is that the degree of approximation of the solution (e.g. the truncation~$n$ of the development in spherical harmonics  and the number of iterations $\sigma$ in the resolution of \eqref{eq:91} can be chosen by the user as needed and not fixed once and for all when constructing the theory.


\renewcommand*\bibfont{\small}

\setlength\bibsep{0pt} 



\newpage

\begin{appendices}{\normalsize}


\section{Determination of $\mean{\mca{V}_{1}}_{l}$ for $\mca{V}_{1} \ne 0$}
  \label{anx:J2_astuces}

Begin to expand the disturbing function due to zonal harmonics~$J_{n}$ in Hill-Whittaker variables~\citep{Kaula_1961aa, Kaula_1966aa}, 
\begin{equation}
  \label{eq:Geopot_Jn_EO}
    \mca{R} = {} \frac{\mu}{r}  
    \sum_{n \geq 2}^{}
    \sum_{p = 0}^{n}
    \fracp*{R_{\varoplus}}{r}^{n}
    J_{n}
    F_{n,0,p} (I) \cos \left[ (q+n-2p) \nu + (n-2p) g + n \frac{\pi}{2} \right] 
\end{equation}
with~$F_{n,0,p} (I)$ the standard inclination functions related to the Kaula's inclination functions~$\widetilde{F}_{n,0,p} (I) = (-1)^{\trunc{(n+m+1)/2}} {F}_{n,0,p} (I)$ \citep[see][]{Gooding_2008aa}.

In order to isolate easily the secular and periodic terms, we can introduce the elliptic motion functions as defined in~\eqref{eq:_Phi_nk}, and we develop them by using Fourier series of the true anomaly in the same way than~\citet{Brouwer_1959aa}. However, we propose here to involve the Hansen-like coefficients~$Y_{q}^{n,k}(e)$ \citep{Brumberg_1994aa} which permits to have a more general, compact and closed form representation :%
\begin{equation}
  \label{eq:dev_fourier_v}
  \Phi_{n,k} {} 
  = \sum_{q=-\infty}^{+\infty} Y_{q}^{n,k}(e) \exp\ci q \nu
\end{equation}
These coefficients are very interesting. In case where~$n < 0$, the~$Y$-elements can be expressed in closed form and the sum over~$q$ is bounded by~$\intervalleff{-n+k}{n+k}$. Indeed, they are null for~\linebreak[2]{$0 \le -n < k$},%
\begin{equation}
\label{eq:Y_1_a}
Y_{q}^{n,k} = {} \beta^{K_{+}} \frac{(1-\beta^2)^{2n}}{(1+\beta^2)^{n}} 
\sum_{s=0}^{-n-K_+} \binom{-n}{s} \binom{-n}{s+K_{+}} \beta^{2s}
\end{equation}
with~$\beta = e/(1 + \eta)$ and~$K_{+} = k-q \ge 0$.

More over, we can deduce from~\eqref{eq:dev_fourier_M} the properties:%
\begin{align}
  Y_{q}^{n,k} 
  = {} Y_{-q}^{n,-k}
  = {} Y_{0}^{n,k-q}
  = {} Y_{q-k}^{n,0}
\end{align}
Hence, rewriting \eqref{eq:Geopot_Jn_EO} as%
\begin{subequations}
  \begin{align}
  \label{eq:Geopot_Jn_EO_b}
  & \mca{R} 
  = {} \sum_{n \geq 2}^{}
  \sum_{p = 0}^{n}
  \sum_{q = -n+1}^{n-1}
  \eta \fracp*{a}{r}^{2} \mca{A}_{n,p,q}
  \cos \left[ (q+n-2p) \nu + (n-2p) g + n \frac{\pi}{2} \right]
  \\
  & \mca{A}_{n,p,q}
  = {} \frac{\mu}{a} 
  \fracp*{R_{\varoplus}}{a}^{n}
  J_{n} \, \frac{1}{\eta} Y_{q}^{-n+1,0}(e) \, F_{n,0,p} (I) 
  \end{align}
\end{subequations}
the secular part is%
\begin{align}
  \label{eq:Geopot_Jn_EO_sec}
    \mca{R}_{sec} 
    & = {} \frac{1}{(2\pi)^{2}}\int_{0}^{2\pi} \mca{R} \diff{l} \diff{g} 
    = {} \frac{1}{(2\pi)^{2}} {\int_{0}^{2\pi} \int_{0}^{2\pi} }  \frac{1}{\eta} \fracp*{r}{a}^{2} \mca{R} \diff{\nu} \diff{g}
    \nonumber
    \\
    & = {} \sum_{p = 0}^{n} (-1)^{p} \mca{A}_{2p,p,0}
\end{align}
and the periodic part%
\begin{equation}
  \label{eq:Geopot_Jn_EO_per}
  \begin{split}
    \mca{R}_{per} 
    & = {}
    \sum_{n \geq 2}^{}
    \sum_{p = 0}^{n}
    \sum_{q = -n+1}^{n-1}
    \left[ \eta \fracp*{a}{r}^{2} - \delta_{2p}^{n} \delta_{0}^{q} \right]
    \mca{A}_{n,p,q}
    \\
    & \hspace{1em} \times \cos \left[ (q+n-2p) \nu + (n-2p) g + n \frac{\pi}{2} \right] 
    \end{split}
\end{equation}
From the last equation, it is easy to show that the generating function modeling the short periods term due to the zonal harmonic at the order one can be given by:%
\begin{align}
  \label{eq:12}
    \mca{V}_{1}
    & = {-} \frac{1}{\omega_{0}} \int \mca{R}_{per} \diff{l}
    \nonumber
    \\
    \begin{split}
    & = {-} \frac{1}{\omega_{0}} 
    \sum_{n \geq 2}^{} 
    \sum_{p = 0}^{n} 
    \sum_{q = -n+1}^{n-1}
    \mca{A}_{n,p,q} \left\lbrace
      \frac{1-\delta_{n-2p}^{-q}}{q+n-2p + \delta_{n-2p}^{-q}}
      \right.
      \\ & \left. \hspace{2em}
      \times \sin \left[ (q+n-2p) \nu + (n-2p) g + n \frac{\pi}{2} \right]
      + 
      \delta_{0}^{n-2p} \delta_{0}^{q} \phi
    \right\rbrace
  \end{split}
\end{align}
with~$\phi = \nu - l$ the equation of the center.

We can now proceed to the computation of the mean value of~$\mca{V}_{1}$ with respect to the mean anomaly~$l$ needed in the coupling term \eqref{eq:1}. Because~$\phi$ contains purely periodic terms, so~$\mean*{\phi}_{l} = 0$, the only contribution comes from the averaging over~$l$ of the trigonometric term~$\sin (\alpha\nu + \beta g)$. By isolating~$\nu$ and~$g$, we get
\begin{equation}
  \label{eq:14}
  \mean*{\sin (\alpha\nu + \beta g)}_{l}
  = {} \mean{\sin \alpha\nu}_{l} \cos \beta g
  + \mean{\cos \alpha\nu}_{l} \sin \beta g
\end{equation}
As sine is an odd function~$\mean*{\sin \alpha\nu}_{l} = 0$ and according to the definition~\eqref{eq:dev_fourier_M}, equation \eqref{eq:14} reduces to the simple value:%
\begin{align}
  \label{eq:15}
  \mean*{\sin (\alpha\nu + \beta g)}_{l}
  & 
  = {} \frac{1}{\eta} Y_{0}^{2,\alpha} \sin \beta g
\end{align}
Hence, 
\begin{equation}
  \label{eq:16}
  \begin{split}
  \mean*{\mca{V}_{1}}_{l} 
  & = {-} \frac{1}{\omega_{0}} 
    \sum_{n \geq 2}^{} 
    \sum_{p = 0}^{n} 
    \sum_{q = -n+1}^{n-1}
    \mca{A}_{n,p,q} 
    \frac{1-\delta_{n-2p}^{-q}}{q+n-2p + \delta_{n-2p}^{-q}} 
    \\ & \hspace{2em}
    \times \frac{1}{\eta}  Y_{0}^{2,q+n-2p} \sin (n-2p) g
    \end{split}
\end{equation}

\section{Proof of the recurrence $\mca{V}_{2}^{(\sigma)}$}
  \label{anx:proof_rec}

Let us prove that if the solution~\eqref{eq:103_a_bis} works for the order~$\sigma$, then it works for the order~$\sigma +1~$. 

Inserting~\eqref{eq:103_a_bis} into~\eqref{eq:91b} leads to%
\begin{footnotesize}
  \begin{align}
    \label{eq:105}
    \dpartial{\mca{V}_{2}^{(\sigma+1)}}{E}  
    & = {-} \sum_{}^{\bullet} \left[
      \vphantom{\sum_{1}^{1}}
      (-1)^{\sigma} \mcat{A}_{n,\ldots,q,q'}^{(\sigma+1)}
      \exp\ci \Theta_{n,\ldots,0,q'}
      \sum_{\mathclap{s=-(\sigma+2)}}^{\sigma+2} \zeta_{q,s}^{\prime (\sigma+1)}\exp\ci (q+s)E
    \right]
    \;,
  \end{align}
\end{footnotesize}
\\
with%
\begin{subequations}
  \label{eq:106}
  \begin{align}
    \label{eq:106a}
    & \mcat{A}_{n,\ldots,q,q'}^{(\sigma+1)} 
    = \overline\varepsilon_{n,\ldots,q'}\mcat{A}_{n,\ldots,q,q'}^{(\sigma)}
    \;,
    \\
    \label{eq:106b}
    & \zeta_{q,s}^{\prime (\sigma+1)} 
    = \ci \left( \zeta_{q,s}^{(\sigma)} -\frac{e}{2}\zeta_{q,s-1}^{(\sigma)} -\frac{e}{2} \zeta_{q,s+1}^{(\sigma)} \right)
    \;.
  \end{align}
\end{subequations}
\textcolor{black}{
Let us make two remarks. 
Firstly, we consider in our process that an element~$\zeta_{q,s}^{\prime (\sigma)}$ is null if no value has been assigned in previous iterations. 
Secondly, by imposing the constraint}
\begin{equation}
  \label{eq:104}
  \zeta_{q,-q}^{(\sigma)} 
  = \frac{e}{2}\left(\zeta_{q,-q-1}^{(\sigma)} +\zeta_{q,-q+1}^{(\sigma)}\right)
  \;,
\end{equation}
we ensure that~$\lpartial{\mca{V}^{(\sigma+1)}}{E}$ contains no terms independent of~$E$:%
\begin{equation}
  \label{eq:109}
  \zeta_{q,-q}^{\prime (\sigma+1)} = 
  \ci \left( \zeta_{q,-q}^{(\sigma)} -\frac{e}{2}\zeta_{q,-q-1}^{(\sigma)} -\frac{e}{2} \zeta_{q,-q+1}^{(\sigma)} \right) 
  = 0
  \;,
\end{equation}
and so~$\mean{\mca{V}^{(\sigma)}}_{l} = 0$.

Finally, we derive from the integration of \eqref{eq:105} the correction at the order~${\sigma +1}$:%
\begin{small}
  \begin{align}
    \label{eq:Gen_SP_Moon_exp}
    \mca{V}_{2}^{(\sigma+1)} 
    & = {-} \sum_{}^{\bullet}
    \left(
      \frac{(-1)^{\sigma+1}}{\ci}
      \mcat{A}_{n,\ldots,q,q'}^{(\sigma+1)} 
      \sum_{\mathclap{s=-(\sigma+2)}}^{\sigma+2} \zeta_{q,s}^{(\sigma+1)} 
      \exp\ci\Theta_{n,\ldots,q+s,q'}
    \right)
  \end{align}
\end{small}

  \section{Derivatives of the generating functions}
  \label{anx:derive_gen}

In this part, we give all the partial derivatives with respect to the keplerian elements $(a,e,I,h,g,l)$ of the generating functions $\mca{V}_{1,\Jd}$, $w_{1,\Jd}$, $\mca{V}_{2,3b}$ and $\mcat{W}_{2,3c}$, required in the canonical transformations.

\subsection{Partial derivatives of~$\mca{V}_{1,\Jd}$}
  \label{sec:dW1}

Derivatives of $\mca{V}_{1,\Jd}$ with respect to the kelperian elements are those given in \citet{Brouwer_1959aa}:%
\begin{small}
\begin{subequations}
  \label{eq:dS_lgh}
  \begin{align}
    \label{eq:dS1sdl}
    \dpartial{\mca{V}_{1,\Jd}}{l} 
    & = 2 \gamma_{2} \eta L
    \left\lbrace
      \left(
        -1 + 3 c^{2}
      \right)
      \left[
        \fracp*{a}{r}^{3} \eta^3  - 1
      \right]
      + 3 s^2
      \fracp*{a}{r}^{3} \cos (2g + 2\nu)
    \right\rbrace
    \;,
    \\
    \label{eq:dS1sdg}
    \dpartial{\mca{V}_{1,\Jd}}{g} 
    & = 6 \gamma_{2} s^2 G
    \left[
      \cos (2g + 2\nu) + e \cos (2g + \nu) + \frac{e}{3} \cos (2g + 3\nu)
    \right]
    \;,
    \\
    \label{eq:dS1sdh}
    \dpartial{\mca{V}_{1,\Jd}}{h} 
    & = 0
    %
    \\
    \label{eq:dS1sda}
    \dpartial{\mca{V}_{1,\Jd}}{a} 
    & = {-} 2 \frac{\gamma_{2}}{a} \mca{V}_{1,\Jd}
    \;,
    \\
    \label{eq:dS1sde}
    \begin{split}
    \dpartial{\mca{V}_{1,\Jd}}{e} 
    & = {} \gamma_{2} G
    \Bigg\{
    2 \left(
    -1 + 3 c^{2}
    \right)
    \left(
    \Gamma + 1
    \right)
    \sin\nu
    \;,
    \\ & \hspace{1em} 
    - 3 s^2 \left[
    \left(
    \Gamma  - 1
    \right)
    \sin (2g + \nu)
    - \left(
    \Gamma + \frac{1}{3}
    \right)
    \sin (2g + 3\nu)
    \right]
    \Bigg\}
    \;,
    \end{split}
    \\
    \label{eq:dS1sdI}
    \begin{split}
    \dpartial{\mca{V}_{1,\Jd}}{I}  
    & = 6 \gamma_{2} G c s \left[
    -2 \left( \phi + e\sin\nu \right) \right. 
    \\ & \left. \hspace{1em} 
    + \sin (2g + 2\nu) + e \sin (2g + \nu) 
    + \frac{e}{3} \sin (2g + 3\nu)
    \right]
    \;.
    \end{split}
  \end{align}
\end{subequations}
\end{small}
\\
with~$\Gamma = {} \frac{a}{r} \left( \frac{a}{r} \eta^{2} + 1 \right)$.

   \subsection{Partial derivatives of~$w_{1,\Jd}$}
  \label{sec:dV1}

Since~$w_{1,\Jd}$ is only independent of~$h$ and~$l$,%
\begin{subequations}
  \label{eq:dT_lgh}
  \begin{align}
    \label{eq:dT1sda}
    \dpartial{w_{1,\Jd}}{a} 
    & = {-} w_{1,\Jd} \left(
    \frac{1}{\varpi_{g}} \dpartial{\varpi_{g}}{a} + \frac{5}{a}
    \right)
    \;,
    \\
    \label{eq:dT1sde}
    \dpartial{w_{1,\Jd}}{e} 
    & = {-} w_{1,\Jd} \left(
    \frac{1}{\varpi_{g}} \dpartial{\varpi_{g}}{e} - \frac{7 e}{\eta^2} - \frac{2}{e}
    \right)
    \;,
    \\
    \label{eq:dT1sdI}
    \dpartial{w_{1,\Jd}}{I} 
    & = {-} w_{1,\Jd} \left(
    \frac{1}{\varpi_{g}} \dpartial{\varpi_{g}}{I} - \frac{4 c (7 - 15 s^2)}{s (14 - 15 s^2)}
    \right)
    \;,
    \\
    \label{eq:T_LP_g}
    \dpartial{w_{1,\Jd}}{g} 
    & = 6 \frac{\omega_{0}}{\varpi_{g}}  \gamma_{2}^{2} e^2 G s^2 \left( 14-15 s^{2} \right) \cos 2g 
    \;,
    \\
    \label{eq:T_LP_h}
    \dpartial{w_{1,\Jd}}{h} 
    & = \dpartial{w_{1,\Jd}}{l} = {} 0 
    \;,
  \end{align}
\end{subequations}
Note that these relations yield to those of \citet{Brouwer_1959aa} for $\varpi_{g} = \omega_{g,\Jd}$.

   \subsection{Partial derivatives of~$w_{coup}$}
   \label{sec:dw_coup}
   
   The generating function $w_{coup}$ is independent of~$h$ and~$l$. We have:%
   \begin{subequations}
   	\label{eq:dwcoup}
   	\begin{align}
   	\label{eq:dwcsda}
   	\dpartial{w_{coup}}{a} 
   	& = {} \frac{2}{\varpi_{g}} \left[
   	\dpartial{\omega_{g,3b}}{a} - \frac{\omega_{g,3b}}{\varpi_{g}} 
   	\left( \dpartial{\varpi_{g}}{a} + \frac{2}{a} \right) 
   	\right]
   	\gamma_{2}\,s\,G\,\Gamma_{3} \sin 2 g
   	\;,
   	\\
   	\label{eq:dwcsde}
   	\dpartial{w_{coup}}{e} 
   	& = {} \frac{2}{\varpi_{g}} \left[
   	\left( \dpartial{\omega_{g,3b}}{e} - \frac{\omega_{g,3b}}{\varpi_{g}} \dpartial{\varpi_{g}}{e} \right) \Gamma_{3}
   	+ e \frac{\omega_{g,3b}}{\varpi_{g}} \left(
   	\frac{2 (2+\eta)}{\left(1+\eta\right)^2} + \frac{3}{\eta^2}
   	\right) 
   	\right]
   	\gamma_{2}\,G\,s \sin 2 g
   	\;,
   	\\
   	\label{eq:dwcsdI}
   	\dpartial{w_{coup}}{I} 
   	& = {} \frac{2}{\varpi_{g}} \left[
   	s \dpartial{\omega_{g,3b}}{I} - \frac{\omega_{g,3b}}{\varpi_{g}} 
   	\left( s \dpartial{\varpi_{g}}{I} - c \right)
   	\right]
   	\gamma_{2} G\,\Gamma_{3} \sin 2 g
   	\;,
   	\\
   	\label{eq:dwcsdg}
   	\dpartial{w_{coup}}{g} 
   	& = 4 \gamma_{2}\,s\,G\,\Gamma_{3} \frac{\omega_{g,3b}}{\varpi_{g}} \cos 2 g
   	\;,
   	\\
   	\label{eq:dwcsdh}
   	\dpartial{w_{coup}}{h} 
   	& = \dpartial{w_{coup}}{l} = {} 0 
   	\;,
   	\end{align}
   \end{subequations}
	with
\begin{align}
\label{eq:def_Gamma_3}
& \Gamma_{3}
= {} \frac{(1-\eta)(1+2\eta)}{1+\eta}
\end{align}

\subsection{Partial derivatives of~$\mca{V}_{2}$}
  \label{sec:dU_sp}

  Since we have chosen to represent~$\mca{V}_{2}$ by a series (see~\eqref{eq:Taylor_W1sp}):%
  \begin{equation}
    \label{eq:dWspsdEO}
    \mca{V}_{2} 
    = \mca{V}_{2}^{(0)}+ \sum_{\sigma \geq 1} \mca{V}_{2}^{(\sigma)}
    \;,
  \end{equation}
  these derivatives are deduced from~$\mca{V}^{(\sigma)}$.

  \paragraph{\pucedingk Derivatives of~$\mca{V}^{(\sigma)}$}
  ~~ \\
  From~\eqref{eq:103_a_bis} and \eqref{eq:103_sun_ci}, we get%
  \begin{subequations}
    \allowdisplaybreaks
    \begin{align}
      \label{eq:dUrsdx}
      \dpartial{\mca{V}^{(\sigma)}_{n,\ldots,q,q'}}{h} & = {}  
      m\, \mca{V}^{(\sigma)}_{n,\ldots,q,q'}
      \;,
      \\
      \dpartial{\mca{V}^{(\sigma)}_{n,\ldots,q,q'}}{g} & = {} 
      (n-2p)\, \mca{V}^{(\sigma)}_{n,\ldots,q,q'}
      \;.
    \end{align}
  \end{subequations}
  Since our generating functions involves the satellite's eccentric anomaly~$E$, the~$l$-derivative is%
  \begin{align}
    \dpartial{\mca{V}^{(\sigma)}_{n,\ldots,q,q'}}{l} 
    & = {} \frac{r}{a} \dpartial{\mca{V}^{(\sigma)}_{n,\ldots,q,q'}}{E}
    \;,
    \nonumber
    \\
    \label{eq:dUrsdl}
    & =  {-} (-1)^{\sigma}
    \frac{r}{a} \mcat{A}_{n,\ldots,q,q'}^{(\sigma)} 
    \sum_{s=-(\sigma+1)}^{\sigma+1} (q+s)\zeta_{q,s}^{(\sigma)} \exp\ci \Theta_{n,\ldots,q+s,q'}
    \;.
  \end{align}
  For the metric elements,%
  \begin{align}
    \label{eq:dUrsdEOm}
    \dpartial{\mca{V}^{(\sigma)}_{n,\ldots,q,q'}}{(a,I)} 
    & =  {-} \frac{(-1)^{\sigma}}{\ci}
    \dpartial{\mcat{A}_{n,\ldots,q,q'}^{(\sigma)}}{(a,I)} 
    \sum_{s=-(\sigma+1)}^{\sigma+1} 
    \zeta_{q,s}^{(\sigma)}(e)
    \exp\ci \Theta_{n,\ldots,q+s,q'}
    \;.
  \end{align}
  Given that~$\mca{V}^{(\sigma)}$ depends on~$e$ both explicitly and implicitly through~${E(e,l)}$, with use of
  \begin{align}
    \label{eq:dAEsde_exp}
    \dpartial{E}{e}
    = {}  \frac{a}{r} \sin E
    = {} \frac{1}{2\ci} \frac{a}{r}  \left[ \exp(\ci E) - \exp(-\ci E) \right]
    \;,
  \end{align}
  we obtain%
  \begin{small}
    \begin{align}
      \begin{split}
        \dpartial{\mca{V}^{(\sigma)}_{n,\ldots,q,q'}}{e} 
        {} & = {-} \frac{(-1)^{\sigma}}{\ci}
        \sum_{s=-(\sigma+1)}^{\sigma+1} \left[ 
          \mcat{A}_{n,\ldots,q,q'}^{(\sigma)} \left(
            \dpartial{\zeta_{q,s}^{(\sigma)}}{e}
            +\ci (q+s) \dpartial{E}{e} \zeta_{q,s}^{(\sigma)}
          \right)
        \right. \\ {} & \left. {} \hspace{6em}
          + \dpartial{\mcat{A}_{n,\ldots,q,q'}^{(\sigma)}}{e} 
          \zeta_{q,s}^{(\sigma)} 
        \right]
        \exp\ci \Theta_{n,\ldots,q+s,q'}
        \;.
      \end{split}
    \end{align}
  \end{small}

  \paragraph{\pucedingk Derivatives of~$\mcat{A}_{n,\ldots,q,q'}^{(\sigma)}$}
  ~~ \\
  From~\eqref{eq:def_As_wrt_A0}, we can compute derivatives of~$\mcat{A}_{n,\ldots,q,q'}^{(\sigma)}$ by recurrence:%
  \begin{align}
    \allowdisplaybreaks
    \label{eq:127}
    \dpartial{\mcat{A}_{n,\ldots,q,q'}^{(\sigma)}}{(a,e,I)} =  {} 
    \sigma \dpartial{\overline\varepsilon_{n,\ldots,q'}}{(a,e,I)} \mcat{A}_{n,\ldots,q,q'}^{(\sigma-1)}
    {} + \overline\varepsilon_{n,\ldots,q'}^{\sigma}\dpartial{\mcat{A}_{n,\ldots,q,q'}^{(0)}}{(a,e,I)}
    \;,
  \end{align}
  with%
  \begin{equation}
    \label{eq:dA0dx}
    \dpartial{\mcat{A}_{n,\ldots,q,q'}^{(0)} }{(a,e,I)} 
    = {} \frac{1}{(q + \delta_{0}^{q}) \omega_{0}} 
    \left[
      \dpartial{\mcat{A}_{n,\ldots,q,q'}}{(a,e,I)}
      - (q + \delta_{0}^{q}) \mcat{A}_{n,\ldots,q,q'}^{(0)}\dpartial{\omega_{0}}{(a,e,I)} 
    \right] 
  \end{equation}
  The differential of~$\varepsilon_{n,\ldots,q'}^{}$ and~$\overline\varepsilon_{n,\ldots,q'}^{}$ are given by:%
  \begin{subequations}
    \begin{align}
    \label{eq:dvarepsdx_sp}
    \dpartial{\varepsilon_{n,\ldots,q'}^{}}{(a,e,I)} 
    & = (n-2p) \dpartial{\omega_{g}}{(a,e,I)} + m \dpartial{\omega_{h}}{(a,e,I)}
    \\
    \dpartial{\overline \varepsilon_{n,\ldots,q'}^{}}{(a,e,I)} 
    & = \frac{1}{\omega_{0}}
    \left(
    \dpartial{\varepsilon_{n,\ldots,q'}^{}}{(a,e,I)}
    - \overline\varepsilon_{n,\ldots,q'}^{} \dpartial{\omega_{0}}{(a,e,I)} 
    \right)
    \;,
    \end{align}
  \end{subequations}
  with the partial derivatives of~$\omega_{0}$, $\omega_{g}$ and~$\omega_{h}$ defined in the Appendice~\ref{anx:derivative_pulsations}.

  \paragraph{\pucedingk Derivatives of~$\zeta_{q,s}^{(\sigma)}$}
  ~~ \\
  Derivatives of~$\zeta_{q,s}^{(\sigma)}$ can be computed by means of recurrence relation. Using~\eqref{eq:def_zeta_os} for~$s \neq q$, we get%
  \begin{align}
    \label{eq:131_a}
    \begin{split}
      \dpartial{\zeta_{q,s}^{(\sigma+1)}}{e} & = 
      \frac{1}{(q+s)}
      \left( 
        \dpartial{\zeta_{q,s}^{(\sigma)}}{e} -\frac{e}{2}\dpartial{\zeta_{q,s-1}^{(\sigma)} }{e} 
        -\frac{e}{2} \dpartial{\zeta_{q,s+1}^{(\sigma)} }{e}
      \right. \\ {} & \left. {} \hspace{4em}
        -\frac{1}{2} \zeta_{q,s-1}^{(\sigma)} -\frac{1}{2}\zeta_{q,s+1}^{(\sigma)}
      \right)
      \;,
    \end{split}
  \end{align}
  and for~$s = -q$%
  \begin{align}
    \label{eq:131_b}
    \dpartial{\zeta_{q,-q}^{(\sigma+1)}}{e} =
    \frac{1}{2} \left(
      \zeta_{q,-q-1}^{(\sigma+1)}
      + \zeta_{q,-q+1}^{(\sigma+1)} + e\dpartial{\zeta_{q,-q-1}^{(\sigma+1)}}{e}
      + e\dpartial{\zeta_{q,-q+1}^{(\sigma+1)}}{e}
    \right) 
    \;.
  \end{align}
  Concerning the initialization $\lpartial{\zeta_{q,0}^{(0)}}{e}$, according to~\eqref{eq:def_zeta_o0}, we have%
  \begin{subequations}
    \label{eq:130}
    \begin{empheq}[left=\empheqlbrace]{alignat=6}
      &\dpartial{\zeta_{q,0}^{(0)}}{e}   = 0 
      && ,\quad
      \dpartial{\zeta_{q,-1}^{(0)}}{e}  = \frac{1}{2}\delta^q_{1} 
      && ,\quad 
      \dpartial{\zeta_{q,1}^{(0)}}{e}  =\frac{1}{2}\delta^{q}_{-1}
      && ,\quad \mbox{ si } q \neq 0
      \\
      &\dpartial{\zeta_{0,0}^{(0)}}{e}   =0 
      && ,\quad
      \dpartial{\zeta_{0,-1}^{(0)}}{e} =-\frac{1}{2} 
      && ,\quad 
      \dpartial{\zeta_{0,1}^{(0)}}{e}  = \frac{1}{2}
      && ,\quad \mbox{ si } q=0 
    \end{empheq}
  \end{subequations}

  \subsection{Partial derivatives of~$\mcat{W}_{2,3b}$}
  \label{sec:dW_lp}

  Let us pose%
  \begin{align}
    \label{eq:Wlp_C}
    \mca{C}_{n,\ldots,q,q'}
    = {} \frac{\mcat{A}_{n,\ldots,q,q'}}{\varepsilon_{n,\ldots,q'}}
    \;.
  \end{align}
  such that the generating function eliminating the long-periodic terms~$\mcat{W}_{2,3b}$ writes:%
  \begin{equation}
    \label{eq:121_bis}
    \mcat{W}_{2,3b} = {}
    2 \ci \sum^{\bullet\bullet}
    \mca{C}_{n,\ldots,0,q'}
    \exp\ci \Theta_{n,\ldots,0,q'}
    \;.
  \end{equation}
  The partial derivatives with respect to~$(l, g, h)$ are simple to obtain:%
  \begin{subequations}
    \label{eq:dWlpsdx}
    \begin{align}
      \label{eq:dWlpsdh}
      \dpartial{\mcat{W}_{2,3b}}{h}  
      & =  {-} 2 \sum^{\bullet\bullet}
      m\, \mca{C}_{n,\ldots,0,q'}
      \exp\ci \Theta_{n,\ldots,0,q'}
      \;,
      \\
      \label{eq:dWlpsdg}
      \dpartial{\mcat{W}_{2,3b}}{g}  
      & =  {-} 2 \sum^{\bullet\bullet}
      (n-2\,p)\, \mca{C}_{n,\ldots,0,q'}
      \exp\ci \Theta_{n,\ldots,0,q'}
      \;,
      \\
      \label{eq:dWlpsdl}
      \dpartial{\mcat{W}_{2,3b}}{l}  
      & = {} 0
      \;,
    \end{align}
  \end{subequations}
  while those with respect to the metric elements~$(a,e,I)$ require more attention:%
  \begin{align}
    \label{eq:dWlpsdX}
    \dpartial{\mcat{W}_{2,3b}}{(a,e,I)} 
    & = {} 2 \ci \sum^{\bullet\bullet}
    \dpartial{\mca{C}_{n,\ldots,0,q'}}{(a,e,I)} 
    \exp\ci \Theta_{n,\ldots,0,q'}
    \;,
  \end{align}
  with%
  \begin{align}
    \label{eq:Wlp_dC}
    \dpartial{\mca{C}_{n,\ldots,0,q'}}{(a,e,I)} 
    & = {} \frac{1}{\varepsilon_{n,\ldots,q'}} \left(
      \dpartial{\mcat{A}_{n,\ldots,0,q'}}{(a,e,I)}
      - \mca{C}_{n,\ldots,0,q'} \dpartial{\varepsilon_{n,\ldots,q'}}{(a,e,I)} 
    \right)
    \;.
  \end{align} 
  Derivatives of~$\mca{A}_{n,\ldots,q,q'}$ are%
  \begin{subequations}
    \allowdisplaybreaks
    \label{eq:dAdx}
    \begin{align}
      \dpartial{\mcat{A}_{n,\ldots,q,q'}}{a} & = {}
      \frac{n}{a} \mcat{A}_{n,\ldots,q,q'}
      \;,
      \\
      \dpartial{\mcat{A}_{n,\ldots,q,q'}}{e} & = {}
      \left( \frac{\mcat{A}_{n,\ldots,q,q'}}{Z_{q}^{n+1,n-2p}} \right)
      \dpartial{ Z_{q}^{n+1,n-2p}}{e} 
      \;,
      %
      \\
      \dpartial{\mcat{A}_{n,\ldots,q,q'}}{I} & = {}
      \left(  \frac{\mcat{A}_{n,\ldots,q,q'}}{F_{n,m,p}} \right)
      \dpartial{F_{n,m,p}}{I}
      \;.
      %
      {}
    \end{align}
  \end{subequations}
  and for $\varepsilon_{n,\ldots,q'}^{}$:%
    \begin{align}
    \label{eq:dvarepsdx_lp}
    \dpartial{\varepsilon_{n,\ldots,q'}^{}}{(a,e,I)} 
    & = (n-2p) \dpartial{\varpi_{g}}{(a,e,I)} + m \dpartial{\varpi_{h}}{(a,e,I)}
    \end{align}
  where $\varpi_{g}$ and $\varpi_{h}$ defined in \eqref{eq:pulsation_coup}. Partial derivatives of the pulsations are established in Appendice~\ref{anx:derivative_pulsations}.

\section{Derivatives of the pulsations}
\label{anx:derivative_pulsations}

Derivatives of~$\vecbf{x} = (a,e,I)^\intercal$ with respect to~$\vecbf{Y} = (L,G,H)^\intercal$ are
\begin{equation}
\label{eq:Jaco_EOvsED}
{J} 
= \dpartial{\vecbf{x}}{\vecbf{Y}} 
= \begin{pmatrix}
\dfrac{2}{n_0\,a} & 0 & 0 \\ 
\dfrac{\eta^2}{n_0\,a^2\,e} & - \dfrac{\eta}{n_0\,a^2\,e} & 0 \\ 
0 & \dfrac{c}{n_0\,a^2\eta\,s} & - \dfrac{1}{n_0\,a^2\eta\,s}
\end{pmatrix} 
\end{equation}
Denoting~$J_{i,j} = \lpartial{x_{i}}{Y_{j}}$, we have 
\begin{subequations}
  \label{eq:dJsdx}
  \begin{align}
  & \dpartial{J}{a} 
  = {-}\frac{1}{2a} 
  \begin{pmatrix}
  - J_{a,L} & 0 & 0 \\ 
  J_{e,L} & J_{e,G} & 0 \\ 
  0 & J_{I,G} & J_{I,H}
  \end{pmatrix} 
  \\
  & \dpartial{J}{e} 
  = {-}\frac{1}{e \eta^2} 
  \begin{pmatrix}
  0 & 0 & 0 \\ 
  (1+e^2) J_{e,L} & J_{e,G} & 0 \\ 
  0 & -e^2 J_{I,G} & -e^2 J_{I,H}
  \end{pmatrix} 
  \\
  & \dpartial{J}{I} 
  = \frac{1}{s} 
  \begin{pmatrix}
  0 & 0 & 0 \\ 
  0 & 0 & 0 \\ 
  0 & J_{I,H} & J_{I,G}
  \end{pmatrix} 
  \end{align}
\end{subequations}
Given that~$\cpoiss{\vecbf{y}}{\vecbf{x}} = \lpartial{\vecbf{x}}{\vecbf{Y}}$, derivatives of the pulsation can be written
\begin{align}
\label{eq:13}
\dpartial{}{x_{k}} \left( \derive{\vecbf{y}}{t} \right)
& = {} \sum_{j = 1}^{3} 
\left(
\frac{\partial^{2}{x_{j}}}{\partial{x_{k}}\,\partial{\vecbf{Y}}}
\dpartial{\mca{M}}{x_{j}}
+ \dpartial{x_{j}}{\vecbf{Y}} \frac{\partial^{2}{\mca{M}}} {\partial{x_{k}} \partial{x_{j}}}
\right)
\end{align}
We give in Table~\ref{tab:derive_eff_sec_EO} the derivatives of mean motion and secular variations due to~$\Jd$. 
Those associated to the secular part of the third body, $\lpartial{\omega_{y,3b}}{\vecbf{x}}$, can be determined by using the expression~\eqref{eq:21} and the partial derivatives%
\begin{align}
\label{eq:derive_H3secc}
&\frac{\partial^{i+j+k}{\mca{H}_{3b,sec}}}{\partial^{i}{a}\,\partial^{j}{e}\,\partial^{k}{I}}
= {-} \sum_{p \ge 1}^{}
\binom{2p}{i} \frac{i !}{a^{i}} \mca{B}_{p} \dpartial[j]{{F}_{2p,0,p}(I)}{I} \dpartial[k]{Z_{0}^{2p+1,0}(e)}{e}
\end{align}
{
  \renewcommand{\arraystretch}{1.5}
  
  \begin{table}[H]
    \captionsetup{width=1\textwidth}  
    \centering 
    \begin{small}
      \begin{tabular}{%
          *{1}{K{5em}}
          *{1}{K{12em}}
          *{1}{K{8em}}
          *{1}{K{6em}} 
        } 
        \toprule 
        
        & \multicolumn{1}{c}{$a$}
        & \multicolumn{1}{c}{$e$}
        & \multicolumn{1}{c}{$I$} \\
        
        \cmidrule( r){1-1}
        \cmidrule(l ){2-4}

        \lpartial{\gamma_{2}}{[\cdot{}]}
        & -2 \frac{\gamma_{2}}{a}
        &  4 \frac{e}{\eta^{2}} \gamma_{2}
        & 0 
        \\
        
        \cmidrule( r){1-1}
        \cmidrule(l ){2-4}
        
        \lpartial{\omega_{0}}{[\cdot{}]}
        & {-} \frac{3}{2} \frac{\omega_{0}}{a}
        & 0
        & 0 
        \\
        
        \cmidrule( r){1-1}
        \cmidrule(l ){2-4}
        
        \lpartial{\omega_{l,\Jd}}{[\cdot{}]}
        &
        {-}\frac{3}{2} \frac{\omega_{0}}{a} 
        \left\lbrack 1 + 14\,\gamma_2 \eta \left( 1 - 3\,c^2 \right) \right\rbrack
        &
        18\,\omega_{0} \gamma_2 \frac{e}{\eta} \left( 1 - 3\,c^2 \right)
        &
        36\,\omega_{0} \gamma_2 \eta \,c \,s
        \\
        \lpartial{\omega_{g,\Jd}}{[\cdot{}]}
        &
        {-} 21 \frac{\omega_{0}}{a} \gamma_2 \left( 1 - 5\,c^2 \right)
        &
        24\,\omega_{0} \gamma_2 \frac{e}{\eta^2} \left( 1 - 5\,c^2 \right)
        &
        60\,\omega_{0} \gamma_2 \,c \,s
        \\
        \lpartial{\omega_{h,\Jd}}{[\cdot{}]}
        &
        {-} 42\frac{\omega_{0}}{a} \gamma_2 \,c
        &
        48\,\omega_{0} \gamma_2 \frac{e}{\eta^2} \,c
        &
        {-} 12\,\omega_{0} \gamma_2 \,s
        \\
        \bottomrule 
      \end{tabular}
    \end{small}
    
    \caption{Partial derivatives of~$\gamma_2$, the mean motion~$\omega_{0}$ and the secular variations~$(\omega_{l,\Jd}, \omega_{g,\Jd}, \omega_{h,\Jd})$ with respect to~$(a, e, I)$. We have put~$\eta = \sqrt{1-e^2}$, $c = \cos I$ and~$s = \sin I$.}
    \label{tab:derive_eff_sec_EO}
  \end{table} 
}


  \section{Trigonometric transformation}
  \label{anx:trigo_transfo}

  In this appendix, we present a method to convert the determining functions related to the disturbing body from exponential to trigonometric form. The method is similar to that we have used in~\citet[][see Section 3]{Lion_2011ac}. Since this kind of transformation is tedious but can easily lead to algebraic errors, we give the main results to establish the trigonometric expression of the Moon's long-periodic and the short-periodic generating function (much harder than for the Sun).

  \subsection{Symmetries}
  \label{sanx:func_spe_sym}

  To begin, the eccentricity functions:~$X_{q}^{n,m} (e)$, $Z_{q}^{n,m} (e)$, $\zeta_{q,s}^{(\sigma)}(e)$, and the inclination functions:~$F_{n,m,p}(I)$, $U_{n,m,s} (\epsilon)$, admit several symmetries. Particularly, we have for~$m < 0$ the following properties:%
  \begin{subequations}
    \label{eq:Fonc_spe_ex_sym}
    \begin{alignat}{3}
      \label{eq:Znms_sym_a}
      Z_{-s}^{n,-m}(e) 
      & = {} Z_{s}^{n,m}(e)
      && \;, \quad \left[ n, m, s \in \Z \right]
      \\
      \label{eq:Xnms_sym_a}
      X_{-s}^{n,-m} (e)
      & = {} X_{s}^{n,m} (e)
      && \;, \quad \left[ n, m, s \in \Z \right]
      \\
      \label{eq:Zeta_sym_a}
      \zeta_{-q,-s}^{(\sigma)} (e)
      & = {} (-1)^{\sigma} (1-2\delta_{0}^{q}) \zeta_{q,s}^{(\sigma)} (e)
      && \;, \quad \left[ q, s \in \Z ; \, \sigma \in \N \right]
    \end{alignat}
  \end{subequations}
  and
  \begin{subequations}
    \label{eq:Fonc_spe_inc_sym}
    \begin{alignat}{3}
      \label{eq:Fnmk_sym_a}
      F_{n,-m,n-p}(I) 
      & = {} (-1)^{n-m} \frac{(n-m)!}{(n+m)!} F_{n,m,p}(I)
      && \;, \quad \left[ n, p \in \N ; \, m \in \Z \right]
      \\
      \label{eq:Unmk_sym_a}
      U_{n,-m,-s} (\epsilon) 
      & = {} (-1)^{s-m} U_{n,m,s} (\epsilon)
      && \;, \quad \left[ n \in \N ; \, m, s \in \Z \right]
    \end{alignat}
  \end{subequations}
  Note that the last symmetry can be obtained from~\eqref{eq:Wigner_U2d} and the relation \citep[e.g.][]{Wigner_1959aa, Sneeuw_1992aa}%
  \begin{align}
    \label{eq:dnmk_sym_a}
    d_{n,-m,-s} (\epsilon) 
    & = {} (-1)^{s-m} \frac{(n-m)! (n+s)!}{(n+m)! (n-s)!} d_{n,m,s} (\epsilon)
    \;.
  \end{align}
  Consider now three polynomial functions~$f$, $f'$ and~$g$ defined by%
  \begin{subequations}
    \label{eq:def_fandg}
    \begin{alignat}{3}
      \label{eq:def_fnmpq_sat}
      & f
      && = f_{n,m,p,q} 
      && = {} (n-2p) \alpha_{1} + m \alpha_{2} + p \alpha_{3} + q \alpha_{4}
      \\
      \label{eq:def_fnmpq_3c}
      & f^{\prime} 
      && = f_{n,m',p',q'}^{\prime} 
      && = {} (n-2p') \alpha_{1}^{\prime} + m' \alpha_{2}^{\prime} + p' \alpha_{3}^{\prime} + q' \alpha_{4}^{\prime}
      \\
      \label{eq:def_g_vs_f_fp}
      & g^{\pm}
      && = g_{n,m,m',p,p',q,q'}^{\pm} 
      && = f \pm f^{\prime}
    \end{alignat}
  \end{subequations}
  with the~$\alpha_{j}$ and~$\alpha_{j}^{\prime}$ some arbitrary real constants. 
  \\
  There results that we have the symmetries%
  \begin{subequations}
    \label{eq:sym_fandg}
    \begin{alignat}{3}
      \label{eq:sym_g_a}
      & g_{n,-m,m',n-p,p',-q,q'}^{} 
      && = - g_{n,m,-m',p,n-p',q,-q'}^{}
      && = - g^{+}
      \\
      \label{eq:sym_g_b}
      & g_{n,-m,-m',n-p,n-p',-q,-q'}^{} 
      && = - g_{n,m,m',p,p',q,q'}^{}
      && = - g^{-}
    \end{alignat}
  \end{subequations}
  In this way, we can deduce easily from the Table of correspondence~\ref{tab:def_behaviour_func} the symmetries with respect to the indices of the functions~$\Psi, \Psi^{\prime}, \Theta, \varepsilon$ involved in the development of our determining functions.

  \begin{table}[!htb]
    \captionsetup{width=0.95\textwidth}  
    \centering 
    \begin{small}
      \begin{tabular}{
          *{1}{R{1em}}
          *{1}{L{6em}}@{\,}*{1}{C}@{\,}*{1}{R{6em}}
          *{1}{L{5em}}@{\,}*{1}{C}@{\,}*{1}{R{5em}}
        } 
        \toprule 
        \textrm{} 
        & \multicolumn{3}{c}{\multirow{1}{*}{Moon}} 
        & \multicolumn{3}{c}{\multirow{1}{*}{Sun}}  \\
        
        \cmidrule( r){1-1}
        \cmidrule(lr){2-4}
        \cmidrule(l ){5-7}
        
        \Psi^{}
        & \Psi_{n,m,p,q} & \equiv & f_{n,m,p,q}
        & \Psi_{n,m,p,q} & \equiv & f_{n,m,p,q}  
        \\
        \Psi^{\prime}
        & \Psi_{n,m',p',q'}^{\prime} & \equiv  & f_{n,m',p',q'}^{\prime}  
        & \Psi_{n,m,p',q'}^{\prime} & \equiv & f_{n,m,p',q'}^{\prime}  
        \\

        \cmidrule( r){1-1}
        \cmidrule(lr){2-4}
        \cmidrule(l ){5-7}

        \Theta
        & \Theta_{n,m,m',p,p',q,q'}^{} & \equiv &  g_{n,m,m',p,p',q,q'}^{}  
        & \Theta_{n,m,p,p',q,q'}^{} & \equiv & g_{n,m,m,p,p',q,q'}^{}  
        \\
        \varepsilon
        & \varepsilon_{n,m,m',p,p',q'}^{} & \equiv &  g_{n,m,m',p,p',0,q'}^{}  
        & \varepsilon_{n,m,p,p',q'}^{} & \equiv & g_{n,m,m,p,p',0,q'}^{}  
        \\
        \bottomrule 
      \end{tabular}
    \end{small}
    \caption{Matching between the functions~$\Psi, \Psi^{\prime}, \Theta, \varepsilon$ and the functions~$f, f^{\prime}, g$ defined in \eqref{eq:def_fandg} according to the disturbing body. Note that in the case where the orbital elements of the satellite and the disturbing body are referred with respect to the same orbital plane (e.g. expansion of the solar disturbing function), $m'$ is not involved but~$m$.}
    \label{tab:def_behaviour_func}
  \end{table}


  \subsection{From exponentials to trigonometric form: implementation principle}
  \label{sanx:conv_principe}

  The main steps to convert an exponential expression to trigonometric form are outlined below:%
  \begin{enumerate}[label=\arabic*)]
    \setlength{\itemsep}{0pt}

  \item Split the sum over~$-n \leq  m \leq n$ into two parts such that~$m$ runs from~$0$ to~$n$. To avoid double counting of~$m = 0$, we must introduce the factor~$(2 - \delta_{0}^{m})/2$. Proceed the same if there is a summation over~$-n \leq  m' \leq n$;
  \item For each terms, if the second index of~$F_{n,k,p}(I)$ is negative, change the indices~$p$ by~$n-p$, $q$ by~$-q$ and ~$s$ by~$-s$ if this is involved. Same for~$F_{n,k,p'}(I')$, replace~$p'$ by~$n-p'$ and~$q'$ by~$-q'$.
  \item Substitute each inclination functions having a negative value as a second index by their symmetry relations given in~\eqref{eq:Fonc_spe_inc_sym};
  \item Substitute each eccentricity function having a negative value as a third index by their symmetry relations given in~\eqref{eq:Fonc_spe_ex_sym};
  \item With the help of Table~\ref{tab:def_behaviour_func}, subsitute each function~$\Theta$ and~$\varepsilon$ by their associated symmetry~\eqref{eq:sym_fandg} if the second index is negative;
  \item Isolate the terms with the same phase, then factorize and convert the exponentials to trigonometric form.
  \end{enumerate}

  \subsection{Long-periodic generating function}
  \label{sanx:lp_det_func}

  Starting from the generating function~$\mcat{W}_{2,3b}$ defined in~\eqref{eq:Ulp_exp} and applying the step~$1$, we have%
  \bgroup
  \begin{footnotesize}
    \allowdisplaybreaks
    \begin{align}
      \label{eq:Gen_LP_moon_etape1}
      \begin{split}
        \mcat{W}_{2,3b} & = {} 
        - \frac{\mu'}{a'}
        \sum_{}^{\circ\circ}
        \frac{ \Delta_{0}^{m,m'} }{2 \ci} (-1)^{m-m'} \fracp*{a}{a'}^{n}
        Z_{0}^{n+1,n-2p}(e) X_{q'}^{-(n+1),n-2p'} (e')
        \\ 
        {} & {} \hspace{1em} \times
        \left[
          \frac{(n-m')!}{(n+m)!} F_{n,m,p}(I) F_{n,m',p'}(I') \frac{ U_{n,m,m'}(\epsilon)}{\varepsilon_{n,m,m',p,p',q'}}
          \exp\ci \Theta_{n,m,m',p,p',0,q'}
        \right. \\ {} & {} \left.
          \hspace{2em} + {}
          \frac{(n+m')!}{(n+m)!} F_{n,m,p}(I) F_{n,-m',p'}(I') \frac{U_{n,m,-m'}(\epsilon)}{\varepsilon_{n,m,-m',p,p',q'}}
          \exp\ci \Theta_{n,m,-m',p,p',0,q'}
        \right. \\ {} & {} \left.
          \hspace{2em}+ {}
          \frac{(n-m')!}{(n-m)!} F_{n,-m,p}(I) F_{n,m',p'}(I') \frac{U_{n,-m,m'}(\epsilon)}{\varepsilon_{n,-m,m',p,p',q'}}
          \exp\ci \Theta_{n,-m,m',p,p',0,q'}
        \right. \\ {} & {} \left.
          \hspace{2em}+ {}
          \frac{(n+m')!}{(n-m)!} F_{n,-m,p}(I) F_{n,-m',p'}(I') \frac{U_{n,-m,-m'}(\epsilon)}{\varepsilon_{n,-m,-m',p,p',q'}}
          \exp\ci \Theta_{n,-m,-m',p,p',0,q'}
        \right] 
      \end{split}
    \end{align}
  \end{footnotesize}
  \egroup
  \\
  with~$\Delta_{0}^{m,m'} = (2 - \delta_{0}^{m}) \, (2 - \delta_{0}^{m'})/2$. Note that the symbols~$\delta_{0}^{q}$ and~$\Delta_{0}^{m,m'}$ are not affected by the changes of sign.
  \\
  Step~$2$ gives
  \begin{footnotesize}
    \begin{equation}
      \label{eq:Gen_LP_moon_etape2}
      \begin{split}
        \mcat{W}_{2,3b} & = {} 
        -\frac{\mu'}{a'}
        \sum_{}^{\circ\circ}
        \frac{ \Delta_{0}^{m,m'} }{2 \ci} (-1)^{m-m'} \fracp*{a}{a'}^{n}
        \\ 
        {} & {} \times
        \left[
          \frac{(n-m')!}{(n+m)!} Z_{0}^{n+1,n-2p}(e) X_{q'}^{-(n+1),n-2p'} (e') F_{n,m,p}(I) F_{n,m',p'}(I')
        \right. \\ {} & {} \left.
          \hspace{2em} \times
          \frac{U_{n,m,m'}(\epsilon)}{\varepsilon_{n,m,m',p,p',q'}}
          \exp\ci \Theta_{n,m,m',p,p',0,q'}
        \right. \\ {} & {} \left.
          \hspace{1em} + {}
          \frac{(n+m')!}{(n+m)!}  Z_{0}^{n+1,n-2p}(e) X_{-q'}^{-(n+1),-(n-2p')} (e') F_{n,m,p}(I) F_{n,-m',n-p'}(I')
        \right. \\ {} & {} \left.
          \hspace{2em} \times
          \frac{U_{n,m,-m'}(\epsilon)}{\varepsilon_{n,m,-m',p,p',q'}}
          \exp\ci \Theta_{n,m,-m',p,n-p',0,-q'}
        \right. \\ {} & {} \left.
          \hspace{1em} {}
          + \frac{(n-m')!}{(n-m)!}  Z_{0}^{n+1,-(n-2p)}(e) X_{q'}^{-(n+1),n-2p'} (e') F_{n,-m,n-p}(I) F_{n,m',p'}(I')
        \right. \\ {} & {} \left.
          \hspace{2em} \times
          \frac{U_{n,-m,-m'}(\epsilon)}{\varepsilon_{n,-m,m',p,p',q'}}
          \exp\ci \Theta_{n,-m,m',n-p,p',0,q'}
        \right. \\ {} & {} \left.
          \hspace{1em} {}
          + \frac{(n+m')!}{(n-m)!} Z_{0}^{n+1,-(n-2p)}(e) X_{-q'}^{-(n+1),-(n-2p')} (e') F_{n,-m,n-p}(I) F_{n,-m',n-p'}(I')
        \right. \\ {} & {} \left.
          \hspace{2em} \times
          \frac{U_{n,-m,-m'}(\epsilon)}{\varepsilon_{n,-m,-m',p,p',q'}}
          \exp\ci \Theta_{n,-m,-m',n-p,n-p',0,-q'}
        \right]
      \end{split}
    \end{equation}
  \end{footnotesize}
  and it results from the steps~$3$ to~$6$:%
  \bgroup
  \begin{equation}
    \label{eq:Gen_LP_moon_etape3}
    \begin{footnotesize}
      \begin{split}
        \mcat{W}_{2,3b} = {} & 
        -\frac{\mu'}{a'}
        \sum_{}^{\circ\circ}
        \frac{ \Delta_{0}^{m,m'} }{2\ci} \mca{A}_{n,m,m',p,p',0,q'}
        \left[
          \frac{U_{n,m,m'}(\epsilon)}{\varepsilon^{-}}
          \left(
            \exp\ci (\Theta_{}^{-}) -  \exp (-\ci \Theta_{}^{-})
          \right)
        \right. \\ {} & {} \left. \hspace{3em} 
          + {}
          (-1)^{n-m'} \frac{U_{n,m,-m'}(\epsilon)}{\varepsilon^{+}}
          \left(
            \exp\ci (\Theta_{}^{+}) -  \exp (-\ci \Theta_{}^{+})
          \right)
        \right]
      \end{split}
    \end{footnotesize}
  \end{equation}
  \egroup
  Making appear the sine, we get the trigonometric development~\eqref{eq:Gen_LP_moon_trig}.

  \subsection{Short-periodic generating function}
  \label{sanx:sp_det_func}

  Starting from the generating function~$\mcat{W}_{2,3b}$ defined in~\eqref{eq:103_a_bis}, step~$1$ gives%
  \begin{equation}
    \label{eq:Gen_CP_moon_etape1}
    \begin{scriptsize}
      \begin{split}
        \mca{V}_{}^{(\sigma)} & = {} 
        - (-1)^{\sigma}  \frac{\mu'}{a'}
        \sideset{}{} \sum_{}^{\circ}
        \frac{ \Delta_{0}^{m,m'} }{2 \ci} \frac{(-1)^{m-m'}}{(q+\delta_{0}^{q}) \omega_{0}}  \fracp*{a}{a'}^{n}
        \sum_{s=-(\sigma+1)}^{\sigma+1} 
        \zeta_{q,s}^{(\sigma)}(e) Z_{q}^{n+1,n-2p}(e) X_{q'}^{-(n+1),n-2p'} (e')
        \\ 
        {} & {} \hspace{1em}  \times 
        \left[
          \frac{(n-m')!}{(n+m)!} \overline\varepsilon_{n,m,m',p,p',q'}^{\sigma} F_{n,m,p}(I) F_{n,m',p'}(I') U_{n,m,m'}(\varepsilon)
          \exp\ci \Theta_{n,m,m',p,p',q+s,q'}
        \right. \\ {} & {} \left. \hspace{2em} 
          + {} \frac{(n+m')!}{(n+m)!} \overline\varepsilon_{n,m,-m',p,p',q'}^{\sigma} F_{n,m,p}(I) F_{n,-m',p'}(I') U_{n,m,-m'}(\varepsilon)
          \exp\ci \Theta_{n,m,-m',p,p',q+s,q'}
        \right. \\ {} & {} \left. \hspace{2em} 
          + {} \frac{(n-m')!}{(n-m)!} \overline\varepsilon_{n,-m,m',p,p',q'}^{\sigma} F_{n,-m,p}(I) F_{n,m',p'}(I') U_{n,-m,m'}(\varepsilon)
          \exp\ci \Theta_{n,-m,m',p,p',q+s,q'}
        \right. \\ {} & {} \left. \hspace{2em} 
          + {} \frac{(n+m')!}{(n-m)!} \overline\varepsilon_{n,-m,-m',p,p',q'}^{\sigma} F_{n,-m,p}(I) F_{n,-m',p'}(I') U_{n,-m,-m'}(\varepsilon)
          \exp\ci \Theta_{n,-m,-m',p,p',q+s,q'}
        \right]
      \end{split}
    \end{scriptsize}
  \end{equation}
  Focus now our attention on step~$2$, and particularly on the coefficient~$(q+\delta_{0}^{q})$. 
  As~$\mca{V}_{}^{(\sigma)}$ is formulated so that it can automatically handle cases for which~$ q = 0~$ and~$ q \neq 0$, we must to slightly modify this element if we want to effectively use the symmetry relations after changing~$q$ by~$-q$ and to keep a compact form.

  Make this change for~$q \ne 0$ is not a problem and the coefficient can be rewritten in the form~$-(q+\delta_{0}^{q})$. 
  However, this trick for~$q=0$ can not work because we would get the value~$-1$, while the expected value is~$1$.
  To restore the correct sign, we make appear the factor~$(1-2\delta_{0}^{q})$, without consequence on the final result. In fact, this factor was not choose by chance. This will be offset with the factor related to the~$zeta$-functions~\eqref{eq:Zeta_sym_a}.

  To sum up, when we apply the change of indice~$q$ by~$-q$ on the relevant members of~\eqref{eq:Gen_CP_moon_etape1}, we also need to make the following substitution:%
  \begin{subequations}
    \begin{empheq}[left={%
        \dfrac{1}{q+\delta_{0}^{q}} \to 
        -\dfrac{1-2\delta_{0}^{q}}{q+\delta_{0}^{q}} = 
      }\empheqlbrace]{alignat=1}
      \label{eq:148} 
      1 & \quad , \quad q=0
      \nonumber
      \\
      -\frac{1}{q} & \quad ,  \quad q \ne 0
      \nonumber
    \end{empheq}
  \end{subequations}
  and we find at step~$2$:%
  \begin{equation}
    \label{eq:Gen_CP_moon_etape2}
    \begin{fssizeadapt}
      \begin{split}
        & \mca{V}_{}^{(\sigma)} = {}
        - (-1)^{\sigma} \frac{\mu'}{a'}
        \sideset{}{}\sum_{}^{\circ}
        \frac{ \Delta_{0}^{m,m'} }{2 \ci} \frac{(-1)^{m-m'}}{(q+\delta_{0}^{q}) \omega_{0}} \fracp*{a}{a'}^{n}
        \\ 
        {} & {} \hspace{1em}  \times
        \sum_{s=-(\sigma+1)}^{\sigma+1}
        \left[
          \frac{(n-m')!}{(n+m)!} \overline\varepsilon_{n,m,m',p,p',q'}^{\sigma} \zeta_{q,s}^{(\sigma)}(e) Z_{q}^{n+1,n-2p}(e) X_{q'}^{-(n+1),n-2p'} (e') 
        \right. \\ {} & {} \left.
          \hspace{2em} \times
          F_{n,m,p}(I) F_{n,m',p'}(I') U_{n,m,m'}(\varepsilon)
          \exp\ci \Theta_{n,m,m',p,p',q+s,q'}
        \right. \\ {} & {} \left.
          \hspace{1em} + {}
          \frac{(n+m')!}{(n+m)!} \overline\varepsilon_{n,m,-m',p,-p',-q'}^{\sigma} \zeta_{q,s}^{(\sigma)}(e) Z_{q}^{n+1,n-2p}(e) X_{-q'}^{-(n+1),-(n-2p')} (e')
        \right. \\ {} & {} \left.
          \hspace{2em} \times
          F_{n,m,p}(I) F_{n,-m',n-p'}(I') U_{n,m,-m'}(\varepsilon)
          \exp\ci \Theta_{n,m,-m',p,n-p',q+s,-q'}
        \right. \\ {} & {} \left.
          \hspace{1em} {}
          -(1-2\delta_{0}^{q}) \frac{(n-m')!}{(n-m)!} \overline\varepsilon_{n,-m,m',n-p,p',q'}^{\sigma} \zeta_{-q,-s}^{(\sigma)}(e) Z_{-q}^{n+1,-(n-2p)}(e) X_{q'}^{-(n+1),n-2p'} (e') 
        \right. \\ {} & {} \left.
          \hspace{2em} \times
          F_{n,-m,n-p}(I) F_{n,m',p'}(I') U_{n,-m,m'}(\varepsilon)
          \exp\ci \Theta_{n,-m,m',n-p,p',-(q+s),q'}
        \right. \\ {} & {} \left.
          \hspace{1em} {}
          -(1-2\delta_{0}^{q}) \frac{(n+m')!}{(n-m)!} \overline\varepsilon_{n,-m,-m',n-p,n-p',-q'}^{\sigma} \zeta_{-q,-s}^{(\sigma)}(e) Z_{-q}^{n+1,-(n-2p)}(e) X_{-q'}^{-(n+1),-(n-2p')} (e') 
        \right. \\ {} & {} \left.
          \hspace{2em} \times
          F_{n,-m,n-p}(I) F_{n,-m',n-p'}(I') U_{n,-m,-m'}(\varepsilon)
          \exp\ci \Theta_{n,-m,-m',n-p,n-p',-(q+s),-q'}
        \right] \;.
      \end{split}
    \end{fssizeadapt}
  \end{equation}
  Then, performing step~$3$ to~$6$ we get%
  \begin{widetext}
    \begin{equation}
      \label{eq:Gen_CP_moon_etape4}
      \begin{split}
        \mca{V}_{}^{(\sigma)} = {} & 
        - (-1)^{\sigma}
        \sideset{}{}\sum_{}^{\circ}
        \frac{ \Delta_{0}^{m,m'} }{2 \ci} \frac{\mca{A}_{n,m,m',p,p',q,q'}}{(q+\delta_{0}^{q}) \omega_{0}} 
        \sum_{s=-(\sigma+1)}^{\sigma+1} \zeta_{q,s}^{(\sigma)}(e)
        \\ 
        {} & {} \hspace{1em} \times
        \left\lbrace
          \vphantom{(-1)^{n-m'}}
          \overline\varepsilon_{}^{\,-,\sigma} U_{n,m,m'}(\varepsilon) 
          \left[
            \exp \ci \Theta_{}^{-}
            - \exp \left( -\ci \Theta_{}^{-}\right)
          \right]
        \right. \\ {} & {} \left.
          \hspace{2em}  {}
          + (-1)^{n-m'} \overline\varepsilon_{}^{\,+,\sigma} U_{n,m,-m'}(\varepsilon)
          \left[
            \exp \ci \Theta_{}^{+}
            -\exp \left( -\ci \Theta_{}^{+}\right)
          \right]
        \right\rbrace
      \end{split}
    \end{equation}
  \end{widetext}
  which is equivalent to~\eqref{eq:Gen_CP_moon_trig}.

\section{Data}
\label{Anx:Data}

\begin{table}[h!] 
  
  \sisetup{
    table-number-alignment = left,
    table-figures-integer = 1, 
    table-figures-decimal = 0,
    table-sign-mantissa,
    table-sign-exponent,
    exponent-product =  \times, 
    output-exponent-marker 
  } 
  
  \captionsetup{width=1\textwidth}  
  \centering 
  
  \begin{footnotesize}
    
    \begin{tabular}{
        *{1}{K{5em}}
        *{1}{S[table-format = -11.15e2]}
      } 

      \toprule 
      
      \multicolumn{1}{c}{\textrm{Symbol}} & \multicolumn{1}{c}{Earth}\\ 
      
      \cmidrule(rl){1-1}
      \cmidrule(rl){2-2}
      
      \mu [\ctegravm] 	
      & 398600.44150E9
      \\
      R_{\varoplus} [m] 	
      & 6378136.460E0
      \\
      \Jd
      & .10826264572318E-02
      \\

      \bottomrule 
    \end{tabular}
  \end{footnotesize}
  \caption{Parameters of the Earth from the gravity field model Eigen-5C.}
  \label{tab:earth_para}
\end{table}

\begin{table}[h!] 
  
  \sisetup{
    table-number-alignment = left,
    table-figures-integer = 1, 
    table-figures-decimal = 0,
    table-sign-mantissa,
    table-sign-exponent,
    exponent-product =  \times, 
    output-exponent-marker 
  } 
  
  \captionsetup{width=1\textwidth}  
  \centering 
  
  \begin{footnotesize}
    
    \begin{tabular}{
        *{1}{K{5em}}
        *{1}{S[table-format = -6.12e2]}
        *{1}{S[table-format = -11.15e2]}
      } 

      \toprule 
      
      \multicolumn{1}{c}{\textrm{Symbol}} & \multicolumn{1}{c}{Moon}& \multicolumn{1}{c}{Sun}\\ 
      
      \cmidrule(rl){1-1}
      \cmidrule(rl){2-2}
      \cmidrule(rl){3-3}
      
      \mu [\ctegravm] 	
      & 4902.801076E9
      & 132712442099.0e9
      \\

      \cmidrule(rl){1-1}
      \cmidrule(rl){2-2}
      \cmidrule(rl){3-3}

      a_{0} [\si{\metre}]
      & 383397.0E3 
      & 149598140.0E3
      \\
      e_{0}
      & 0.05556452E0
      & 0.016715e0
      \\
      I_{0} [\si{\degree}]
      & 5.15665e0 
      & 23.4393e0
      \\ 

      \cmidrule(rl){1-1}
      \cmidrule(rl){2-2}
      \cmidrule(rl){3-3}

      \Omega_{0} [\si{\degree}]
      & 125.04455501e0
      & \textrm{Not defined}
      \\
      \omega_{0} [\si{\degree}]
      & 83.35324312e0
      & 282.937340e0
      \\
      M_{0} [\si{\degree}]
      & 134.96340251e0
      & 357.52910918e0
      \\

      \cmidrule(rl){1-1}
      \cmidrule(rl){2-2}
      \cmidrule(rl){3-3}

      \dot\Omega_{} [\radpersec]
      & -0.106969620630E-07
      & \textrm{Not defined}
      \\
      \dot\omega_{} [\radpersec]
      & 0.332011088218E-07
      &  0.951001308674908D-11
      \\
      \dot{M}_{} [\radpersec]
      & 0.263920305313E-05
      & 0.199096875237661D-06
      \\
      \bottomrule 
    \end{tabular}
  \end{footnotesize}
  \caption{Orbital elements for modeling the Moon's (resp. Sun) apparent motion about the Earth are given with respect to the ecliptic plane (resp. equatorial plane) and refered from the epoch~$J2000.0$ \citep[see][section 3.5 case (b.3) and section 3.6 case (a)]{Simon_1994aa}.}
  \label{tab:lunisol_para}
\end{table}

\end{appendices}

\end{document}